\documentclass[a4paper,11pt]{article}

\usepackage{subcaption}
\usepackage{standalone}
\usepackage{pgfplots}
\pgfplotsset{compat=1.16}

\usepackage{jheppub} %
\usepackage{derivative}
\DeclareDifferential{\Dd}{\mathrm{D}}[style-notation=multiple, style-notation-*=single
]

\usepackage{amssymb}
\usepackage{slashed}

\usepackage{iftex}
\ifLuaTeX
\usepackage{fontspec} %
\else
\usepackage[utf8]{inputenc} %
\DeclareUnicodeCharacter{2212}{-} %
\usepackage[LGR,T1]{fontenc} %
\usepackage{textgreek}
\fi

\usepackage[utf8]{inputenc} %
\DeclareUnicodeCharacter{2212}{-} %
\usepackage[LGR,T1]{fontenc} %
\usepackage{textgreek}

\usepackage{lineno}
\usepackage{csquotes}
\usepackage{float}
\usepackage{xcolor}
\usepackage{cancel}
\usepackage{bm}
\usepackage[shortlabels]{enumitem} %
\usepackage{listings} %
\usepackage{mathtools}
\usepackage{cleveref}
\usepackage{ytableau}
\usepackage{comment}
\usepackage{amstext} %
\usepackage{nicematrix}
\usepackage{xspace} %

\newcommand{\dotwo}{\tfrac{d}{2}}

\newcommand{\I}{{\rm I}}

\def\det{\mathop{\rm det}\nolimits}

\def\Tr{\mathop{\rm Tr}\nolimits}
\def\bbra{{\langle\kern-2.5pt\langle}}
\def\kket{{\rangle\kern-2.5pt\rangle}}
\def\Bbra{{\Big\langle\kern-3.5pt\Big\langle}}
\def\Kket{{\Big\rangle\kern-3.5pt\Big\rangle}}

\DeclarePairedDelimiter\abs{\lvert}{\rvert}

\DeclarePairedDelimiterX\braket[2]{\langle}{\rangle}{#1\,\delimsize\vert\,\mathopen{}#2}
\DeclarePairedDelimiter\expval{\langle}{\rangle}
\DeclarePairedDelimiter\floor{\lfloor}{\rfloor}

\newcommand   \half{\frac 1 2}
\newcommand   \lptl{\raise .8ex\hbox{$^\leftarrow$} \hspace{-9pt} \partial}
\newcommand   \rptl{\raise .8ex\hbox{$^\rightarrow$} \hspace{-9pt} \partial}
\newcommand   \lrptl{\raise .8ex\hbox{$^\leftrightarrow$} \hspace{-9pt} \partial}

\newcommand   \SO    {\mathrm{SO}}
\newcommand   \gO    {\mathrm{O}}

\DeclareMathOperator{\arccosh}{arccosh}

\newcommand   \cC {\mathcal{C}}
\newcommand   \cD {\mathcal{D}}

\newcommand   \cF {\mathcal{F}}

\newcommand   \cL {\mathcal{L}}
\newcommand   \cM {\mathcal{M}}

\newcommand   \cO {\mathcal{O}}
\newcommand   \cP {\mathcal{P}}

\newcommand   \cR {\mathcal{R}}

\newcommand{\be}{
  \begin{equation}
  \begin{aligned}
}
\newcommand{\ee}{
  \end{aligned}
  \end{equation}
}

\makeatletter
\DeclareRobustCommand\bea{\@ifnextchar[{\@@bea}{\@bea}}
\def\@@bea[#1]#2\eea{\begin{subequations}\begin{align}#2\end{align}\label{#1}\end{subequations}}
\def\@bea#1\eea{\begin{subequations}\begin{align}#1\end{align}\end{subequations}}
\makeatother

\newcommand\Pochhammer[2]{{\left(#1\right)_{#2}}}

\usepackage{tikz}
\usepackage{silence} %
\usepackage{array}   %
\newcolumntype{L}{>{$}l<{$}} %
\makeatletter
\edef\savedcodes{\catcode`\noexpand\_=\the\catcode`\_}
\edef\@tempa{\csname opt@newtxmath.sty\endcsname}
\def\@tempb{{subscriptcorrection}}
\expandafter\expandafter\expandafter
  \in@\expandafter\@tempb\expandafter{\@tempa}
\ifin@
  \catcode`\_=12 %
\fi
\makeatother
\usepackage{tensor}
\savedcodes

\title{The $T^{\mu\nu}$ of the conformal scalars} %

\author[\Delta]{(Kit|Ludo) Fraser-Taliente}%

\affiliation[\Delta]{Rudolf Peierls Centre for Theoretical Physics, University of Oxford, Oxford, OX1 3PU, UK}

\addTwoExplicitEmails{\tt (\href{mailto:cristofero.fraser-taliente@physics.ox.ac.uk}{cristofero}|\href{mailto:ludovic.fraser-taliente@physics.ox.ac.uk}{ludovic}).fraser-taliente@physics.ox.ac.uk}

\abstract{%
We construct the unique primary energy-momentum tensor $T^{\mu\nu}$ for the conformal free scalar with scaling dimension $\Delta=\tfrac{d}{2}-\zeta$ as a sum of Gegenbauer polynomials. 
For integer $\zeta$, the sum truncates at order $\zeta$, compactly reproducing all known results; for the nonlocal case of real $\zeta$, it is an infinite sum, with a two-parameter extension that reflects the nonuniqueness of the nonlocal geometric coupling.
We find $T^{\mu\nu}$ by imposing off-shell conservation and tracelessness, and then directly solving the primary condition in momentum space.
In the integer $\zeta$ case, we reproduce the known two-point function, and confirm the match with the $T^{\mu\nu}$ computed from Juhl's formulae for the GJMS operators (the Weyl-covariant upgrades of $(-\partial^2)^\zeta$), an equality following from the descent of Weyl covariance to conformal invariance.
}


\newcommand{\thalf}{\tfrac{1}{2}}
{%
}%

\newcommand{\TtextOrPDF}{\texorpdfstring{$T$}{T}\xspace}

\usepackage{empheq} %

\pgfmathsetmacro\MathAxis{height("$\vcenter{}$")} %

\begin{document}
\maketitle
\flushbottom

\newcommand{\di}{\chi} %
\newcommand{\gflat}{\delta}
\newcommand{\WeylOp}{{\Delta}}
\newcommand{\mg}{M}

\newpage

\section{Summary}

Perhaps the simplest family of CFTs is that of the conformal scalars of dimension $\Delta\equiv \dotwo - \zeta$,
\begin{equation}
    \expval{\phi(x) \phi(y)} = \frac{C_\phi}{\abs{x-y}^{2\left(\tfrac{d}{2}-\zeta\right)}},
\end{equation}
with all higher-point functions determined by Wick contractions.
For positive integer $\zeta$, these are the free higher-derivative CFTs with flat-space Euclidean action $S=\int_x \half \phi (-\partial^2)^\zeta \phi$, and so equation of motion $(-\partial^2)^\zeta \phi =0$.
Being CFTs, each theory must possess an important operator: the unique conserved and traceless energy-momentum (EM) tensor $T^{\mu\nu}$. 
We like this operator for two reasons: first, when contracted with the conformal Killing vectors it yields the Noether currents for the conformal symmetries; second, its correlators are interesting observables.
An explicit construction, however, has been lacking.

\makeatletter
\newcommand{\propitem}[2]{%
  \item[#1]%
  \protected@edef\@currentlabel{#1}%
  \label{#2}%
}
\makeatother
In this paper, we construct for arbitrary integer $\zeta$ the unique operator $T^{\mu\nu}(x)$ that is
\begin{enumerate}[label=P.\arabic*,itemsep=0.1em,leftmargin=2cm]
    \propitem{P\textsubscript{sy}}{prop1} symmetric, $T^{[\mu\nu]} = 0$; 
    \propitem{P\textsubscript{co}}{prop2} conserved, $\partial_\mu T^{\mu\nu} = -(\partial^\nu \phi) (-\partial^2)^\zeta \phi \overset{\text{eom}}{=} 0$;
    \propitem{P\textsubscript{tr}}{prop3} traceless, $\gflat_{\mu\nu} T^{\mu\nu} =-(\dotwo -\zeta) \phi (-\partial^2)^\zeta \phi \overset{\text{eom}}{=} 0$; 
    \propitem{P\textsubscript{pr}}{prop4} a global conformal primary, $\hat{K}_\rho T^{\mu\nu}|_{x=0}=0$.
\end{enumerate}
By $\overset{\text{eom}}{=}$ we mean an identity that holds inside correlation functions -- i.e. when using the equation of motion (\enquote{on shell}).
We further constrain $\partial_\mu T^{\mu\nu}$ and $\gflat_{\mu\nu}T^{\mu\nu}$ to take the off-shell forms given in \ref{prop2} and \ref{prop3}, thereby fixing the trivial ambiguities of normalization and mixing of $T^{\mu\nu}$ with the \enquote{zero operator} $\gflat^{\mu\nu}\phi (-\partial^2)^\zeta \phi \overset{\text{eom}}{=} 0$. 
Those particular forms are chosen to match the Weyl-covariant construction below.

The first three properties are standard for a CFT EM tensor; requiring it to also be primary makes it unique, and so we call it \textit{the} EM tensor.  
Let us present it immediately:
\begin{subequations}
\begin{align} \label{eq:fullT}
     T^{\mu\nu} &= -\left(\frac{\gflat^{\mu\nu}}{2}-\zeta \frac{\cP^{\mu\nu}}{-\partial^2}\right) \cO + \left(\gflat^{\mu(\rho}\gflat^{\sigma)\nu}-\frac{\cP^{\mu\nu}}{-\partial^2}\gflat^{\rho\sigma}\right) U_{\rho\sigma} + \frac{\cD_{\mathrm{TT}}^{\mu\nu\rho\sigma}}{(-\partial^2)^2} R_{\rho\sigma},
\end{align}
where $\cP^{\mu\nu} =(\partial^\mu \partial^\nu-\gflat^{\mu\nu} \partial^2)/(d-1)\sim \partial^2$ %
and $\cD_{\mathrm{TT}}^{(\mu\nu)(\rho\sigma)} \overset{\eqref{eq:DQdef}}{\sim} \partial^4$ are known operators.
They have unit and vanishing trace respectively, and are transverse:
\begin{equation}
   \gflat_{\mu\nu} \frac{\cP^{\mu\nu}}{-\partial^2}= 1, \quad \gflat_{\mu\nu} \cD_{\mathrm{TT}}^{\mu\nu\rho\sigma} = 0, \quad \partial_\mu \cP^{\mu\nu} = 0, \quad \partial_\mu \cD_{\mathrm{TT}}^{\mu\nu\rho\sigma} = 0.
\end{equation}
These identities mean that only $\cO$ and $U$ are required to ensure properties (\ref{prop1}, \ref{prop2}, \ref{prop3}); $R$ is an \enquote{improvement} term that makes $T^{\mu\nu}$ primary (and, as it happens, local, for integer $\zeta$).
Defining $\lptl/\rptl$ to act on the left/right $\phi$ respectively, the operators $\cO$, $U$ and $R$ are
\begin{align}
        \cO &= \phi (-\partial^2)^\zeta \phi\\
        U_{\rho\sigma} &\equiv -\phi \frac{\lrptl_\rho \lrptl_\sigma}{4}  \frac{(-\lptl^2)^{\zeta }-(-\rptl^2)^{\zeta} }{(-\lptl^2)-(-\rptl^2)} \phi\\
        R_{\rho\sigma} &\equiv \sum_{k=2}^\infty \frac{k-1}{k-1-\dotwo} (-\partial^2)^{k-1}\left[ \phi\frac{\lrptl_\rho \lrptl_\sigma}{4}  \mg^{\zeta-k} C_{\zeta-k}^{(k)}(\cos\theta)\phi\right] \label{eq:Rdef}
\end{align}
\end{subequations}%
for $\lrptl_\rho \equiv \lptl_\rho-\rptl_\rho$, $\mg=\sqrt{\lptl^2 \rptl^2}$, $\cos\theta = \frac{\lptl \cdot \rptl}{\mg}$, and $C_{\zeta-k}^{(k)}$ the Gegenbauer polynomial of degree $\zeta-k$; this ensures that 
$\mg^{\zeta-k} C_{\zeta-k}^{(k)}(\cos\theta)$ is a polynomial of degree $\zeta-k$ in $\lptl^2 \rptl^2$ and $\lptl \cdot \rptl$. 

In this paper, we motivate and prove our construction, and then demonstrate a few consequences in the case of integer $\zeta$, which we now summarise:
\begin{enumerate}
    \item The $R_{\rho\sigma}$ sum truncates at $k=\zeta$, as $C_{-1,-2,\cdots}^{(k)} = 0$. 
    Then the unique $T$ is local -- where a \textit{local} operator is defined as one expressible as a finite polynomial in fields and their derivatives. 
    \textit{This means that all the inverse Laplacians in \eqref{eq:fullT} cancel}.
\item The poles in \eqref{eq:Rdef} manifestly show that for $d=2,4,6,\cdots,2\zeta-2$, it is not possible to simultaneously satisfy the four properties (\ref{prop1}--\ref{prop4}); equivalently, in those dimensions, it is not possible to make the CFT of a scalar of dimension $\dotwo -\zeta$ into a Weyl-covariant theory, as we will discuss in \cref{sec:WeylConf}.
\item The two-point function's normalization $C_T$ is fixed to be
\begin{equation}
    \expval{T^{\mu\nu}(x) T^{\rho\sigma}(0)} = \frac{C_T}{S_d^2} \frac{1}{(x^2)^d} I^{\mu\nu, \rho\sigma}(x), \quad C_T =  \zeta \frac{\Pochhammer{\dotwo +2}{\zeta-1}}{\Pochhammer{1-\dotwo}{\zeta-1}} \frac{d}{d-1},
\end{equation}
where $S_d \equiv \frac{2 \pi^{d/2}}{\Gamma(d/2)}$ is the surface area of the $(d-1)$-dimensional unit sphere.
We also explicitly confirm the value \eqref{eq:CTphiphi} of the OPE coefficient $C_{T\phi\phi} \propto \Delta$ fixed by the conformal Ward identities.
\item This $T^{\mu\nu}$ is the same as the \enquote{metrical} energy-momentum tensor: the one found by varying the action $S=\half \int_x \phi \WeylOp_{\zeta} \phi$ with respect to $g_{\mu\nu}$, where $\WeylOp_{\zeta}$ is the GJMS operator, the Weyl-covariant upgrade of $(-\partial^2)^\zeta$.
This must be true because Weyl covariance descends to conformal invariance: a Weyl-covariant theory restricted to flat space should be a CFT -- and the covariant operator in such a theory giving the response to metric variations should become the CFT's primary $T^{\mu\nu}$.
\end{enumerate}

This formula \eqref{eq:fullT} matches the known computations \cite{Stergiou:2022qqj} in all cases (despite it being in Lorentzian mostly-plus signature), including the standard result \cite[(39)]{Stergiou:2022qqj} for the canonical free scalar,
\begin{align}\label{eq:TForzetaOne}
T^{\mu\nu}_{\zeta=1} &= \partial^\mu \phi \partial^\nu \phi - \half \gflat^{\mu\nu} (\partial \phi)^2 - \frac{d-2}{4}\cP^{\mu\nu} \phi^2.
\end{align}

One fruitful thread in physics has been analytically continuing integer-valued data to take continuous values, as in the Wilson-Fisher $4-\epsilon$ expansion and the large-$N$ expansions.
By taking arbitrary $\zeta\in \mathbb{R}$, we can define a continuous family of nonlocal CFTs, which will also be nonunitary for $\zeta>1$. 
For noninteger $\zeta$, there still must exist a Noether current for spacetime translations, but it is not unique because the nonlocal Weyl-covariant action is not unique.
Our computations straightforwardly extend to the nonlocal case, and we find a two-parameter extension of \eqref{eq:Rdef} involving associated Legendre functions, $R_{\rho\sigma}^{\text{ext}}$ \eqref{eq:RdefHom}.
Each possible $T^{\mu\nu}$ is a nonlocal operator, in a way that we shall make precise in \cref{sec:nonlocal}.

\subsection{A sketch of the strategy and the paper}\label{sec:sketchStrategy}

We construct our $T^{\mu\nu}$ by explicitly solving the four conditions \ref{prop1}--\ref{prop4}.
This is easiest in momentum space, where the conditions (\ref{prop1},\ref{prop2},\ref{prop3}) are straightforward to satisfy; this fixes $\cO$ and $U_{\rho\sigma}$, and the prefactor $\cD_{\mathrm{TT}}$ of $R_{\rho\sigma}$.
The remaining freedom is in the symmetric tensor $R_{\rho\sigma}$ itself -- it is fixed only up to an arbitrary scalar function $Y$, $R_{\rho\sigma} \sim \phi \lrptl_\rho \lrptl_\sigma Y(-\lptl^2,-\rptl^2, -\lptl\cdot\rptl)\phi$.
Imposing the primary condition $\hat{K}_\rho T_{\mu\nu}(x=0) = 0$ and requiring locality for integer $\zeta$ then fixes $R_{\rho\sigma}$ to \eqref{eq:Rdef}.
To do this, we find expressions for the action of $\hat{K}_\rho$ on arbitrary expressions of the form
\begin{equation}
    \phi A_{\mu\nu} f(-\lptl^2, -\lptl \cdot \rptl, -\rptl^2) \phi, \quad A_{\mu\nu} \in \{\gflat_{\mu\nu}, -\lptl_\mu \lptl_\nu, -\rptl_\mu \rptl_\nu,- \lptl_{(\mu} \rptl_{\nu)}\}
\end{equation}
as differential operators acting on the functions $f$.
Computing $\hat{K}_\rho T_{\mu\nu}=0$ for our candidate $T$ yields a collection of explicit inhomogeneous partial differential equations for $Y$ (a function of three scalar variables), which we solve using Gegenbauer polynomials to find \eqref{eq:Rdef}.
On a heuristic level, the Gegenbauer polynomials arise for the following reason. 
The variables that we have access to (via derivatives) are the momenta of the two $\phi$ fields.
The natural variables for the primary condition are the angle between them and their magnitudes; as we discuss in \cref{sec:whereGegenbauersArise}, the Gegenbauer polynomials translate between these two bases.
This procedure completely fixes the form of $T^{\mu\nu}$ for integer $\zeta$. %

The structure of this paper is as follows.
\begin{enumerate}[start=2]
    \item In \cref{sec:confScalars} we review standard facts about conformal scalars, nonlocality, and the distinction between Weyl and conformal invariance. This allows us to justify our four properties.
    \item In \cref{sec:construction} we construct the non-primary $T^{\mu\nu}$.
    \item In \cref{sec:primaryconstruction} we make it primary using the procedure just described.
    \item In \cref{sec:properties} we verify this $T^{\mu\nu}$'s properties: its locality, its two-point function $\expval{T^{\mu\nu} T^{\rho\sigma}}$, and its three-point function $\expval{T^{\mu\nu} \phi\phi}$, which gives the OPE coefficient for integer $\zeta$.
    \item In \cref{sec:Juhl} we review the known construction of the Weyl-covariant upgrade of $(-\partial^2)^\zeta$, the GJMS operators $\WeylOp_\zeta$, and verify that their linearized variation gives $T^{\mu\nu}$.
\end{enumerate}
Various useful results are packaged in the appendices:
\cref{app:conventions} contains a review of our conventions;
\cref{sec:separatedPoints} discusses the normal ordering that is otherwise kept implicit throughout, with the aim of justifying the use above of the equation of motion in correlators with fields apparently at coincident points (i.e. $\expval{\text{``}{\phi(x)(-\partial^2)^\zeta \phi(x)}\text{''}} = \expval{:\phi(x)(-\partial^2)^\zeta \phi(x):}=0$);
\cref{app:Gegenbauer} contains a review of Gegenbauer polynomials;
\cref{app:Kbhat} contains explicit formulae for the action of the special conformal transformations $\hat{K}_\rho$ on $\phi$ bilinears;
\cref{app:OSform} contains the explicit Osborn-Vacca-Zanusso form of our tensor;
\cref{app:GJMSidentities} inductively proves various identities used in \cref{sec:Juhl};
\cref{app:code} gives \lstinline|Mathematica| code for our $T^{\mu\nu}$.

We now give our conclusion and discussion.

\subsubsection{Conclusion and discussion}

In this paper we computed the energy-momentum tensor for a free scalar of an arbitrary scaling dimension, where it exists, and discussed how it descends from an explicitly Weyl-covariant field theory. 

This $T^{\mu\nu}$ can now be used as a starting point for perturbation theory, where one perturbs around solvable theories -- these are often generalized free field theories (GFFs) \cite{Fraser-Taliente:2024hzv}.
For example, we can now construct the EM tensors of the interacting $\Box^k$ $\gO(N)$ theories \cite{Gliozzi:2016ysv, Gracey:2017erc,Safari:2017tgs,Safari:2017irw} in the $d=4k-\epsilon$ expansion: %
this would generalize \cite[(5.14),\S B]{Cappelli:1990yc} for the canonical $k=1$ Wilson-Fisher theory. 

More subtly, some ways of solving for the data of local CFTs in perturbation theory require perturbing around these nonlocal CFTs -- particularly in the large-$N$ expansion.
For example, in the large-$N$ $\gO(N)$ model, the DREG+$\delta$ regulator is particularly useful.
With this regulator, the UV behaviour of the propagator is improved by shifting the scaling dimension of the auxiliary field $\sigma$ by $\delta$.
This shift can only be implemented consistently at the level of the action, which is done by adding nonlocal kinetic terms $K_\delta^{-1} \propto (-\partial^2)^\zeta$ with $\zeta=\dotwo-2+\delta$ \cite[\S 3.13]{Fraser-Taliente:2025udk},
\begin{equation}
    S_{\gO(N),\delta} \equiv \int_x \frac{Z_\phi}{2} \phi^i \left(-\partial^2 -\frac{Z_g g \mu^{\delta}}{\sqrt{N}} \sigma \right) \phi^i + \half \int_x \sigma (K^{-1}_{\delta} - \mu^{2\delta} K^{-1}_{\delta=0})\sigma.
\end{equation}
Accordingly, the EM tensor becomes nonlocal for finite $\delta$,
\begin{equation}
T^{\mu\nu}_{\gO(N)} = \sum_i Z_\phi T^{\mu\nu}_{\phi_i,\zeta=1} +(T^{\mu\nu}_{\sigma,\Delta_\sigma = 2-\delta}-\mu^{2\delta} T^{\mu\nu}_{\sigma,\Delta_\sigma = 2}) - \gflat^{\mu\nu}  \half \frac{Z_\phi Z_g \mu^\delta}{\sqrt{N}} \phi^i \phi^i \sigma, 
\end{equation}
though we know that it must be local in the limit $\delta\to 0$.
Careful treatment of these terms resolves \cite{fraser-talientePreparation2026} the issue of an apparent \enquote{$Z_T$ renormalization of the stress tensor} in the large-$N$ CFT literature \cite{Diab:2016spb,Giombi:2016fct,Tarnopolskiy:2017zgz,Diab:2017azu}, which was thought to be necessary to match the computations of \cite{Petkou:1995vu}.

A natural next step is an extension of our construction to fermions \cite[\S VI]{Stergiou:2022qqj}.
Doing so corresponds to constructing the linearized variation of the fractional Dirac operators, most recently discussed in \cite{maalaouiConformalFractionalDirac2025}.
Likewise, an extension to conformal vector fields would be desirable: some initial steps towards this were presented in \cite[\S 3.3]{Heydeman:2020ijz}.
The immediate motivation for both is to confirm that in the conformal Gross-Neveu and QED theories the \enquote{$Z_T$ renormalization} is unnecessary.
Indeed, in any large-$N$ or supersymmetric field theories \cite{Fraser-Taliente:2024hzv} where we expand around a generalized free field, we ought to be careful to include these terms if we are regulating our quantum corrections using an analytic regularization.
All of this will be discussed further in \cite{fraser-talientePreparation2026}.

Another use of our construction would be a direct proof of the conformal invariance of the long-range theories, by proving the Ward identities as discussed in \cite[\S 6.3]{Paulos:2015jfa} (this was done in the case $d=2$ in \cite[\S 5.4]{Basa:2020cyn} and \cite{Basa:2024gzj}).
The toolkit here could also be used to explicitly construct the operators dual to the higher-spin tower in the bulk dual of $\gO(N)$ theory.
We also checked consistency of our integer-$\zeta$ EM tensor with that obtained by varying the local GJMS operators, which are found via Juhl's formulae.
It is possible that our explicit construction of the flat-space $T^{\mu\nu}$ could suggest a way to write down an explicit formula for the Weyl-covariant $\WeylOp_\zeta$; that in turn should reduce to \eqref{eq:EinsteinGJMS} in the limit of an Einstein manifold.
Likewise, it could be relevant for any explicit construction of the fractional-dimension GJMS operators (\cite{caseGJMSOperatorsGeometry2025}: \cite{changFractionalLaplacianConformal2010,caseFractionalGJMSOperators2014,Gonzalez:2016kjr}). 
We note that there is no canonical definition of such an operator for spacetimes that are not $\mathbb{R}^d$ or $S^d$ \cite[\S 7]{Gonzalez:2016kjr}; this was reflected in the additional ambiguity of $T^{\mu\nu}$ that we found in the nonlocal cases. 
It would be nice to understand: first, this two-parameter freedom in the nonlocal $T^{\mu\nu}$ from a geometric perspective; and second, whether the analytic continuation of the integer-$\zeta$ case (i.e. setting $C_1=0=C_2$), which appears to be the natural and minimal solution, is special in any way.

\acknowledgments

(K|L)FT is supported by the (Gould Watson|Dalitz) Scholarship from the University of Oxford and (Lady Margaret Hall|Wadham College).
We thank Petr Kravchuk, Grisha Tarnopolsky, John Wheater, and Bernardo Zan for illuminating discussions.

For the purpose of open access, the authors have applied a CC BY public copyright licence to any Author Accepted Manuscript (AAM) version arising from this submission. %

\section{Conformal scalars, nonlocality, and Weyl-covariant theories} \label{sec:confScalars}

In this section we review conformal scalars on flat space and their extension to Weyl-covariant field theories. 
We also discuss locality and nonlocality, and justify the four properties we demanded of $T^{\mu\nu}$.

Not only are the conformal scalars interesting in their own right as possibly the simplest continuous family of CFTs, but they also arise generically at the leading order of large-$N$ theories \cite{Fraser-Taliente:2025udk,Fraser-Taliente:2024hzv}, as well as in the effective actions of mixed-dimensional theories \cite{Fraser-Taliente:2024lea}.
We study the free theory here, though the interacting theories have been studied for some time \cite{Honkonen:1990mr} \cite{Paulos:2015jfa,Behan:2017dwr,Behan:2017emf,Gliozzi:2017hni,Behan:2018hfx,Slade:2016yer,Kazakov:2018qbr,Behan:2025ydd,Benedetti:2020rrq,Benedetti:2024mqx,Chai:2021arp,Safari:2021ocb,Benedetti:2021wzt,Giombi:2022gjj,Harribey:2022esw,Behan:2023ile,Shen:2023srk,Li:2024uac,Benedetti:2024wgx,Giombi:2024zrt,Herzog:2024zxm,Fraser-Taliente:2025udk,Fraser-Taliente:2024rql,Bianchi:2024eqm,Benedetti:2025nzp,vanVliet:2025swv,Guo:2025edk}.
They have scaling dimension $\Delta\equiv \dotwo-\zeta$, and so flat-space correlators
\begin{equation}
    \expval{\phi(p) \phi(q)} = \frac{1}{p^{2\zeta}} \delta(p+q), \quad \expval{\phi(x) \phi(y)} = \frac{C_\phi}{\abs{x-y}^{2(\dotwo-\zeta)}},
\end{equation}
where %
$C_\phi\equiv 1/\cF_{d-2\zeta,0} = \frac{\Gamma \left(\frac{d}{2}-\zeta \right)}{\pi ^{d/2} 2^{2 \zeta } \Gamma (\zeta )}$. 
Being a free theory, all higher correlators can be obtained by Wick contractions (for operators $\phi^n$ with coincident $\phi$s, we must normal order, $:\phi^n:$, such that in correlators we can omit internal contractions).
The flat-space Euclidean action is
\begin{equation}
    S= + \half \int_x \, \phi(x) (-\partial^2)^\zeta \phi(x),
\end{equation}
up to total derivatives (see \cref{app:conventions} for conventions).
The equation of motion is then
\begin{equation}
    (-\partial^2)^\zeta \phi(x) = 0,
\end{equation}
and it can be applied in correlation functions when there are no other field insertions at the point $x$ (we explain why this is not a problem for our conservation and tracelessness conditions in \cref{sec:separatedPoints}). 
As mentioned in the introduction, we do not automatically apply the equation of motion (go \enquote{on shell}) in this paper, as we wish to keep track of \enquote{zero operators} like $:\phi(-\partial^2)^\zeta \phi:$.
Accordingly, our conditions \ref{prop2} and \ref{prop3} are off-shell conditions that track the zero operator on the right-hand side.

The \enquote{fractional Laplacian} $(-\partial^2)^\zeta$ is most easily defined in momentum space,
\begin{equation}
    \half \int_x \phi (-\partial^2)^\zeta \phi =  \half \int_k \phi(k) k^{2\zeta} \phi(-k),
\end{equation}
and is manifestly the $\zeta$-th power of the Laplacian in the (local) limit of integer $\zeta=k$, where these theories are called the $\Box^k$ CFTs. %
In position space, it is defined by an explicitly nonlocal subtracted hypersingular integral, which for $0<\zeta<1$ is
\begin{equation}
 (-\partial^2)^\zeta \phi \equiv \lim_{r\to 0} \frac{1}{\cF_{d+2\zeta,0}}\int_{y, \, \abs{x-y}>r} \frac{\phi(y) -\phi(x)}{\abs{x-y}^{d+2\zeta}}. \label{eq:hypersingular}
\end{equation}
For $\zeta >1$, more derivative terms are required on the right-hand side \cite[\S A.3]{Fraser-Taliente:2025udk}.

Not only are these CFTs generically nonlocal in a sense we shall make precise below, but for $\zeta>1$ they will be non-reflection positive (in Euclidean signature) or equivalently nonunitary (in Lorentzian signature)\footnote{Of course, they will also be nonunitary for arbitrary $\zeta$ in $d\notin \mathbb{Z}$, due to the evanescent operators with negative norm \cite{Hogervorst:2015akt,DiPietro:2017vsp,Ji:2018yaf}.}, due to the violation of the standard scalar unitarity bound $\Delta=\dotwo-\zeta \geq \dotwo -1$ \cite[(5.61)]{Osborn2019cftlectures}. 
Worse still, when written in a Hamiltonian formulation there is an Ostrogradsky instability: the Hamiltonian is linear in one or more of the conjugate momenta, and so is unbounded below \cite[\S 6]{Chalabi:2022qit}.
This does not prevent us from defining these theories in Euclidean space, and in any case they are indeed conformal field theories, in that they satisfy the CFT axioms \cite[\S 5]{Paulos:2015jfa} (see also \cite[\S 4.2]{Gubser:2017vgc} for comments on making this action bounded below),
and have been studied in depth (for example, see \cite[\S 5]{Paulos:2015jfa} for the proof of conformality, and \cite{Heemskerk:2009pn,Fitzpatrick:2011dm} for the conformal block decomposition and the $\phi\times \phi$ OPE). %

\subsection{Review: local operators and local CFTs}\label{sec:nonlocal}

\enquote{Local} is a relative term (consider, for example, mutually nonlocal operators such as the order and disorder operators in the Ising model). 
It is also possible to probe a local theory with nonlocal operators, and vice versa.
We want to distinguish these, and make concrete the idea of a local theory, so let us proceed carefully. 
Our aim will be to carefully define what we mean by the set of local operators in a CFT, and then to define what we mean by a local CFT.

We begin by defining an arbitrary CFT with the following data\footnote{This certainly suffices to obtain all correlation functions of point operators, but for $d\geq 2$ extended (i.e. line, surface, etc.) operators may be required for a full definition \cite[\S 6.2]{Belin:2016yll}. 
We do not explore this here.}:
\begin{enumerate}
    \item A (possibly infinite) list of primaries $P_\mathrm{CFT} = \{\cO_i\}$, where each primary is specified by its $\text{Spin}(d+1,1)\times G$ representation: a scaling dimension $\Delta_i$, a spin representation, and an internal symmetry group $G$ representation.
    \item The values of all OPE coefficients $C_{ijk}$.
\end{enumerate}
The list of primaries must satisfy various consistency conditions (exploring this is the work of the conformal bootstrap). 
The one that we will use here is that this list must be closed under the OPE.

\subsubsection{Local and nonlocal operators}\label{sec:localnonlocal}

Then we \textit{define} a single operator $\phi \in P_\mathrm{CFT}$ to be a local operator.
The set of local operators is then defined as the set of operators that are local relative to $\phi$ (compactly, we say they are $\phi$-local).
Then we define the set of local primaries to be the OPE algebra generated by $\phi$ -- that is, the set of operators that, starting with $\phi$, are generated by repeatedly applying the OPE:
\begin{equation}\label{eq:PlocalDef}
   P^\mathrm{CFT}_\text{$\phi$-local} \quad = \quad\text{local primaries} \quad = \quad \text{ the closure under OPE of the set } \{\phi\}.
\end{equation}
This set does not include infinite sums of local operators.
We could \textit{choose} to put some nonlocal operators into $P_\mathrm{CFT}$, so $P_\mathrm{CFT} \supsetneq P^\mathrm{CFT}_\text{$\phi$-local}$.
However, it should not be necessary to do so to self-consistently define a CFT, since the OPE of local operators will never produce a nonlocal operator: hence it is always consistent to truncate to just local operators: $P_\mathrm{CFT} = P^\mathrm{CFT}_\text{$\phi$-local}$.

Combining the set $P^\mathrm{CFT}_\text{$\phi$-local}$ with their descendants, we obtain the complete set of local operators $L^\mathrm{CFT}_\text{$\phi$-local}$.
In the one-field case, since the OPE is essentially a Taylor (or more properly Laurent) expansion, the field-theory realization of these operators consists only of polynomials in the field $\phi$ and its derivatives.
Suppressing contractions of derivatives, in our free one-field case:
\begin{equation}
    L^\mathrm{CFT}_\text{$\phi$-local} = \text{local operators} \quad = \quad \{:\partial^{a_1}\phi \, \partial^{a_2} \phi \cdots \,\partial^{a_k} \phi:, \quad a_i,k \in \mathbb{N}_0\}. \label{eq:Ldef}
\end{equation}
In this paper we study bilinears: the local bilinears are those which have two $\phi$ fields and are polynomial in $\lptl$ and $\rptl$.
Evidently the operator in the action, $\phi(-\partial^2)^\zeta \phi$, will not be in this $L^\mathrm{CFT}_\text{$\phi$-local}$: for generic $\zeta$ it is manifestly a nonlocal operator, as \eqref{eq:hypersingular} does not depend only on the value of $\phi$ and its spatial derivatives at a single point.

We then define a nonlocal action as one that cannot be written as the integral of a local operator: evidently our action is generically nonlocal, as it can only be written as an integral of a bilocal action.
This also means in our case\footnote{In interacting theories, it is slightly subtler: in the DREG$+\delta$ regularization of the large-$N$ CFTs, the action is nonlocal, but nonetheless yields a local CFT at the critical point, with a local EM tensor. 
Hence, locality of the Landau-Ginzburg action is not a requirement for the CFT to be local.
Similarly, one can have an action that appears nonlocal but can be made local by adding back in local degrees of freedom: consider the action of massless QED after integrating out the fermions. %
} that there is no local EM tensor for generic real $\zeta$, in the sense that a conserved $T^{\mu\nu}$ does not appear in the OPE of the local field $\phi$.
However, some EM tensor must exist, being the Noether current for spacetime translations: it is nonpolynomial in derivatives, and so is a nonlocal operator.
This leads to our definition of a local CFT.

\subsubsection{Local and nonlocal theories}

The general definition\footnote{
It is worth noting that generic QFTs can be nonlocal to differing degrees. 
For generic $\zeta$, our CFTs are nonlocal in a special way: a kind of holographic locality. 
According to the Caffarelli-Silvestre extension theorem, for $\zeta = \half\sqrt{d^2 + 4m^2 L^2}$ we can equate their action with the action of the AdS theory of a \textit{local} scalar of mass $m^2$ \cite[\S 2]{Frassino:2019yip} \cite[(1.34)]{Rychkov:2016iqz} (see also \cite[\S 5.3]{Paulos:2015jfa} or \cite{LaNave:2016nxa,LaNave:2017lwf}). 
Our CFTs also admit a slightly different local realization: the restriction $\phi(x)=\Phi(x,z=0)$ of a $d+p$-dimensional canonical free field $\Phi(x,z)$ to a $d$-dimensional defect, for $\Delta=\frac{d+p-2}{2}$ (taking the codimension $p$ to be continuous). %
} of a local CFT is that it contains a local conserved EM tensor \cite[\S H.1]{Poland:2018epd}.
We can now make this statement more precise: 
a CFT is local with respect to a particular basis of point operators (which we indicate by its seed $\phi$) if a traceless conserved spin-2 EM tensor $\tilde{T}^{\prime\mu\nu}$ exists in the closure of the OPE of $\phi$:
\begin{equation}
    \text{$\phi$-local CFT}  \quad \leftrightarrow \quad \tilde{T}^{\prime\mu\nu} \in L^\mathrm{CFT}_\text{$\phi$-local}
\end{equation}
That operator need not be primary, but must be traceless and conserved in order for the theory to be a CFT (preempting the next section, we call it $\tilde{T}^{\prime\mu\nu}$ to indicate this). 
This definition means that our theories are local for integer $\zeta$, but nonlocal for generic $\zeta$.

Somewhat surprisingly, it is sometimes possible to make a nonlocal theory local with a different choice of $\phi$ in \eqref{eq:PlocalDef}.
This is because if we choose a different operator to be our local field, then we obtain a different set of nonlocal operators. 
For example, consider the free theory with $\zeta = 2\alpha +k$ for any integer $k$. 
If we choose the field-theoretic operator $\psi=(-\partial^2)^{\alpha} \phi$ to be our local operator, the CFT defined by the closure of $\psi \times \psi$ is indeed a $\psi$-local CFT, with an energy-momentum tensor -- precisely the $\Box^k$ theory (see \cite[II.B]{Basa:2019ywr} \cite[\S 4.1]{Basa:2020cyn} for more on this).

This dependence of properties of a CFT on $P_\mathrm{CFT}$ should be familiar from other contexts, such as that of unitarity: for example, it is now understood that the unitarity of, for instance, the Wilson-Fisher model in integer dimensions depends on  $P_\mathrm{CFT}$ \cite[\S 2.5]{Henriksson:2025kws}. 
In generic $d$ there exist evanescent operators which have negative norm in arbitrary $d$. However, exactly for integer $d$, they do not appear in the OPE of the fundamental fields and so can be removed from $P_\mathrm{CFT}$, ensuring the unitarity of the theory.
Likewise, in the $d=2$ $\gO(n)$ model, for generic $n$ there are negative-norm states which decouple in the limit of $n\to 1,2$ \cite{Gorbenko:2020xya}.
Crucially, in each case there are a set of operators, closed under the OPE, which give a unitary CFT.

\subsection{The energy-momentum tensor}\label{sec:EMtensor}
From the fact that the theory is assumed to be globally conformally invariant\footnote{
Given a particular flat-space QFT, it may be (or admit an uplift to become) a member of any of the following sets: QFT $\supset$ Poincar\'e-invariant QFT $\supset$ (Poincar\'e+scale)-invariant QFT (SFT) $\supset$ CFT $\supset$ Weyl-covariant FT (possibly with Weyl anomalies) $\supset$ Weyl-covariant FT (no Weyl anomalies) $\supset$ topological QFT (TQFT) $\supset$ \{empty theory\}.
We do not consider the possibility of scale invariance without conformal invariance, in which case $\tilde{T}\indices{^\mu_\mu} = \partial_\mu J^\mu$ for some $J$, but $J^\mu \neq \partial_\nu L^{[\mu\nu]}$ for any $L$: it does not arise here.
}, there should exist a Noether current: some conserved energy-momentum tensor $\tilde{T}^{\mu\nu}$, which we can choose to be symmetric (since it should be spin-$2$) and (by conformality) traceless \cite[(4.4)]{Osborn2019cftlectures}:
\begin{equation}\label{eq:onshellTtilde}
    \tilde{T}^{[\mu\nu]}=0, \quad \partial_\mu \tilde{T}\indices{^\mu^\nu} \overset{\text{eom}}{=} 0,\quad \tilde{T}\indices{^\mu_\mu} \overset{\text{eom}}{=} 0.
\end{equation}
This operator is not unique, for three reasons:
\begin{itemize}
    \item The overall normalization is not fixed.
    \item It can mix with any \enquote{zero operators}, i.e. operators which vanish by the equations of motion. 
    For integer $\zeta$, $\gflat^{\mu\nu} :\phi (-\partial^2)^\zeta \phi:$ is the only local spin-2 operator with the correct number of derivatives that is zero on-shell\footnote{For noninteger $\zeta$, there is a slight subtlety: this $T^{\mu\nu}$ need not be local, so there are additional primaries with which it can mix. We ignore this until \cref{sec:homogeneous}.}.
    \item The transverse-traceless terms $\propto \cD_{\mathrm{TT}}^{\mu\nu\rho\sigma}X_{\rho\sigma}$ in $\tilde{T}^{\mu\nu}$ are unfixed.
\end{itemize}
We can fix the first two ambiguities straightforwardly for integer $\zeta$. 
Off-shell, the only structures compatible with \eqref{eq:onshellTtilde} are
\begin{equation}
   \partial_\mu \tilde{T}\indices{^\mu^\nu} = A_1\, \partial^\nu \phi (-\partial^2)^\zeta\phi  + A_2 \, \phi \,\partial^\nu (-\partial^2)^\zeta \phi,\quad \tilde{T}\indices{^\mu_\mu} =  B \, \phi (-\partial^2)^\zeta \phi.
\end{equation}
So we can consider
\begin{equation}
    \tilde{T}^{\prime\mu\nu}= -\frac{\dotwo-\zeta}{B -A_2 d}(\tilde{T}^{\mu\nu}-A_2 \gflat^{\mu\nu} \phi (-\partial^2)^\zeta \phi),
\end{equation}
which satisfies \ref{prop3} and -- up to an overall factor which is not fixed by this argument but (when we actually do the construction) turns out to be $1$ -- also \ref{prop2}.
At the moment, these are just arbitrary choices which we are free to make, but as we shall see in \cref{sec:TfromWeyl}, we choose them to match the Weyl-covariant theory.

The transverse-traceless terms can clearly be modified by adding a linear combination of descendants.
Hence, one way of removing this ambiguity is by identifying the actual primary operator that has scaling dimension $d$: adding \textit{improvement terms}, which means subtracting off the linear combinations of descendants to leave only the (unique) primary.
In generic dimension $d$, this should be possible, and we call
\begin{equation}
T^{\mu\nu}=\tilde{T}^{\prime\mu\nu} + \text{improvement terms}
\end{equation}
\textit{the} energy-momentum tensor.
In this paper our task is to construct this object.
However, in certain dimensions this might not be possible: removing the required linear combinations of descendants might modify the trace from zero.
If it is not possible, then the conserved symmetric $\tilde{T}^{\mu\nu}$ that \textit{is} traceless is not a primary operator.
When working in generic $d$, this issue shows up as a pole in $d$ in $T^{\mu\nu}$.
In that case, despite the theory being a well-defined CFT, it is not possible to couple it to a metric such that it is \textit{Weyl-covariant}.

\subsection{Review: Weyl covariance \texorpdfstring{$\implies$}{implies} conformal invariance} \label{sec:WeylConf}

In the physics literature, the distinction between conformal invariance and Weyl covariance is now standard \cite{Stergiou:2022qqj,Farnsworth:2017tbz}.
\begin{itemize}
    \item Conformal invariance: relates different correlation functions on the same manifold, usually taken to be flat space ($g=\gflat$).
    \item Weyl covariance: relates correlation functions on different physical manifolds; %
    the first with metric $g$, and the second with metric $g' \equiv \Omega^2 g$.
    This effectively makes physical scale a local quantity \cite{Jia:2023gmk}. %
    For example, for Weyl-covariant scalar operators $\phi_i$,
    \begin{equation}
       \expval{\phi_1(x) \phi_2(y) \cdots}_{\Omega^2 g}= \Omega(x)^{-\Delta_1} \Omega(y)^{-\Delta_2} \cdots \expval{\phi_1(x) \phi_2(y) \cdots}_{g} \label{eq:WeylCov}.
    \end{equation}
    It is Weyl covariance that permits, for example, obtaining spatial sphere ($S^{d-1} \times \mathbb{R}$) or sphere ($S^d$) correlators from flat space ($\mathbb{R}^d$) correlators \cite{Farnsworth:2017tbz}. 
\end{itemize}
If a CFT is Weyl-covariant, then trivially, upon setting $g=\gflat$ the Weyl covariance descends to conformal invariance and the Weyl-covariant operators become conformal primaries.
More precisely, the conformal isometries that leave the flat metric $g=\gflat$ invariant are compositions of a specific type of Weyl transformation and then a diffeomorphism that restores the flat metric; i.e. $\text{Conf} \subset \text{Diff} \times \text{Weyl}$.
Since in the usual QFT setting all theories possess diffeomorphism invariance, clearly the flat space limit of a Weyl-covariant theory must be a CFT \cite{Zanusso:2023vkn} (this is also termed Zumino's theorem \cite[\S III]{Nakayama:2019xzz}).
See also \cite[\S II.A]{Stergiou:2022qqj}.
Similarly, a Weyl-covariant operator transforms exactly as a conformal primary under the combination of the restricted Weyl transformation and compensating diffeomorphism.

The converse does not hold: when lifting a CFT to curved space, there may exist obstructions to making it Weyl-covariant. 
For unitary CFTs in $d\le 10$ it has been proven that we can always lift conformal invariance to Weyl covariance \cite{Farnsworth:2017tbz}\footnote{First, they showed that conformal invariance means that there exists a flat-space EM tensor such that $T\indices{^\mu_\mu}|_{g=\gflat} \overset{\text{eom}}{=}0$. 
Then they enumerated the operators $O_i = (\text{curvature} \times \text{field})$ that could appear in $T\indices{^\mu_\mu}$ for arbitrary $g$, and ruled them out using (i) the unitarity bounds on scaling dimensions and (ii) the fact that Weyl transformations commute. 
Hence, $T\indices{^\mu_\mu} \overset{\text{eom}}{=} 0$ as an operator equation, indicating Weyl covariance. 
In even dimensions, the operators proportional to the identity, $O_i=(\text{curvature}\times 1)$, are still allowed, and lead to the Weyl anomalies, as we discuss in \cref{sec:anomaliesvsobstructions}.\label{footnote:whatFarnsworthDid}}.
However, our CFTs are not unitary, and indeed we know that for certain even $d$ the lift to Weyl covariance is not possible (see \cite{Nakayama:2016dby} for a simple walkthrough of such a theory in the case $d=2$). 
That is, for some nonunitary theories it may not be possible to couple the CFT to a background metric $g_{\mu\nu}$ such that the conformal symmetry descends from a local Weyl symmetry \cite[\S 4]{Farnsworth:2021ycg}.
A conformal transformation is a Weyl transformation (+ a diffeomorphism), so any obstruction to making a \enquote{good $T^{\mu\nu}$} in the CFT (a conserved, symmetric, traceless EM tensor that is also primary) is also an obstruction to making a good Weyl-covariant energy-momentum tensor, which would be required to make the CFT Weyl-covariant.%

It is these obstructions that are captured by the poles in integer $d$ of \eqref{eq:Rdef}.
Consider $d=2$; manifestly, the only free scalar that can be made Weyl-covariant is the $\zeta=1$ canonical free scalar -- all others have poles $\propto 1/(d-2)$ in their stress tensors, and so cannot be defined\footnote{Recalling that the Virasoro group is captured by a subgroup of $\mathrm{Diff} \times \{\Box \Omega = 0 \, | \, \Omega \in \text{Weyl}\}$, this reflects the fact that only the standard Ising model gives a theory with Virasoro invariance; the other \enquote{long-range minimal models} are only invariant under the global conformal group \cite{Rajabpour:2011qr} \cite{Behan:2025ydd}. 
}.

\subsubsection{\TtextOrPDF{} from a Weyl-covariant action}\label{sec:TfromWeyl}

If our CFT can indeed be made Weyl-covariant, in the sense of having correlation functions that transform sensibly under Weyl transformations, then the metrical EM tensor of that Weyl-invariant action should give precisely the correct conformal operator in the flat limit \cite{Stergiou:2022qqj,Iosifidis:2025sjx}.
Let us explain how to obtain the metrical EM tensor. 
Assume that we know the fully Weyl-covariant action of a free scalar of dimension $\Delta\equiv \dotwo-\zeta$ on a Riemannian manifold without boundary $(X,g)$,
\begin{equation}
    S_W^g[\phi] = \half \int_X \odif[d]{x} \sqrt{g} \,\phi \WeylOp_{\zeta}^g \phi,
\end{equation}
where $\WeylOp_{\zeta}^g$ is a Weyl-covariant fractional Laplacian, or GJMS operator, which satisfies
\begin{align}
    \WeylOp_{\zeta}^{ g} (\phi) &= \Omega^{d-\Delta} \WeylOp_{\zeta}^{\Omega^{2} g} \left(\Omega^{-\Delta} \phi\right), \quad \Delta \equiv \dotwo -\zeta,
\end{align}
for all\footnote{This definition is taken from \cite[(2.6)]{Gonzalez:2016kjr}, who uses the mathematical convention of calling Weyl covariance \enquote{conformal covariance}.
This operator is also self-adjoint \cite[Corollary]{grahamScatteringMatrixConformal2001}. %
} $\phi(x), \log\Omega(x) \in C^{\infty}(X)$.
This operator is unique for $\zeta \in \mathbb{N}$, but not for $\zeta \in \mathbb{R}$ \cite[\S 7]{Gonzalez:2016kjr}; however, the ambiguity disappears in the limit $X\to \mathbb{R}^d$.
We shall presume that we are working in dimensions where this operator exists: for $\zeta \in \mathbb{N}$, that is $d \neq 2, 4, \cdots, 2(\zeta-1)$; for $\zeta \in \mathbb{R}$, that is $d \notin 2\mathbb{N}$.

This action is manifestly invariant under a Weyl transformation of the geometry, i.e.
\begin{equation}\label{eq:SWeylCov}
  S_W^{\Omega^2g}[\Omega^{-\Delta}\phi] =   \half \int_X \odif[d]{x} \, \Omega^d\sqrt{g}\,  (\Omega^{-\Delta}\phi) \WeylOp_{\zeta}^{\Omega^2 g}(\Omega^{-\Delta}\phi) = S_W^g[\phi].
  \end{equation}
Note that to show full Weyl covariance (without anomalies) of a quantum field theory we ought to demonstrate that the unit-normalised measure is Weyl-covariant instead,
\begin{equation}\label{eq:WeylCovariantMeasure}
\mu_g[\phi] \equiv \frac{\cD_g{\phi}\, e^{-S^g_W[\phi]}}{Z[g]}  = \mu_{\Omega^2 g}[\Omega^{-\Delta} \phi].
\end{equation}
We shall simply assume this for the moment, and discuss it in the next section.
So, 
\begin{equation}
    \expval{\phi(x) \phi(y) \cdots}_{\Omega^2 g} = \int \mu_{\Omega^2 g}[\phi] \, \phi(x)\phi(y) \cdots  = \int \mu_{\Omega^2 g}[\Omega^{-\Delta} \phi] \, \Omega^{-\Delta}\phi(x) \Omega^{-\Delta}\phi(y) \cdots,
\end{equation}
just by replacing the dummy variables $\phi \mapsto \Omega^{-\Delta} \phi$.
Then we can use \eqref{eq:WeylCovariantMeasure} to find
\begin{equation}
     \expval{\phi(x) \phi(y) \cdots}_{\Omega^2 g} = \Omega(x)^{-\Delta} \Omega(y)^{-\Delta} \cdots \expval{\phi(x) \phi(y) \cdots}_{g};
\end{equation}
so the theory is indeed Weyl-covariant as defined in \eqref{eq:WeylCov}.

Actually constructing $S_W$ (i.e. $\WeylOp_\zeta^g$) is a problem, as it must include curvature couplings in order to be Weyl-covariant\footnote{Not to mention also all of the $\phi$-independent curvature counterterms that yield contact terms in higher-point $T$ correlators.}.
For fractional $\zeta$, as mentioned above, $S_W$ is no longer unique.
We postpone further discussion of $S_W$ to \cref{sec:Juhl}, where we give Juhl's construction of the generalized GJMS operator for arbitrary $g$, and consider now only its first variation, $T_W^{\mu\nu}$.

$\WeylOp_{\zeta\in\mathbb{N}}$ is unique (henceforth we drop the ${}^g$), so the metrical Euclidean Weyl-covariant EM tensor must be
\cite[\S 4]{Osborn2019cftlectures}:
\begin{equation}\label{eq:TdefDeriv}
    T_W^{\mu\nu} \equiv -\frac{2}{\sqrt{g}} \fdv{S_W}{g_{\mu\nu}} = +\frac{2}{\sqrt{g}} \fdv{S_W}{g^{\mu\nu}}.
\end{equation}
Note that this expression has the opposite sign compared to the Lorentzian expression -- see \cref{app:conventions}.
This metrical EM tensor satisfies
\begin{subequations}\label{eq:properties1to3forTW}
\begin{align}
    T_W^{[\mu\nu]}&=0,\\
    \nabla_\mu T_W^{\mu\nu} &=-(\nabla^\nu \phi) \WeylOp_{\zeta} \phi,\\
    g_{\mu\nu} T_W^{\mu\nu} &=-(\dotwo -\zeta) \phi \WeylOp_{\zeta} \phi + \text{curvature terms if $d$ is even.} \label{eq:plusCurvature}
\end{align}
\end{subequations}
off-shell (i.e. without imposing the equations of motion) due to, respectively: the symmetry of $g_{(\mu\nu)}$, the diffeomorphism Ward identity, and \eqref{eq:SWeylCov}\footnote{
For classical proofs of these invariances: Weyl and diffeomorphism invariance tell us that
\begin{align}
    \delta_\sigma S_W = \int_x \underbrace{\frac{1}{\sqrt{g}} \fdv{S_W}{g_{\mu\nu}}}_{-\half T_W^{\mu\nu}} \delta_\sigma g_{\mu\nu} + \fdv{S_W}{\phi} \delta_\sigma \phi =0, \quad 
       \delta_\xi S_W = \int_x \frac{1}{\sqrt{g}} \fdv{S_W}{g_{\mu\nu}} \delta_\xi g_{\mu\nu} + \fdv{S_W}{\phi} \delta_\xi \phi =0
\end{align}
$\fdv{S_W}{\phi} = +\WeylOp_{\zeta} \phi$, because $\WeylOp_{\zeta}$ is self-adjoint. 
\eqref{eq:properties1to3forTW} then follows after substituting the variation rules
\begin{equation}
\delta_\sigma g_{\mu\nu} \equiv 2\sigma g_{\mu\nu}, \quad 
    \delta_\sigma \phi \equiv -(\dotwo - \zeta) \sigma \phi; \quad 
    \delta_\xi g_{\mu\nu} \equiv \nabla_\mu \xi_\nu + \nabla_\nu \xi_\mu, \quad 
    \delta_\xi \phi \equiv \xi^\mu \nabla_\mu \phi,
\end{equation}
where we have used the active diffeomorphism convention. %
}.
It is from these equations that we take the natural choices to fix the ambiguities of $T^{\mu\nu}$ discussed in \cref{sec:EMtensor}: properties \ref{prop2} and \ref{prop3}.
Hence, $\nabla_\mu T^{\mu\nu}_W$ and $g_{\mu\nu} T_W^{\mu\nu}$ should both vanish \textit{as operators} up to (a) contact terms when in correlators with other operators and (b) possible $c$-number contributions from curvature.

The contact terms are very simple to understand. 
They arise when imposing the equations of motion, and, for $g_{\mu\nu} T_W^{\mu\nu}$, encode the transformation of other operators under an infinitesimal Weyl transformation (this is clear from the linearization of \eqref{eq:WeylCov} for $\delta g_{\mu\nu} = (\Omega^2-1)g_{\mu\nu}$) \cite[\S 1]{Farnsworth:2017tbz}.

\subsubsection{An aside: quantum Weyl anomalies vs. Weyl obstructions}\label{sec:anomaliesvsobstructions}

The $c$-number curvature contributions, indicated by the \enquote{$+\text{curvature terms if $d$ is even}$} in \eqref{eq:plusCurvature}, are more subtle.
These universal curvature terms arise due to a quantum anomaly, termed the Weyl anomaly or trace anomaly: namely, any regularization of this CFT will produce a term that is not expected classically (nonuniversal terms are also permitted, but \enquote{nonuniversal} means that they can be cancelled by adding local counterterms such as $\int_x \nabla^2 R$ to the action, which do not modify the dynamics -- so we shall presume that they have been tuned to vanish).

Essentially, if the Weyl covariance of the quantum effective action $W_\mathrm{eff}[g_{\mu\nu}, J]$ is only broken by a local anomaly $\delta_\sigma W_\mathrm{eff} = \int_x \mathcal{A}[\nabla, g(x)]$, we say that the symmetry has a \textit{Weyl anomaly}.
Despite the presence of an \enquote{anomaly}, Weyl covariance is regarded as a good symmetry \cite[\S 4.2]{Farnsworth:2017tbz}, because the anomalous transformation of correlation functions is encoded in the quantum effective action \cite{Baume:2014rla} (so the Weyl Ward identities still hold in a modified form).

Let us explain this further.
As usual, the anomalies arise due to improper $c$-number transformation of the measure $\mu_g[\phi]$ under Weyl variation of $g$, even though the classical theory, encoded by the action $S_W$, is Weyl-invariant.
To see this, we can consider the infinitesimal version of \eqref{eq:WeylCovariantMeasure} for $\Omega = e^\sigma$, $\sigma \ll 1$:
\begin{equation}
    \mu_{\Omega^2 g}[\Omega^{-\Delta}\phi] -\mu_g[\phi] \propto \int_x \sigma( T\indices{^\mu_\mu}-\expval{T\indices{^\mu_\mu}}) \overset{\text{eom}}{=} 0 + \text{contact terms}.
\end{equation}
This can only hold if $T\indices{^\mu_\mu}|_\text{eom}$ is a $c$-number, i.e. proportional to curvature.
By dimensional analysis, this is only possible in even $d$, and so we recover the standard fact that the Weyl anomalies exist only in even $d$.
Thus, the more precise statement of our tracelessness condition for arbitrary $g$ is that the trace of the Weyl-covariant EM tensor is proportional to the equation of motion, modulo $c$-number curvature contributions. 
We do not expect any quantum corrections to the symmetry and conservation conditions, since diffeomorphism invariance should still hold in the quantum theory.

Generically, other terms (which vanish in flat space) could appear on the right-hand side of $T\indices{^\mu_\mu}$. Schematically, they take the form $O_i=(\text{curvature} \times \text{operators})$, and lead to a flat-space CFT that cannot be made Weyl-covariant -- but in a different way to those studied here.
As discussed in \cref{footnote:whatFarnsworthDid}, \cite[\S 4.1]{Farnsworth:2017tbz} ruled these $O_i$s out for local unitary CFTs in $d\leq 10$.
Luckily, our (sometimes nonunitary, sometimes nonlocal) Gaussian theories are so simple that they cannot arise.

The discussion above should make it clear that \textit{the Weyl obstructions which exist for even $d$ are not the same as the Weyl anomalies which exist in even $d$.}
Principally:
\begin{itemize}
    \item The Weyl anomalies are \enquote{quantum}, since they relate to the regularization of the QFT. 
    That is, all well-defined regularized theories that become this QFT when their cutoffs are removed are forced to have terms of this form.
    \item The Weyl obstructions are \enquote{classical}, in that they come from considerations based on the classical action and classical geometry.
    That is, they arise when asking what operators can be built out of $\phi$, the metric, and their derivatives -- and realizing that $\WeylOp_\zeta$ does not exist, so $S_W$ does not exist.
\end{itemize}

Let us return to the example of the $d=2$, $\zeta=1$ theory, where $\WeylOp_1 \rvert_{d=2} \overset{\eqref{eq:minusYamabe}}{=} -\nabla^2$:
\begin{equation}
    Z[g] \equiv e^{-F[g]}\equiv \int\cD_g \phi\, e^{-S^g_W[\phi]} .
\end{equation}
It has a quantum Weyl anomaly \cite[(16)]{Polyakov:1981rd} (see also \cite[\S 5.A]{DiFrancesco:1997nk})%
\begin{align}
    F[g] =\half \log \det'(-\nabla^2) + \text{local}= -\frac{1}{48\pi} \int \odif[2]{x} \sqrt{g} \, \left[\cR\, (\nabla^2)^{-1} \cR + K\right]. %
\end{align}
Clearly the value of the constant $K$ can be changed by local counterterms $\propto \delta K \int \odif[2]{x} \sqrt{g}$.
However, the other term in $F$ is manifestly nonlocal, and so cannot be removed by adding a local counterterm to the action. 
The trace $T\indices{^\mu_\mu}$ does not vanish, and in fact leads to the central charge $c=1$: %
\begin{equation}
-\frac{2}{\sqrt{g}} g^{\mu\nu} \fdv{F}{g^{\mu\nu}} = \expval{T\indices{^\mu_\mu}} = \frac{c}{24\pi}(\cR +{K}).
\end{equation}
Typically, we tune $K$ such that the trace is zero in flat space (this is connected to the normal ordering discussion of \cref{sec:separatedPoints}), but evidently that value will not work for an arbitrary spacetime.
However, the unique primary EM tensor of this theory does exist; hence there is no obstruction to making the free boson Weyl-covariant -- thankfully for string theory.
There is an identical story in all dimensions $d=2n$: %
all the theories with $\zeta>n$ (negative integer $\Delta$) cannot be made Weyl-covariant due to Weyl obstructions, because the relevant GJMS operators do not exist.
Their propagators in flat space take the form of a polynomial with logarithmic factors $\expval{\phi(x)\phi(y)}\sim (x^2)^{\zeta-n} \log\mu\abs{x-y}$ \cite{Brust:2016gjy}.
However, the theory with $\zeta=n$ ($\Delta=0$), which is a CFT with propagator $\expval{\phi(x)\phi(y)}\propto \log\mu\abs{x-y}$\footnote{Technically, the theories for all $\zeta \ge n$ are logCFTs if we do not remove $\phi$ from $P_\mathrm{CFT}$ (the spectrum).}, \textit{can} be made Weyl-covariant, and has a nonzero Weyl anomaly.

\subsection{\texorpdfstring{$T_W$ and $T$}{T\_W and T} are the same}\label{sec:TsAreSame}

In any case, we take the flat space limit of the Weyl-covariant theory, so any such Weyl anomalies must vanish.
In the flat-space limit, a Weyl-covariant $T^{\mu\nu}_W$ satisfies all four of \ref{prop1}--\ref{prop4}. 
Then, because the conformal primary $T^{\mu\nu}$ is unique, $T_W^{\mu\nu}$ must be the same as it:
\begin{equation}
    T^{\mu\nu}_W \rvert_{g=\gflat} = T^{\mu\nu}.
\end{equation} 
For \ref{prop1}, \ref{prop2}, and \ref{prop3}, this follows immediately from \eqref{eq:properties1to3forTW} evaluated for $g=\gflat$. 
\ref{prop4} descends from Weyl-covariance of $T_W^{\mu\nu}$, and can be understood as follows.
First, $T^{\mu\nu}$ comes from the variation of a Weyl-invariant quantity $S_W$ with respect to the metric, and so must be Weyl-covariant:
under $g_{\mu\nu}\mapsto e^{2\sigma} g_{\mu\nu}$, we should have
\begin{equation}
    T_{W,\mu\nu} \mapsto e^{-(d-2)\sigma} T_{W,\mu\nu}.
\end{equation}
This combines with its transformation under the compensating diffeomorphism \cite[(2.5)]{Farnsworth:2017tbz} to yield exactly the transformation law for a spin-$2$ conformal primary.
Hence, in the flat-space limit $T^{\mu\nu}_W$ will be a primary of the global conformal group $\SO(d+1,1)$, and so must satisfy the usual %
\begin{equation}
    \hat{K}_\rho (T_W^{\mu\nu})|_{g=\gflat, x=0}=0,
\end{equation}
which is precisely \ref{prop4}.

\section{Constructing the non-primary stress tensor}\label{sec:construction}

In this section we will construct the non-primary symmetric, traceless, and conserved stress tensor, which reduces to solving polynomial equations in momentum space.
We then give some technical comments about allowed rewritings in terms of Gegenbauer polynomials, which will set the stage for the solution of the primary condition $\hat{K}_\rho T^{\mu\nu} =0$ in \cref{sec:primaryconstruction}.

\subsection{The momentum space toolkit}\label{sec:momSpace}

We are considering a Gaussian theory, so we know that the stress tensor will be a bilinear $\sim\phi \cF_{\mu\nu}(-i\lptl, -i\rptl) \phi$. 
Given that we work in flat space, to get rid of extra minus signs and reduce clutter, it makes sense to transform to momentum space:
\begin{equation}
\phi(x) \cF_{\mu\nu}(-i\lptl, -i\rptl) \phi(x)=  \int_{p,q} e^{i (p+q)\cdot x} \phi(p) \cF_{\mu\nu}(p, q) \phi(q)
\end{equation}
(where everything remains implicitly normal ordered, as is discussed in detail in \cref{sec:separatedPoints}).
This is not strictly necessary: of course, everything we do in momentum space is equivalent to manipulating the derivatives, with replacement rules
\begin{align}
    p &\rightarrow -i \lptl, \quad & q &\rightarrow -i \rptl.
\end{align}

Total derivative contributions are those proportional to $s\equiv p+q \mapsto -i(\lptl+\rptl)$, because for an arbitrary derivative operator $\cF(\lptl,\rptl)$,
\begin{align}
    \int_{p,q} e^{i (p+q)\cdot x} s^\mu\, \phi(p) \cF \phi(q)&=\phi(x)  (-i(\lptl + \rptl)^\mu) \cF \phi(x) = -i\partial^\mu(\phi(x) \cF \phi(x)).
\end{align}
We also define the natural variable to pair with $s$, the difference of the momenta
\begin{align}
    \di &= p-q \quad \mapsto\quad - i \lrptl = -i( \lptl - \rptl).
\end{align}

We will almost always symmetrise our expressions with respect to $p$ and $q$, which can be done freely:
\begin{equation}
\begin{aligned}
\phi(-\partial^2)^\zeta \phi &= \int_{p,q}e^{ix\cdot (p+q)} \phi(p) p^{2\zeta} \phi(q)=\int_{p,q}e^{ix\cdot (p+q)} \phi(p) q^{2\zeta} \phi(q)\\
&= \int_{p,q}e^{ix\cdot (p+q)} \phi(p) \frac{p^{2\zeta}+q^{2\zeta}}{2} \phi(q).
\end{aligned}
\end{equation}
Doing so removes any ambiguity associated with the different ways of writing a coordinate-space expression in momentum space. 
For example, we could write $\phi (-\partial^2) \phi$ as either $p^2$ or $q^2$, but instead make the unique symmetric choice of $\frac{p^2 + q^2}{2}$. 
It also often reveals additional structure: for example, symmetrisation of
\begin{equation}
    \int_{p,q} e^{i s\cdot x} \phi(p) \frac{p^{2\zeta}}{p^2-q^2} \phi(q) =\half  
        \int_{p,q} e^{i s\cdot x} \phi(p) \frac{p^{2\zeta}-q^{2\zeta}}{p^2-q^2}\phi(q) 
\end{equation}
makes clear that this apparently nonlocal bilinear is manifestly local for integer $\zeta$, thanks to the standard difference-of-powers expansion $a^{n}-b^{n}=(a-b)\sum _{k=0}^{n-1}a^{n-1-k}b^{k}$.

The Gegenbauer \enquote{polynomials}, or rather, Gegenbauer functions, which we give more information about in \cref{app:Gegenbauer}, can be expressed for arbitrary argument as
\begin{equation}
    C_\zeta^{(\alpha)}(x) = -\half \frac{1}{\pi \Gamma(\alpha)} \sum_{k=0}^\infty \frac{(-2 x)^k}{k!} \sin (\pi  \zeta) \Gamma \left(\frac{k-\zeta}{2}\right) \Gamma \left(\frac{k+\zeta}{2}+\alpha \right).
\end{equation}
Taking the limit of integer $\zeta$, this is evidently a degree-$\zeta$ polynomial in $x$, as eventually the $\sin(\pi \zeta)$ kills all higher terms. 

\subsubsection{Two useful projectors}

We can now construct two useful differential operators which will be taken to act only on bilinears; in momentum space they will therefore be functions only of $s$.
The first projector is transverse but not traceless:
\begin{equation}
    \cP^{\mu\nu} \equiv \frac{1}{d-1}(-\partial^2 \gflat^{\mu\nu}+\partial^\mu \partial^\nu).
\end{equation}
Using $s=p+q$ and $\di=p-q$ as defined above, we overload our notation to use the same symbol for this operator in momentum space when it acts on bilinears,%
\begin{equation} \label{eq:PmomentumSpace}
    \cP^{\mu\nu} = \frac{1}{d-1}(s^2 \gflat^{\mu\nu}-s^\mu s^\nu).
\end{equation}
Since we only deal with bilinears, this is always unambiguous.
This $\cP$ is the unique symmetric spin-2 derivative operator satisfying the conditions
\begin{align}
\gflat_{\mu\nu} \cP^{\mu\nu}  &= -\partial^2 \quad \leftrightarrow \quad \gflat_{\mu\nu} \cP^{\mu\nu}  = s^2,\\
\partial_\mu \cP^{\mu\nu} &= 0 \quad\leftrightarrow \quad s_{\mu} \cP^{\mu\nu}  =0.
\end{align}
This means that adding an operator of the form $\cP^{\mu\nu} \cO$ to a candidate EM tensor modifies only the (off-shell) trace, without modifying the (off-shell) divergence.

We also define the unique four-derivative operator (up to scaling), symmetric in both $(\mu\nu)$ and $(\rho\sigma)$, which is transverse and traceless (TT),
\begin{equation}\label{eq:DQconditions} 
    \gflat_{\mu\nu} \cD_{\mathrm{TT}}^{\mu\nu\rho\sigma} = 0 = \gflat_{\rho\sigma} \cD_{\mathrm{TT}}^{\mu\nu\rho\sigma}, \quad
    \partial_\mu \cD_{\mathrm{TT}}^{\mu\nu\rho\sigma} =0 =\partial_\rho \cD_{\mathrm{TT}}^{\mu\nu\rho\sigma}.
\end{equation}
Adding an operator of the form $\cD_{\mathrm{TT}}^{\mu\nu\rho\sigma} \cO_{\rho\sigma}$ to a candidate $T^{\mu\nu}$ cannot modify the trace or divergence (even off-shell).

Explicitly, it is\footnote{This is $\cD_{\mathrm{TT}} \equiv (2-d) \cD_B$ as defined in \cite{Stergiou:2022qqj}; it is neater in our context to remove the factor of $(2-d)$, as it makes our formulae for $T^{\mu\nu}$ more explicitly democratic with respect to the even dimensions; it also ensures the neat proportionality factor of $s^4$ in \eqref{eq:DTTeqPiTT}.}
\begin{align}
    \cD_{\mathrm{TT}}^{\mu\nu\rho\sigma} &= (\gflat^{\mu(\rho}\gflat^{\sigma)\nu} \partial^2 + \gflat^{\mu\nu} \partial^\rho \partial^\sigma + \gflat^{\rho\sigma} \cP^{\mu\nu}-(\gflat^{\mu(\rho} \partial^{\sigma)}  \partial^\nu +\gflat^{\nu(\rho} \partial^{\sigma)}  \partial^\mu))\partial^2 + (d-2)\cP^{\mu\nu} \partial^\rho \partial^\sigma, \label{eq:DQdef}
    \end{align}
with manifest symmetries $\cD_{\mathrm{TT}}^{(\mu\nu)(\rho\sigma)}$.
In momentum space, this is
\begin{equation} \label{eq:DQmomentumSpace}
    \cD_{\mathrm{TT}}^{\mu\nu\rho\sigma} = (\gflat^{\mu(\rho}\gflat^{\sigma)\nu} s^2 + \gflat^{\mu\nu} s^\rho s^\sigma - \gflat^{\rho\sigma} \cP^{\mu\nu}-(\gflat^{\mu(\rho} s^{\sigma)}  s^\nu +\gflat^{\nu(\rho} s^{\sigma)}  s^\mu))s^2 - (d-2)\cP^{\mu\nu} s^\rho s^\sigma.
\end{equation}
This operator has a large kernel,
\begin{equation}
    \cD_{\mathrm{TT}}^{\mu\nu\rho\sigma}M_{\rho\sigma} = 0,\quad \forall \quad M_{\rho\sigma} = f_1 \gflat_{\rho\sigma} + f_2 s_\rho s_\sigma + f_3 s_{(\rho} \di_{\sigma)},
\end{equation}
for arbitrary scalar functions $f_i(s,\di)$. 
Hence, we can freely modify the structure of $R_{\rho\sigma}$ by adding terms with a tensor structure in this kernel.
To be minimal in our variables, and also manifestly symmetric, we take $R_{\rho\sigma} \propto \di_\rho\di_\sigma$.

$\cD_{\mathrm{TT}}$ is an unnormalized version of the standard spin-2 transverse-traceless projector
\begin{align}
    \cD_{\mathrm{TT}}^{\mu\nu\rho\sigma}&= s^4 \, \Pi_\mathrm{TT}^{\mu\nu\rho\sigma}(s)\label{eq:DTTeqPiTT}\\
    \Pi_\mathrm{TT}^{\mu\nu\rho\sigma}(p) &\equiv \Pi^{\mu(\rho} \Pi^{\sigma)\nu}%
     - \frac{1}{d-1} \Pi^{\mu\nu} \Pi^{\rho\sigma},\quad \Pi^{\mu\nu}(p) = \gflat^{\mu\nu} -\frac{p^\mu p^\nu}{p^2}, \label{eq:TTprojector}
\end{align}
which is indeed transverse, traceless, and a projector:
\begin{equation}
    \Pi_\mathrm{TT}^{\mu\nu\rho\sigma} p_\mu = 0, \quad \Pi_\mathrm{TT}^{\mu\nu\rho\sigma} \gflat_{\mu\nu} = 0, \quad \Pi\indices{_{\mathrm{TT}}^{\mu\nu}_{\alpha\beta}} \Pi_\mathrm{TT}^{\alpha\beta\rho\sigma} = \Pi_\mathrm{TT}^{\mu\nu\rho\sigma}, \quad \label{eq:PiTTproperties}
\end{equation}
with the first holding for a contraction on any index, and the second holding also for $\gflat_{\rho\sigma}$.
We have kept our results in terms of $\cD_\mathrm{TT}$, rather than the $\Pi_\mathrm{TT}$ in terms of which they are manifestly simpler, to ensure that the nonlocality of individual terms remains obvious.
\subsection{Solving the first three conditions}
In momentum space, the solution of \ref{prop1}, \ref{prop2}, and \ref{prop3} is immediate.
We simply write a manifestly symmetric momentum-space expression
\begin{equation}
T^{\mu\nu}(x) = \int_{p,q} e^{ix\cdot(p+q)}\phi(p) S^{\mu\nu}(p,q) \phi(q),
\end{equation}
where
\begin{equation}
    S^{\mu\nu}(p,q) = \gflat^{\mu\nu} a_1 + p^{\mu}p^{\nu} a_2 + q^{\mu}q^{\nu} a_3 + (p^{\mu}q^{\nu} + q^{\mu}p^{\nu}) a_4,
\end{equation}
for arbitrary functions $a_i(p^2,q^2, p\cdot q)$.
As discussed above, we can choose $S^{\mu\nu}(p,q) =S^{\mu\nu}(q,p)$, since the two $\phi$s are indistinguishable.
Our Fourier-transformed conditions \ref{prop2} and \ref{prop3} for $T^{\mu\nu}$ must then also be made manifestly symmetric with respect to $p\leftrightarrow q$:
\begin{align}
    \ref{prop2}:\,\quad & s_\mu S^{\mu\nu} = -\frac{p^\nu q^{2\zeta} + q^\nu p^{2\zeta}}{2}, \quad s_\nu S^{\mu\nu} =-\frac{p^\mu q^{2\zeta} + q^\mu p^{2\zeta}}{2} \label{eq:Sderiv};\\
    \ref{prop3}:\, \quad & \gflat_{\mu\nu} S^{\mu\nu} = -\left(\dotwo -\zeta\right) \frac{p^{2\zeta} + q^{2\zeta}}{2}\label{eq:Strace}.
\end{align}
Considering the coefficient of each tensor structure in \eqref{eq:Sderiv} as a separate equation, and using $\di \cdot s=p^2 -q^2$, we solve three algebraic equations for four unknowns $a_i$ to find
\begin{equation}
\begin{aligned}
    S^{\mu\nu}(p,q)&= -\frac{1}{2} \left(p^{2\zeta}+q^{2\zeta}\right) \left(\frac{\gflat^{\mu\nu}}{2} -\zeta  \frac{\cP^{\mu \nu }}{s^2}\right)+\frac{1}{4} \left(\frac{p^{2\zeta}-q^{2\zeta}}{p^2-q^2}\right) \left(\di ^{\mu } \di ^{\nu }-\di ^2 \frac{\cP^{\mu \nu }}{s^2}\right)\\
    & +\left[\cP^{\mu \nu } \left(\di^2 s^2 -(\di \cdot s)^2\right)-\left(s^2 \di ^{\mu } - \di \cdot s\, s^{\mu }\right) \left(s^2 \di ^{\nu }- \di \cdot s\, s^{\nu }\right)\right]\frac{Y(p,q)}{4 s^4},
\end{aligned}
\end{equation}
for an arbitrary scalar function $Y$.
The prefactor in square brackets of the undetermined $Y$ is exactly the transverse-traceless projector, $-\cD_{\mathrm{TT}}^{\mu\nu\rho\sigma}\di_\rho \di_\sigma$, so
\begin{equation}\begin{aligned} \label{eq:candidateTwithGeneralY}
    S^{\mu\nu}(p,q)&= -\frac{1}{4} \left(p^{2\zeta}+q^{2\zeta}\right) \left(\gflat^{\mu\nu} -2 \zeta  \frac{\cP^{\mu \nu }}{s^2}\right)+\frac{1}{4} \left(\frac{p^{2\zeta}-q^{2\zeta}}{p^2-q^2}\right) \left(\di ^{\mu } \di ^{\nu }-\di ^2 \frac{\cP^{\mu \nu }}{s^2}\right)\\
    & +\frac{\cD_{\mathrm{TT}}^{\mu\nu\rho\sigma}}{s^4}\cdot\frac{-\di_\rho \di_\sigma}{4} Y(p,q).
    \end{aligned}
\end{equation}

\subsection{Attack of the Gegenbauers: expanding the non-primary \texorpdfstring{$T$}{T} as a sum} \label{sec:whereGegenbauersArise}

Before solving for $Y$, let us focus for a moment longer on the other terms in \eqref{eq:candidateTwithGeneralY}.
One might ask how we could write our candidate $T^{\mu\nu}$ in a way that makes it more manifestly polynomial in the momenta for integer $\zeta$.
The standard difference-of-powers formula makes it obvious that 
\begin{equation}\label{eq:diffOfPow}
    \frac{P^\zeta-Q^\zeta}{P-Q}, \quad P=p^2, \quad Q=q^2
\end{equation}
is a polynomial in $P$ and $Q$ for integer $\zeta$; but what polynomial is it?
The answer is: a Gegenbauer.
This is the reason that the Gegenbauer polynomials appear in the $R_{\rho\sigma}$ term.

As mentioned in the introduction, the variables that we have immediate access to, by taking derivatives of the $\phi$ fields, are the momenta of the two $\phi$ fields $p$ and $q$.
It will turn out that the more useful variables for solving the primary condition are the following. 
Define $\theta$ to be the angle between $p$ and $-q$, and $\mg$ to be the product of the magnitudes of these two vectors
\begin{equation}\label{eq:cdef}
    \cos\theta \equiv \frac{-p\cdot q}{\sqrt{p^2 q^2}} \equiv c, \quad \mg \equiv \sqrt{p^2 q^2}.
\end{equation}
This is exactly the change-of-variables which we will later use to solve the primary condition; the minus sign is simply for convenience, though it does ensure that $\theta =0$ for $p=-q$.

Motivated by finding the polynomial to which \eqref{eq:diffOfPow} is equal, we temporarily change variables, $P=\mg e^\lambda, Q=\mg e^{-\lambda}$, to find
\begin{equation}
    \frac{P^\zeta-Q^\zeta}{P-Q}=\mg^{\zeta-1} \,\frac{\sinh(\zeta \lambda)}{\sinh \lambda}=\mg^{\zeta-1} \;C_{\zeta-1}^{(1)}\!\left(\cosh \lambda\right).
\end{equation} 
This is a standard fact about Gegenbauer polynomials (though it is more often seen written with a Chebyshev $U$-polynomial, since $C^{(1)}_n(x) = U_n(x)$).
The Gegenbauer's argument is
\begin{equation}
\cosh \lambda = \frac{P+Q}{2\sqrt{PQ}}  = \cos \theta + \frac{s^2}{2\mg},
\end{equation}
which we can expand for large $\mg$ using another standard result \eqref{eq:GegenbauerExpandedInApp},
\begin{equation}\begin{aligned}\label{eq:Gegenbauerexpansion}
C_{n}^{(\alpha)}(\cos\theta + y) &= \sum_{k=0}^\infty \frac{y^k}{k!} \odv[k]{}{w} C^{(\alpha)}_{n}(w) \rvert_{w=\cos\theta} = \sum_{k=0}^\infty \frac{\Pochhammer{\alpha}{k}}{k!} (2y)^k C^{(\alpha+k)}_{n-k}(\cos\theta).
\end{aligned}\end{equation}
Writing $c\equiv \cos\theta$ defined in \eqref{eq:cdef} for compactness, we find
 \begin{equation}\label{eq:diffPowsAsGegenbauers}
     \frac{P^\zeta-Q^\zeta}{P-Q}= \sum_{k=0}^\infty (s^2)^k \mg^{\zeta-k-1} C_{\zeta-k-1}^{(k+1)}(c)=\sum_{k=1}^\infty (s^2)^{k-1} \mg^{\zeta-k} C_{\zeta-k}^{(k)}(c).
 \end{equation}
Since the left-hand side is independent of $p\cdot q$, it is clear that the $c$-dependence of the right-hand side is illusory, and cancels between $s^2=P-2\mg c+Q$ and the Gegenbauer.

Since we can rewrite this term as a Gegenbauer, it is natural to ask if the other structure $P^\zeta + Q^\zeta$ also admits such a rewriting. 
It does:
\begin{equation}\label{eq:sumPowsAsGegenbauers}
\frac{P^\zeta + Q^\zeta}{2} = \mg^\zeta \cosh(\zeta \lambda) = \frac{\zeta}{2} \mg^\zeta \tilde{C}_\zeta^{(0)}\left(\cos \theta + \frac{s^2}{2\mg}\right) = \frac{\zeta}{2} \sum_{k=0}^\infty \frac{1}{k}(s^2)^k \mg^{\zeta-k} C_{\zeta-k}^{(k)}(c),
\end{equation}
where the \enquote{renormalized} Gegenbauer polynomial $\tilde{C}_\zeta^{(0)}$ is defined as a limit of the usual $C_\zeta^{(\alpha)}$ as $\alpha \to 0$, i.e.
\begin{equation}
    \tilde{C}^{(0)}_\zeta(c) \equiv \lim_{\alpha \to 0} \frac{C^{(\alpha)}_\zeta (c)}{\alpha} = \frac{2}{\zeta} T_\zeta(c) = \frac{2}{\zeta}\cosh(\zeta \arccosh(c)),
\end{equation}
where $T_\zeta$ is a Chebyshev polynomial.
This also means that the $k=0$ term in the sum \eqref{eq:sumPowsAsGegenbauers} is well-defined.

Thus, we see that the Gegenbauers turn up when we want to make the polynomial structure manifest. They also explicitly factor out the scale-dependent part, which is captured by $M = \abs{p}\abs{q}$, and each Gegenbauer in the sum depends only on the angle between the momenta.

\section{Solving the primary condition}\label{sec:primaryconstruction}

Defining $\hat{K}_b \equiv b^\rho \hat{K}_\rho$, it now remains to solve the primary condition \ref{prop4}
\begin{equation}
    \hat{K}_b T_{\mu\nu} =0.
\end{equation}
Using the strategy sketched in \cref{sec:sketchStrategy}, we need to turn $\hat{K}_b$ into a differential operator acting on functions of momenta.
Doing so, and applying it to our candidate $T^{\mu\nu}$, we will find an unwieldy set of differential equations for $Y$, which will give $R_{\rho\sigma}$. 
We will consider just one of these equations.
Solving it, using the hints from the previous section that we should rewrite it using Gegenbauers, will lead us to the correct $R_{\rho\sigma}$.

\subsection{Constructing \texorpdfstring{$\hat{K}_b$}{the SCT operator} as a differential operator}

We use exactly the rules of conformal primary operators outlined in \cite[\S 5]{Osborn:2016bev}.
We need to understand the action of $\hat{K}_b$ on spin-2 bilinears of $\phi$. 
Because $\hat{K}_b$ obeys the Leibniz rule, it suffices to consider only monomials of $\phi$ acted on by derivatives, possibly contracted with vectors.

\subsubsection{On monomials}

Acting on $\phi$ at the origin, we have the general rule that
\begin{equation}\begin{aligned}\label{eq:KonMonomial}
    \hat{K}_b (a\cdot \partial)^n \partial^{2k} \phi = &\frac{1}{2} (a\cdot \partial)^{n-2} \partial^{2(k-1)}
    \left[
    n(2(\Delta+2 k+n-1) (a\cdot b)(a\cdot \partial) \right.\\
    &\left.-(n-1) a^2 (b \cdot \partial)) \partial^2- 2 k (2k + 2\Delta -d) (a\cdot \partial )^2 (b\cdot \partial) 
    \right] \phi
\end{aligned}
\end{equation}
for $a$ an arbitrary vector. 
This reduces to \cite[(5.3-4)]{Osborn:2016bev} for $k=0,1$, and can be proven from them by induction on $k$ using \cite[(5.1)]{Osborn:2016bev}\footnote{
    That suffices for integer $n,k$. 
    More generally, we derive it using the momentum-space action of the special conformal generator acting on $f(x,-i\partial_x) \phi(x)$, $\hat{K}_\mu = \tfrac{1}{2i} (2 \Delta x_\mu - x^2 \partial_\mu + 2 x_\mu x^\alpha \partial_\alpha)$ -- the $\frac{1}{2i}$ implements the unusual $\hat{K}_b$ normalization of \cite{Osborn:2016bev} instead of \cite[\S 2.8]{Coriano:2020ccb}.  %
    This is  %
    \begin{equation}
        \hat{K}_\mu \sim -\Delta \pdv{}{p^\mu} + \half p_\mu \pdv{}{p_\alpha,p^\alpha}- p^\alpha \pdv{}{p^\alpha p^\mu},
    \end{equation}
    which immediately gives \eqref{eq:KonMonomial} on application to $a \cdot (i p)^n (-p^2)^k$. \label{footnote:Knorm}
}.
Differentiating with respect to $a^\mu$ (and then again with respect to $a^\nu$), we find 
{\footnotesize
\begin{equation}\begin{aligned}
    &\hat{K}_b \partial_\mu (a\cdot \partial)^n \partial^{2k} \phi = \frac{1}{2}  (a \cdot\partial )^{n-2}  \partial^{2(k-1)}\left[2 \partial ^2 (a \cdot\partial ) \left(b_{\mu } (a \cdot\partial ) (\Delta +2 k+n)-n a_{\mu } (b \cdot\partial )\right)\right. \\
    & \left. +\partial_{\mu } \left\{2 k (a \cdot\partial )^2 (b \cdot\partial ) (2 (\Delta +k)-d)+2 \partial^2 n (a \cdot b ) (a \cdot\partial ) (\Delta +2 k+n) -a^2 \partial^2 (n-1) n (b \cdot\partial )\right\}\right]\phi,
    \\
    &\hat{K}_b \partial_\mu \partial_\nu (a\cdot \partial)^n \partial^{2k} \phi= \frac{1}{2} (a \cdot\partial )^{n-2} \partial^{2(k-1)}  \left[2 \partial^2 (a \cdot\partial ) \left\{b_{\mu } \partial_{\nu } (a \cdot\partial ) (\Delta +2 k+n+1)\right. \right.\\
    &\left.-(b \cdot\partial ) \left(n a_{\mu } \partial_{\nu }+(a \cdot\partial ) \gflat_{\mu  \nu }\right)\right\}
    +
    \partial_{\mu } \left\{2 k \partial_{\nu } (a \cdot\partial )^2 (b \cdot\partial ) (2 (\Delta +k)-d)\right.\\
    &\left.\left.+\partial^2 \left(n (b \cdot\partial ) \left(-2 a_{\nu } (a \cdot\partial )-a^2 (n-1) \partial_{\nu }\right)+2 (a \cdot\partial ) (\Delta +2 k+n+1) \left(b_{\nu } (a \cdot\partial )+n \partial_{\nu } (a \cdot b)\right)\right)\right\}\right]\phi.
\end{aligned}
\end{equation}
}

\subsubsection{On bilinears}
Consider now a bilinear at the point $x=0$ of the form 
\begin{equation}
    B = \phi A_{\mu\nu} (-\lptl^2)^\alpha (-\rptl^2)^\beta (-\lptl \cdot \rptl)^\gamma \phi|_{x=0}, \quad A_{\mu\nu} \in \{\gflat_{\mu\nu}, -\lptl_\mu \lptl_\nu, -\rptl_\mu \rptl_\nu,- \lptl_{(\mu} \rptl_{\nu)}\}.
\end{equation}
Applying $\hat{K}_b$ to this bilinear, and using the \href{https://en.wikipedia.org/wiki/General_Leibniz_rule}{Leibniz/product rule}, we find an expression 
\begin{equation}
    \hat{K}_b \phi A_{\mu\nu} (-\lptl^2)^\alpha (-\rptl^2)^\beta (-\lptl \cdot \rptl)^\gamma \phi\rvert_{x=0} = \phi h_{\mu\nu}(b,-i\lptl, -i\rptl, \alpha,\beta,\gamma)\phi\rvert_{x=0}.
\end{equation}
For simplicity, we transform this to momentum space, and define $P=p^2,Q=q^2,R=p\cdot q$:
\begin{align}
    B_{\mu\nu} &= \int_{p,q} e^{i(p+q)\cdot x} \phi(p) A_{\mu\nu} P^\alpha Q^\beta R^\gamma \phi(q)\rvert_{x=0}, \quad A_{\mu\nu} \in \{\gflat_{\mu\nu}, p_\mu p_\nu, q_\mu q_\nu, p_{(\mu} q_{\nu)}\}.\\
    \hat{K}_b B_{\mu\nu} &= \int_{p,q} \phi(p) H_{\mu\nu}(b, p, q, \alpha,\beta,\gamma)\phi(q).
\end{align}
For example, for $A_{\mu\nu}=\gflat_{\mu\nu}$ we find
{\footnotesize
\begin{equation}
    \begin{aligned}
\frac{H_{\mu\nu}}{ P^\alpha Q^\beta R^\gamma} &= -\half \gflat_{\mu  \nu } P^{-1} Q^{-1} R^{-2} \\
\quad &\times \left[ Q (b  \cdot p) \left((\gamma -1) \gamma  P Q-2 \gamma  P R (2 \beta +\gamma +\Delta -1)-4 \alpha  R^2 (\alpha +\Delta-\dotwo)\right)\right.\\
    \quad &\;\; \left.+P (b  \cdot q) \left((\gamma -1) \gamma  P Q-2 \gamma  Q R (2 \alpha +\gamma +\Delta -1)-4 \beta  R^2 (\beta +\Delta -\dotwo)\right)\right].
    \end{aligned}
\end{equation}
}
Note that after evaluating, there are no inverses $P^{-1} Q^{-1} R^{-2}$ in the final result.
We can therefore obtain the action on  all possible bilinears that are monomials in $P,Q,R$.
To upgrade this to the action on all possible bilinears $B_{\mu\nu}\propto \gflat_{\mu\nu}$, we simply formally replace 
\begin{equation}
    \alpha \mapsto P \odv{}{P}, \quad  \beta \mapsto Q \odv{}{Q}, \quad\gamma \mapsto R \odv{}{R}.
\end{equation}
For each possible structure $A$, we find a different differential operator $\hat{K}^A_b$ which maps from scalar functions to two-index tensors, such that 
\begin{equation}
    \hat{K}_b \int_{p,q}  \phi(p) \left(\sum_{A} A_{\mu\nu} f_A(P,Q,R) \right)\phi(q) =  \int_{p,q}  \phi(p) \left(\sum_{A} [\hat{K}_b^A f_A(P,Q,R)]_{\mu\nu} \right)\phi(q).
\end{equation}
These are given in \cref{app:Kbhat}; in the case of the structure $A_{\mu\nu}=\gflat_{\mu\nu}$, $\hat{K}_b^{\gflat}f_\gflat$ is \eqref{eq:KbonMetric}.

Replacing $\sum_A A_{\mu\nu} f_A$ by our candidate $S_{\mu\nu}(p,q)$ \eqref{eq:candidateTwithGeneralY} with a general $Y(P,Q,R)$, and setting the result to zero, we find a single equation. 
Schematically, it is
\begin{equation}
    \begin{aligned}\label{eq:cis} 
    0=&c_1 \left(b^{\nu} q^{\mu} + b^{\mu} q^{\nu}\right) + c_2 \left(b^{\nu} p^{\mu} + b^{\mu} p^{\nu}\right) + c_3 p^{\mu} p^{\nu} (b \cdot p) + c_4 (b \cdot p) \gflat^{\mu\nu} + c_5 (b \cdot q) \left(p^{\nu} q^{\mu} + p^{\mu} q^{\nu}\right)\\
    & + c_6 (b \cdot q) \gflat^{\mu\nu} + c_7 (b \cdot p) \left(p^{\nu} q^{\mu} + p^{\mu} q^{\nu}\right) + c_8 p^{\mu} p^{\nu} (b \cdot q) + c_9 q^{\mu} q^{\nu} (b \cdot p) + c_{10} q^{\mu} q^{\nu} (b \cdot q),
    \end{aligned}
\end{equation}
where the $c_i$s depend on $P,Q,R$, $d$, and $\zeta$.
In order for $S_{\mu\nu}$ to be primary, each $c_i$ must vanish independently.
It will suffice for the moment to focus on analytically solving only one linear combination of these ten equations -- the simplest.

\subsection{Solving one of the equations}

By inspection of the full form of \eqref{eq:cis}, we notice that $c_1-c_2$ is particularly simple: it is
\begin{equation}
\begin{aligned}\label{eq:Yequation}
0&= \zeta \frac{P^\zeta + Q^\zeta}{P-Q}-(P+Q)\frac{P^\zeta-Q^\zeta}{(P-Q)^2} +
\frac{1}{P+Q+2 R}\left[\frac{d-2 \zeta+2}{2} (P-Q) Y\right.\\
&\left. + R (P-Q) \pdv{Y}{R} +2 Q (P+R) \pdv{Y}{Q} -2 P (Q+R) \pdv{Y}{P}\right]
\end{aligned}
\end{equation}
Considering this as $ J(P,Q,R) + f(P,Q,R) Y+ \cL \,Y =0$, a natural first question is: what is the kernel of the differential operator $\cL$?
We observe that $\cL\, h(s^2, \cos\theta)=0$ for arbitrary functions $h$, which motivates the ansatz
\begin{equation}
    Y(P,Q,R) = \mg^{\zeta-1-\dotwo} Z(s^2, \cos\theta, \mg).
\end{equation}
This ansatz also removes the term $\propto Y$ in \eqref{eq:Yequation}, reducing it to
\begin{align}
    &\zeta \frac{P^{\zeta}+Q^{\zeta}}{(P-Q)^2}-(P+Q)\frac{P^\zeta-Q^\zeta}{(P-Q)^3} +\frac{1}{s^2} \mg^{\zeta-\dotwo-1} \mg \pdv{Z}{\mg}_{s^2,\cos\theta}=0,
\end{align}
which we now want to solve.
Motivated by the discussion in \cref{sec:whereGegenbauersArise}, we change variables again to $\mg,\lambda$, and spot precisely a Gegenbauer polynomial:
\begin{equation}
    s^2 \mg^{\zeta-3} \underbrace{\half \frac{\zeta \cosh \zeta \lambda- \frac{\sinh \zeta \lambda}{\tanh \lambda}}{(\sinh\lambda)^2}}_{C_{\zeta-2}^{(2)}(\cosh \lambda)} + \mg^{\zeta -\dotwo-1} \pdv{Z}{\mg}_{s^2,\cos\theta}=0. \label{eq:Cspotted}
\end{equation}
Now, from the previous section, $R=-\mg\cos\theta$, and $\cosh \lambda = (P+Q)/(2\mg) = \cos \theta +s^2/(2\mg)$.
\begin{equation}
    Z = -s^2\int^\mg \odif{\mg} \, \mg^{\dotwo-2} C_{\zeta-2}^{(2)}(\cos\theta + \tfrac{s^2}{2 \mg}) + c_Z(s^2, \cos\theta),
\end{equation}
where $c_Z(s^2,\cos\theta)$ is the \enquote{constant} of integration.
Expanding this Gegenbauer polynomial for large $\mg$ using \eqref{eq:GegenbauerExpandedInApp}, we find
\begin{align}
    C_{\zeta-2}^{(2)}(\cos\theta + y) &=\sum_{k=2}^\infty (k-1) (2y)^{k-2} C^{(k)}_{\zeta-k}(\cos\theta).
\end{align}
We then switch the integral and sum. 
This is clearly justified for integer $\zeta$, because the sum truncates after $k=\zeta$; for noninteger $\zeta$, we can use this to define a formal power series that technically converges only for $\mg \gg s^2$.
$Y$ becomes
\begin{align}
    Y &= - \mg^{\zeta-1-\dotwo}  \sum_{k=2}^\infty (k-1) (s^2)^{k-1} C_{\zeta-k}^{(k)}(\cos\theta) \int^\mg \odif{\mg} \, \mg^{\dotwo-k} +   \mg^{\zeta-1-\dotwo} c_Z\\
    &= \sum_{k=2}^\infty \frac{k-1}{k-1-\dotwo} (s^2)^{k-1} \mg^{\zeta-k} C_{\zeta-k}^{(k)}(\cos\theta) +  Y_\text{hom}. \label{eq:Yfound}
\end{align}
$Y_\text{hom} =\mg^{\zeta-1-\dotwo} c_Z$ is an as-yet-undetermined function which we will fix by considering the other 9 equations coming from \eqref{eq:cis} -- it will be forced to be zero for integer $\zeta$ by requiring $T^{\mu\nu}$ to be a local operator.

We could also write the nonhomogeneous part as 
\begin{equation}
    Y = \left[1 + \dotwo (s^2)^{\dotwo} \int \odif{s^2} (s^2)^{-1-\dotwo} \, \cdot\, \right] \frac{P^\zeta-Q^\zeta} {P-Q},
\end{equation}
where the integral is taken at constant $(\mg,\cos\theta)$, i.e. constant $(\abs{p}\abs{q},p\cdot q)$, and the integration constant is dropped.

\subsubsection{The homogeneous solutions give extra nonlocal primaries}\label{sec:homogeneous}

By construction, \eqref{eq:Yfound} sets the coefficient of one of the structures in \eqref{eq:cis} to zero.
However, solving this one equation has led to an undetermined function $Y_\text{hom}(p,q) = \mg^{\zeta-\dotwo-1} c_Z(s^2, \cos\theta)$.
For $Y_\text{hom}=0$, it can be verified by a straightforward computation that every $c_i=0$ in $\eqref{eq:cis}$. 
So, if it is nonzero, $Y_\text{hom}$ must be a homogeneous solution to the primary condition,
\begin{equation}\label{eq:KYhom}
    \hat{K}_b \int_{p,q}  \frac{\cD_{\mathrm{TT}}^{\mu\nu\rho\sigma} }{s^4}\phi(p) \frac{\chi_\rho\chi_\sigma}{4}Y_\text{hom}\phi(q)  = 0. 
\end{equation}
Computing \eqref{eq:KYhom}, the coefficient of $b^\mu p^\nu$ (i.e. $c_2$ in \eqref{eq:cis}) is
\begin{equation}
    s^2 \pdv{c_Z}{s^2}_{\cos\theta} %
    =\frac{d}{2} c_Z, \quad \implies c_Z(s^2,\cos\theta)=(s^2)^{\tfrac{d}{2}} V(\cos\theta).
\end{equation}
Hunting for the simplest equation, we see that $c_4-c_6=0$ is
\begin{equation}
    \left(c^2-1\right) V''(c)+ c (d+3) V'(c)+ (\dotwo + \zeta +1) (\dotwo -\zeta +1)V(c)=0
\end{equation}
whose solutions for $V$, and therefore for $Y_\text{hom}$, are the following associated Legendre functions,
\begin{align}
    V(c) &= \left(1-c^2\right)^{-\frac{d}{4}-\frac{1}{4}} \left(C_1 P_{\zeta -\frac{1}{2}}^{\frac{d+1}{2}}(c)+C_2 Q_{\zeta -\frac{1}{2}}^{\frac{d+1}{2}}(c)\right),\\
    Y_{\text{hom}}
    &= (s^2)^{\tfrac{d}{2}} \mg^{\zeta-\dotwo-1}  \frac{C_1 P_{\zeta -\frac{1}{2}}^{\frac{d+1}{2}}(c)+C_2 Q_{\zeta -\frac{1}{2}}^{\frac{d+1}{2}}(c)}{\left(1-c^2\right)^{\frac{d+1}{4}}},
\end{align}
for arbitrary constants $C_1$ and $C_2$\footnote{
Recall that the associated Legendre functions are defined for general parameters by \cite[(14.3.1)]{NIST:DLMF},
and hence an alternative basis of solutions can be obtained using
\begin{equation}
    \frac{P_{\zeta -\frac{1}{2}}^{\frac{d+1}{2}}(c)}{\left(1-c^2\right)^{\frac{d+1}{4}}} = (1-c)^{-\tfrac{d+1}{2}} \,_{2}\tilde{F}_{1}(\thalf-\zeta ,\zeta+\thalf;\tfrac{1-d}{2} ;\tfrac {1-c}{2}).
\end{equation}
Since the associated Legendre $Q$ is a linear combination of $P$s \cite[14.9(i)]{NIST:DLMF}, we find
\begin{equation}
    \tfrac{Y_\mathrm{hom}}{ s^{d}\mg^{\zeta-\dotwo-1}}= D_1 (1-c)^{-\tfrac{d+1}{2}} \,_{2}\tilde{F}_{1}(\thalf-\zeta ,\zeta+\thalf;\tfrac{1-d}{2} ;\tfrac {1-c}{2}) + D_2(1+c)^{-\tfrac{d+1}{2}} \,_{2}\tilde{F}_{1}(\thalf-\zeta ,\zeta+\thalf;\tfrac{d+3}{2} ;\tfrac {1-c}{2}).
\end{equation}
}.
This $Y_\text{hom}$ completely solves \eqref{eq:KYhom}.
However, for all values of $d$ these are not polynomial in momenta, and hence are not local. 
They could only arise from a nonlocal action, so we should discard them in the integer-$\zeta$ cases (where we want a local theory).

One possible interpretation of these two degrees of freedom is as an additional pair of \textit{nonlocal primaries} $\cO_{1,2}^{\mu\nu}$ (for example, light-ray operators in Lorentzian CFT are often described as such \cite{Kravchuk:2018htv}), in the sense that they are spin-2 symmetric primaries of the global conformal group satisfying $\hat{K}_b \cO_i^{\mu\nu} \rvert_{x=0}=0$:
\begin{equation}
    \cO_{1,2}^{\mu\nu}(x) = \int_{p,q} e^{i x\cdot (p+q)} \frac{\cD_{\mathrm{TT}}^{\mu\nu\rho\sigma}}{s^4} \phi(p) \frac{\chi_\rho\chi_\sigma}{4}Y_\text{hom}(p,q)\phi(q) \Big\rvert_{C_{1,2} =1, \, C_{2,1}=0}.
\end{equation}
They are automatically traceless and conserved (even off-shell, thanks to the $\cD_{\mathrm{TT}}$ structure), and so also have dimension $\Delta_{\cO_{1,2}}=d$. 
In the noninteger case $T^{\mu\nu}$ is itself nonlocal (as are $T^{\mu\nu}_W$ and $S_W^g$), so we cannot justify dropping the $\cO_{1,2}$s -- but it would be interesting to find a physical interpretation, either way.
In any case, we can take the \enquote{actual} $T^{\mu\nu}$ in the noninteger case to be given by the following two-parameter extension of $R_{\rho\sigma}$,
\begin{align}\label{eq:RdefHom}
R_{\rho\sigma}^{\text{ext}} = \eqref{eq:Rdef} + \phi \frac{\lrptl_\rho \lrptl_\sigma}{4} (-\partial^2)^{\tfrac{d}{2}} \mg^{\zeta-\dotwo-1}  \frac{C_1 P_{\zeta -\frac{1}{2}}^{\frac{d+1}{2}}(\cos\theta)+C_2 Q_{\zeta -\frac{1}{2}}^{\frac{d+1}{2}}(\cos\theta)}{\left(1-\cos^2\theta\right)^{\frac{d+1}{4}}}\phi.
\end{align}
This gives us three nonlocal symmetric, conserved, traceless, primary operators: $T^{\mu\nu}\rvert_{R=R^{\text{ext}}}$ and $\cO_{1,2}^{\mu\nu}$.

These parameters $C_{1,2}$ give us a two-parameter family of EM tensors, and hence a two-parameter family of Weyl-covariant actions (assuming $d \notin 2\mathbb{N}$). 
Investigating this is beyond the scope of this paper, but there are two immediate questions which suggest themselves.
First, what is the geometric origin of this freedom?
Second, is the natural analytic continuation of the integer-$\zeta$ case ($C_1=0=C_2$) special in any way?

\subsection{Comments on the local solution}

We now focus on the solutions with $Y_\text{hom}=0$ in \eqref{eq:Yfound}.
The fact that $C^{(\alpha)}_{\beta}(x)=0$ for negative integer $\beta$ means that the infinite sum truncates for integer $\zeta$.
The Gegenbauer polynomial $C_{n}^{(\alpha)}$ is a polynomial of degree $n$, which is even for $n$ even and odd for $n$ odd. 
These two facts mean that, if $\zeta$ is an integer, then $\mg^{\zeta -k} C_{\zeta-k}^{(k)}(-p\cdot q/\mg)$ is either zero or it is a polynomial of degree $\zeta-k$ in the variables $p\cdot q$ and $\mg^2=p^2 q^2$.
Hence, for integer $\zeta$, $Y$ is a finite polynomial in $(p+q)^2$, $p\cdot q$, and $p^2 q^2$, and therefore contains a finite number of derivatives.
All dependence on $\dotwo$ is localized to a single place in the sum, making the pole structure in $d$ obvious.
For example, it is now clear that the Weyl-covariant $\zeta=k$ CFT does not exist for $d=2,4,6,\cdots,2(k-1)$, as the pole gives an obstruction to the existence of a primary EM tensor.

We conclude that the full EM tensor is given by
\begin{equation}\label{eq:fullMomSpace}
    \begin{aligned}
T^{\mu\nu}(x) &= \int_{p,q} e^{ix\cdot(p+q)}\phi(p) S^{\mu\nu}(p,q) \phi(q)\\
    S^{\mu\nu}(p,q)&= -\frac{p^{2\zeta}+q^{2\zeta}}{2} \left(\frac{\gflat^{\mu\nu}}{2} - \zeta  \frac{\cP^{\mu \nu }}{s^2}\right)+ \left(\frac{p^{2\zeta}-q^{2\zeta}}{p^2-q^2}\right) \frac{1}{4}\left(\di^\mu \di^\nu -\di ^2 \frac{\cP^{\mu \nu }}{s^2}\right)\\
    &\quad-\cD_{\mathrm{TT}}^{\mu\nu \rho \sigma} \frac{\di_\rho \di_\sigma}{4} \sum_{k=2}^\infty \frac{k-1}{k-1-\dotwo} (s^2)^{k-3} \mg^{\zeta-k} C_{\zeta-k}^{(k)}(\cos\theta),
\end{aligned}
\end{equation}
which can be simplified to
\begin{equation}
    \begin{aligned}
        T^{\mu\nu}(x) &= -\left(\half \gflat^{\mu\nu}-\zeta \frac{\cP^{\mu\nu}}{-\partial^2}\right) \cO + \left(\gflat^{\mu(\rho}\gflat^{\sigma)\nu}- \frac{\cP^{\mu\nu} \gflat^{\rho\sigma}}{-\partial^2}\right) U_{\rho\sigma} + \frac{\cD_{\mathrm{TT}}^{\mu\nu\rho\sigma}}{(-\partial^2)^2} R_{\rho\sigma},\\
        \cO &= \phi (-\partial^2)^\zeta \phi\\
        U_{\rho\sigma}(x) &\equiv \int_{p,q} e^{i x\cdot s} \phi(p) \frac{\di_\rho \di_\sigma}{4}  \frac{p^{2 \zeta }-q^{2\zeta} }{p^2-q^2} \phi(q)\\
        R_{\rho\sigma} &\equiv \sum_{k=2}^\infty \frac{k-1}{k-1-\dotwo} (-\partial^2)^{k-1}\int_{p,q}e^{i x\cdot s} \phi(p) \frac{-\di_\rho \di_\sigma}{4} \mg^{\zeta-k} C_{\zeta-k}^{(k)}(\cos\theta)\phi(q)
\end{aligned}
\end{equation}
$U$ and $R$ are trivially local for integer $\zeta$, though they are acted on by nonlocal differential operators.
Of course, for integer $\zeta$, the overall $T^{\mu\nu}$ must be local, as we shall show in the next section.

Written purely in terms of derivatives, we map $p\mapsto -i \lptl$, $q\mapsto -i \rptl$, $\di \mapsto -i \lrptl$; and $s \mapsto -i\partial_\mu$ sitting outside the operator. 
Defining for convenience
\begin{equation}
    \mg=\sqrt{\lptl^2 \rptl^2}, \quad \cos\theta = \frac{\lptl \cdot \rptl}{\mg},
\end{equation}
we find exactly\footnote{
    We could use the kernel of $\cD_{\mathrm{TT}}$ to simplify $\di_\rho\di_\sigma/4 \mapsto q_\rho q_\sigma$, but this way keeps the symmetry manifest.
    } \eqref{eq:fullT}. 
    For the user's sanity, observe that we can expand $U$ as an infinite sum,
\begin{equation}
U_{\rho\sigma} = -\phi \frac{\lrptl_\rho \lrptl_\sigma}{4}\frac{(-\lptl^2)^{\zeta }-(-\rptl^2)^{\zeta} }{(-\lptl^2)-(-\rptl^2)} \phi = -\sum_{k=1}^\infty (-\partial^2)^{k-1}  \left[\phi \frac{\lrptl_\rho \lrptl_\sigma}{4} \mg^{\zeta-k} C_{\zeta-k}^{(k)}(\cos\theta)\phi\right]
\end{equation}
and extend the sum in $R$ to $k=1$. 
Then the non-$\cO$ part can be rewritten
\begin{equation}\begin{aligned}\label{eq:sumOfDerivsActingonTensors}
       T^{\mu\nu}(x) &= -\left(\half \gflat^{\mu\nu}-\zeta \frac{\cP^{\mu\nu}}{-\partial^2}\right) \cO \\
    + &\sum_{k=1}^\infty\left[\frac{\cP^{\mu\nu}}{-\partial^2} \gflat^{\rho\sigma}-\gflat^{\mu(\rho}\gflat^{\sigma)\nu} + \frac{\cD_{\mathrm{TT}}^{\mu\nu\rho\sigma}}{(-\partial^2)^2} \frac{k-1}{k-1-\dotwo}  \right] (-\partial^2)^{k-1} R^{(k)}_{\rho\sigma}\\
       R^{(k)}_{\rho\sigma} &\equiv \phi\frac{\lrptl_\rho \lrptl_\sigma}{4}  \mg^{\zeta-k} C_{\zeta-k}^{(k)}(\cos\theta)\phi.
       \end{aligned}
\end{equation}
This is manifestly just a sum of nonlocal derivatives acting on the symmetric local operators $R^{(n)}_{(\rho\sigma)}$.
Of course, in the case of noninteger $\zeta$, the position-space version can only be properly defined via \eqref{eq:hypersingular} and its generalization to $\zeta>1$; however, the momentum-space definition works just as well.
These equations are the main result of this work.

\section{Properties of this \texorpdfstring{$T$}{T}} \label{sec:properties}

We now study various properties of our $T^{\mu\nu}$, ignoring the nonlocal extension.
We confirm the locality of this stress tensor for integer $\zeta$, and then check the three-point function coefficient $C_{T\phi\phi}$ -- two computations which are straightforward; and then we provide a partial proof of the two-point function $\expval{TT}$ normalization $C_T$ for integer $\zeta$ -- which is not so straightforward.

\subsection{Verification of locality and a Weyl-inspired rewriting}\label{sec:locality}

Can we prove that the $T^{\mu\nu}$ we have constructed is local for $\zeta \in \mathbb{N}$?
Recall that in \cref{sec:localnonlocal} we defined locality for bilinears as being polynomial in derivatives.
For integer $\zeta$, it is easy to see that most of the terms in the momentum-space expression \eqref{eq:fullMomSpace} are indeed polynomial: in the sum over $k$, the $(s^2)^{k-3}$ means that we only need to keep $k=2$, since otherwise the summands are manifestly local.
The only dangerous terms are
\begin{equation}
    \begin{aligned}
    &4S_{\mu\nu}(p,q) s^2\supset 2 \zeta \left(p^{2\zeta}+q^{2\zeta}\right)  \cP_{\mu \nu } - \left(\frac{p^{2\zeta}-q^{2\zeta}}{p^2-q^2}\right)\di ^2 \cP_{\mu \nu } -\cD_{\mathrm{TT}}^{\mu\nu \rho \sigma} \di_\rho \di_\sigma \frac{\mg^{\zeta-2} C_{\zeta-2}^{(2)}(\cos\theta)}{1-\dotwo}. \label{eq:mustGets2}
\end{aligned}
\end{equation}
Due to the $\cD_{\mathrm{TT}}$, this contains factors of both $(d-1)$ and $(d-2)$. However, using partial fractions, we can make it linear in $1/(d-1)$ and $1/(d-2)$, and we study each one separately.
The coefficients of each must contain an $s^2$ to ensure locality.
The coefficient of $1/(d-2)$ in \eqref{eq:mustGets2} is manifestly proportional to $s^2$:
\begin{equation}\begin{aligned}
    \mg^{\zeta-2} C_{\zeta-2}^{(2)}(\cos\theta)[\di^2(s_\mu s_\nu - s^2 \gflat_{\mu\nu}) + s^2 \di_\mu \di_\nu - \di\cdot s(\di_\mu s_\nu + \di_\nu s_\mu) + \gflat_{\mu\nu}(\di \cdot s)^2] s^2.
    \end{aligned}
\end{equation}
The presence of an $s^2$ prefactor in the $1/(d-1)$ coefficient of \eqref{eq:mustGets2} is harder to see. 
Stripping off a factor of $\cP_{\mu\nu}$, it is
\begin{equation}\begin{aligned}\label{eq:dm1coeff}
    2\zeta (p^{2\zeta} +q^{2\zeta}) - (p-q)^2 \frac{p^{2\zeta} -q^{2\zeta}}{p^{2}-q^{2}} - 8((p\cdot q)^2-p^2 q^2)\mg^{\zeta-2}C_{\zeta-2}^{(2)}(\cos\theta).
    \end{aligned}
\end{equation}
We now rewrite this expression in terms of $s^2, A=\di^2,B=s\cdot \di$, and give a formal polynomial argument that shows the existence of an overall $s^2$ prefactor.
Treating $A$ and $B$ as formal independent variables, take the variable $s^2\to 0$, which yields
\begin{equation}
    2^{-2\zeta}\left[\tfrac{A}{B} \left((A-2 B)^{\zeta }-(A+2 B)^{\zeta }\right)+2 \zeta  \left((A+2 B)^{\zeta }+(A-2 B)^{\zeta }\right)-32 B^2 t^{\zeta-2} C_{\zeta -2}^{(2)}\left(\frac{A}{\sqrt{t^2}}\right)\right],
\end{equation}
where we have defined $t^2 =A^2-4B^2$ for compactness.
From the definition of $C^{(2)}_{\zeta-2}$ that we used in \eqref{eq:Cspotted}, but with $\cosh \lambda = A/\sqrt{t^2}$, this vanishes automatically.
Because \eqref{eq:mustGets2} is a polynomial in $s^2$, $\di \cdot s$, and $\di^2$, it must have an overall factor of $s^2$. 
Thus, $S^{\mu\nu}(p,q)$ is also a polynomial in $p$ and $q$ for integer $\zeta$, and we have proven the locality of our $T^{\mu\nu}$.
This also means that if we want a local $T^{\mu\nu}$ that satisfies \ref{prop1}, \ref{prop2}, \ref{prop3}, but not \ref{prop4}, we only need to keep the $k=2$ term of \cref{eq:Rdef} (one can also, optionally, add $R_{\rho\sigma}'=(-\partial^2)^2 \cO_{\text{local}}$ for any local operator $\cO_{\text{local}}$). 
Such a $T^{\mu\nu}$ is a non-primary operator that vanishes under the application of either $P_\mu$ (i.e. when multiplied by $s_\mu$) or $P_\mu K_\rho$, but not under $K_\rho$ alone.

Alternatively, one can rewrite the $T^{\mu\nu}$ to make locality manifest.
Stergiou, Vacca, and Zanusso defined the specific Weyl-inspired form \cite[(22)]{Stergiou:2022qqj}
\begin{equation}
    T^{\mu\nu} = T^{\mu\nu}_c -\cP^{\mu\nu} \cO_\cP + \cD^{\mu\nu}{}_{\rho\theta} \mathcal{S}^{\rho \theta} + \cD_B^{\mu\nu}{}_{\rho\theta}\tilde{\mathcal{S}}^{\rho \theta}.
\end{equation}
Insisting on this form leads to expressions that are manifestly local for $\zeta\in\mathbb{N}$, but are much uglier than necessary, as we demonstrate in \cref{app:OSform}.
For example, in our language for $\zeta=2$ we find simply
\begin{equation}
    R_{\rho\sigma}\rvert_{\zeta=2} = \frac{\partial^2 (\phi \partial_\rho\partial_\sigma \phi)}{2(d-2)},
\end{equation}
where we have used the kernel of $\cD_{\mathrm{TT}}$ to replace $\lrptl_\rho\lrptl_\sigma\mapsto 4 \rptl_\rho \rptl_\sigma$.

\subsection{Verifying \texorpdfstring{$C_{T\phi\phi}$}{C\_T phi phi}}

The Euclidean three-point correlator $\expval{T \phi \phi}$ is fixed by conformal symmetry and conservation to be \cite[(3.1)]{Osborn:1993cr}
\begin{equation}\label{eq:fixedTphiphi}
    \expval{T_{\mu\nu}(x_1) \phi(x_2) \phi(x_3)} = -C_{T\phi\phi}\frac{(X_{23})_\mu(X_{23})_\nu - \frac{\gflat_{\mu\nu}}{d} (X_{23})^2 }{(x_{12}^2 x_{13}^2)^{\dotwo -1} (x_{23}^2)^{1-\zeta}} , \quad (X_{23})_\mu = \frac{(x_{12})_\mu}{x_{12}^2} -  \frac{(x_{13})_\mu}{x_{13}^2}.
\end{equation}

For integer $\zeta$, we can apply the $\phi \times \phi$ OPE inside this correlator to conclude that this gives the OPE coefficient $C_{T\phi\phi}$: i.e., $\phi \times \phi \ni C_{T\phi\phi} T^{\mu\nu}$.
For noninteger $\zeta$, $T^{\mu\nu}$ does not appear in the $\phi\times \phi$ OPE because it is a nonlocal operator: hence this $C_{T\phi\phi}$ is just the coefficient of the three-point function. 
However, we keep the name $C_{T\phi\phi}$ in both cases.

\subsubsection{Independent calculation of \texorpdfstring{$C_{T\phi\phi}$}{C\_T phi phi}}

Now, let us set $x_2=0$ in \eqref{eq:fixedTphiphi}.
For an infinitesimal scaling transformation $\epsilon_\mu = \epsilon x_\mu$, $\delta_\epsilon \phi(x) = \epsilon(\Delta + x^\mu \partial_\mu)\phi(x)$; then the associated Ward identity relates the variation of the two-point function to two infinitesimal surface integrals around the operator insertions.
Considering only the contribution from the point $x_2=0$, we have
\begin{equation}
    \expval{\delta_\epsilon\phi(0) \phi(x_3)}  = \lim_{r\to 0} -\epsilon \oint_{S_{x_2}^r} \odif{S^\mu} r^\nu \expval{T_{\mu\nu}(x_1) \phi(0) \phi(x_3)}, \, \quad r_\mu \equiv (x_1-x_2)_\mu.
\end{equation}
Evaluating each side is straightforward and yields
\begin{equation}
    \begin{aligned}
    \frac{\epsilon (\Delta+0) C_\phi}{(x_{23}^2)^\Delta} &= \lim_{r\to 0} -\epsilon \int\odif{\Omega_{d-1}} r^{d-2} r^\mu r^\nu \expval{T_{\mu\nu}(x_1) \phi(0) \phi(x_3)} \\
    &= C_{T\phi\phi} \epsilon \int\odif{\Omega_{d-1}}  \frac{\frac{d-1}{d}}{(x_{13}^2)^{\dotwo -\zeta}}.
    \end{aligned}
\end{equation}
Hence, the structure constant (also derived by \cite[(6.20)]{Osborn:1993cr}) should be
\begin{equation} \label{eq:CTphiphi}
    C_{T\phi\phi}=\frac{1}{S_d} \frac{d}{d-1} \Delta C_\phi,
\end{equation}
where the position-space normalization of $\phi$ is $C_\phi=\frac{\Gamma \left(\frac{d}{2}-\zeta \right)}{\pi ^{d/2} 2^{2 \zeta } \Gamma (\zeta )}$. 
We now confirm this.

\subsubsection{Calculating \texorpdfstring{$C_{T\phi\phi}$}{C\_T phi phi} directly}

Transforming \eqref{eq:fixedTphiphi} to momentum space and substituting \eqref{eq:CTphiphi}, we see that we should obtain \cite[(3.29)]{Diab:2016spb}
\begin{equation}\label{eq:FourierOfCTpp}
    \expval{z^\mu z^\nu T_{\mu\nu}(0)\phi(p) \phi(q)} = 2\zeta \frac{(p \cdot z)^2}{(p^2)^{\zeta +1}} \delta(p+q),
\end{equation}
where we have taken the EM tensor $T_{\mu\nu}(p) = \int_k \phi(k) S_{\mu\nu}(k,p-k) \phi(p-k)$ to be at zero momentum, $p\to 0$, and contracted with a null vector $z^2=0$ for simplicity.

Carefully taking the limit $s\to 0$ and $\di=2k-p\to 2k$ in our expression \eqref{eq:fullMomSpace}, we find
\begin{equation}\begin{aligned}
       S_{\mu\nu}(k,-k) &= -\frac{1}{2} k^{2\zeta} \gflat^{\mu  \nu }+\zeta k^{\mu } k^{\nu } k^{2(\zeta-1)} ,
       \end{aligned}
\end{equation}
which is in perfect agreement with \eqref{eq:FourierOfCTpp}:
\begin{equation}
\expval{z^\mu z^\nu T_{\mu\nu}(0)\phi(p) \phi(q)} = 2 \int_k \frac{\delta(p+k)}{p^{2\zeta}} \frac{\delta(q-k)}{q^{2\zeta}}\frac{\zeta (k\cdot z)^2}{k^{2(1-\zeta)} }.
\end{equation}
This confirms \eqref{eq:CTphiphi}.

\subsection{Rearranging the \texorpdfstring{$C_T$}{C\_T} integral}

In this section, we attempt to compute the normalization of the two-point function of $T^{\mu\nu}$, $C_T$.
We will not be able to do so for arbitrary $\zeta$, but for any explicit integer $\zeta$ we find a sum which can be computed exactly, and it matches the known result (see the added note of \cite{Osborn:2016bev}, which used \cite{Fitzpatrick:2011dm,Guerrieri:2016whh}; for other proofs see \cite[\S 4]{Gliozzi:2017hni} and \cite[\S 10]{Dowker:2017qkx})
\begin{equation}
    C_{T,2\zeta}= \zeta \frac{\Pochhammer{\dotwo +2}{\zeta-1}}{\Pochhammer{1-\dotwo}{\zeta-1}} C_{T,S}= \zeta \frac{\Pochhammer{\dotwo}{\zeta+1}}{\Pochhammer{-1-\dotwo}{\zeta+1}} C_{T,S},
\end{equation}
with $C_{T,S}$ being the $\zeta=1$ free scalar result
\begin{equation}
    C_{T,S} = \frac{d}{d-1}.
\end{equation}

Recall that $C_T$ is defined by
\begin{equation}
    S_d^2 \expval{T^{\mu\nu}(x) T^{\rho\sigma}(0)} = C_T \frac{1}{(x^2)^d} I^{\mu\nu, \rho\sigma}(x)
\end{equation}
where $S_d=2\pi^{\dotwo}/\Gamma(\dotwo)$ and $I$ is the inversion tensor for symmetric traceless tensors, 
\begin{equation}
    I^{\mu\nu, \rho\sigma} = \half \left(I^{\mu\rho} I^{\nu\sigma} + I^{\mu\sigma} I^{\nu\rho}\right)  -\frac{1}{d} \gflat^{\mu\nu} \gflat^{\rho\sigma}, \quad I^{\mu\nu} = \gflat^{\mu\nu} - 2\frac{x^\mu x^\nu}{x^2}.
\end{equation}
To find $C_T$ we must therefore compute
\begin{align}
    \expval{T^{\mu\nu}(x) T^{\rho\sigma}(y)} &= \int_{p,q,r,s} e^{ix\cdot(p+q)+iy\cdot(r+s)} S^{\mu\nu}_{pq} S^{\rho\sigma}_{rs} \expval{:\phi(p)\phi(q): :\phi(r)\phi(s):}\notag \\
    &= 2\int_{p,q} e^{i(x-y)\cdot(p+q)} S^{\mu\nu}_{pq} S^{\rho\sigma}_{pq} \frac{1}{p^{2\Delta}} \frac{1}{q^{2\Delta}},
\end{align}
where we have used $S^{\rho\sigma}_{(-p)(-q)} = S^{\rho\sigma}_{pq}$.

We simplify our computation by moving to momentum space and using the transverse-traceless projector \eqref{eq:TTprojector},
\begin{equation}
    \int_k \Pi_\mathrm{TT}^{\mu\nu\rho\sigma}(p) S_{\mu\nu}(k,p-k) S_{\rho\sigma}(k,p-k).
\end{equation}
Our $T^{\mu\nu}$, being primary, will have a vanishing two-point function with descendants of any other primary. 
Thus, we can freely delete all total derivative contributions ($\propto s$) in one of the two $T^{\mu\nu}$s without changing the value of the integral.
Then, for any integer $\zeta$, the sum is finite, and we can freely pull it out of the integral (if we were also considering noninteger $\zeta$, this would need to be justified).
We can then use \eqref{eq:PiTTproperties} to remove many of the remaining terms.
Finally, we use the projection property of $\cD_\mathrm{TT}\rvert_{s=p} = p^4 \Pi_\mathrm{TT}(p)$, with $\chi=k-(p-k)$,
\begin{equation}
 \Pi_\mathrm{TT,\mu\nu\rho\sigma}  \cD_\mathrm{TT}^{\mu\nu\alpha\beta} \chi_{\alpha}\chi_{\beta} \cD_\mathrm{TT}^{\rho\sigma\gamma\delta} \chi_\gamma \chi_\delta = p^8 \Pi_\mathrm{TT}^{\mu\nu\rho\sigma} \chi_\mu \chi_\nu \chi_\rho \chi_\sigma =  16 p^4 \tfrac{d-2}{d-1} ((k\cdot p)^2 -k^2 p^2)^2.
\end{equation}
The ultimate integral that we need to verify is still quite nontrivial, and the remainder of this section is a partial \textit{attempt} to verify the identity
\begin{equation}
\begin{aligned}\label{eq:desiredIresult} 
    I&= \sum_{m=1}^\infty \frac{(p^2)^{m-3}}{m-1-\dotwo} \int_k \frac{((k\cdot p)^2 - k^2 p^2)^2}{(k^2)^\zeta ((p-k)^2)^\zeta} \mg^{2\zeta -m -1} C_{\zeta -1}^{(1)} (c) C_{\zeta-m}^{(m)}(c)\\
    & \overset{?}{=}\frac{-1}{(4\pi)^{\dotwo}} p^d \frac{\Gamma(-\dotwo) \Gamma(\dotwo)^2}{8\Gamma(d-1)} \left[\zeta \frac{\Pochhammer{\dotwo +2}{\zeta-1}}{\Pochhammer{1-\dotwo}{\zeta-1}}\right],
    \end{aligned}
    \end{equation}
where $\theta$ is now the angle between $k$ and $k-p$,
\begin{equation}
c = \cos\theta \equiv -\frac{k \cdot (p-k)}{\sqrt{k^2 (p-k)^2}} = \frac{k^2 - kp \cos\varphi}{\sqrt{k^2 (p^2 -2k p \cos\varphi +k^2)}},
\end{equation}
with $\varphi$ the angle between $k$ and $p$, $k\equiv \sqrt{k^2}$, and $p \equiv \sqrt{p^2}$. 
Aligning the $z$-axis of our coordinates with $p$, we can write
\begin{equation}
    \int_k = \underbrace{\int \odif{\Omega_{d-2}}}_{=S_{d-1}
    } \int k^{d-1} \odif{k} \, (\sin\varphi)^{d-2} \odif{\varphi} = S_{d-1}\int_0^\infty \odif{k}\, k^{d-1} \int_{-1}^1 \odif{\cos\varphi} \, (\sin\varphi)^{d-3},
\end{equation}
with $k_\mu p^\mu =k p \cos \varphi$.
Changing variables %
from $(k,\cos\varphi)$ to $(c,y)$, where  $\mg=p^2/(y-c)$,
    \begin{equation}
    k^2 p^2 \sin^2 \varphi = \mg^2 \sin^2\theta, \quad \frac{\odif{k} \odif{\cos\varphi}}{(p-k)^2} =\frac{\odif{c} \odif{y}}{p\sqrt{y^2-1}(y-c)},
    \end{equation}
such that
\begin{equation}
  I =(p^2)^{d/2} \frac{S_{d-1}}{(2\pi)^d}  \sum_{m=1}^\infty \int_{-1}^1 \odif{c} (\sin\theta)^{d+1} 2^{-d+m-2} \int_1^\infty \odif{y} \frac{(y-c)^{-d+m-3}}{\sqrt{y^2-1}} \frac{C_{\zeta-1}^1(c) C_{\zeta-m}^m(c) }{m-1-\dotwo},
\end{equation}
and we note the new bounds on the $y$ integral.

The $y$ integral is then an associated Legendre function \cite[(14.3.1)]{NIST:DLMF}, 
\begin{equation}
    \int_1^\infty \odif{y} \frac{(y-c)^{-d+m-3}}{\sqrt{y^2-1}} = \sqrt{\frac{\pi }{2}} \left(1-c^2\right)^{\frac{1}{4} (-2 d+2 m-5)} \Gamma (d-m+3) P_{-\half}^{-d+m-\frac{5}{2}}(-c),
\end{equation}
and the full integral reduces to
\begin{equation}
  I =(p^2)^{d/2} \frac{S_{d-1}}{(2\pi)^{d-\half}} \sum_{m=1}^\infty 2^{m-d-3} \Gamma(d-m+3) \int_{-1}^1 \odif{c} (\sin\theta)^{m-\frac{3}{2}} \frac{C_{\zeta-1}^1(c) C_{\zeta-m}^m(c)  P_{-\half}^{-d+m-\frac{5}{2}}(-c)}{m-1-\dotwo}.
\end{equation}

\subsection{Computing \texorpdfstring{$C_T$}{C\_T} for integer \texorpdfstring{$\zeta$}{zeta}}
We will not be able to present a full proof of \eqref{eq:desiredIresult}, but we are able to simplify it to a finite sum of determinants for integer $\zeta$, which are easy to check in each case.
We do not attempt the computation of the normalization for the two-parameter extension.
The construction is unenlightening, and its only upshot is an interesting integral of two Gegenbauer functions and a Legendre function.
The strategy is \enquote{stupid}: we simply perform each integral in the sum.

\subsubsection{Calculating each integral}

By laborious inspection of $\zeta \le 10$, we notice that if we rewrite the equation as
\begin{equation}
  I  = \frac{\pi^2 (p^2)^{d/2}}{4\Gamma(\zeta)\Gamma(-1-d)\sin(\tfrac{\pi d}{2})} \frac{S_{d-1}}{(4\pi)^d}\sum_{m=1}^\zeta \frac{(-1)^m \Gamma (-d+2 m-3) f_{(\zeta-m,m)}(d(d+5-2m))}{\left(m-1-\dotwo \right) \Gamma (m) \Gamma (-m+\zeta +1)} ,
\end{equation}
then $f_{(n,m)}(w=d(d+5-2m))$ so defined is a polynomial in its argument $w$ for integer $n,m$ -- and crucially is otherwise $d$-independent.
Explicitly writing out the definition of this $f$:
\begin{equation}
\begin{aligned}\label{eq:fdef}
   &f_{(\zeta-m,m)}(d(d+5-2m)) = F_{(\zeta-m,m,d)} \int_{-1}^1 \odif{c} (\sin\theta)^{m-\frac{3}{2}} \frac{C_{\zeta-1}^1(c) C_{\zeta-m}^m(c)  P_{-\half}^{-d+m-\frac{5}{2}}(-c)}{m-1-\dotwo}.\\
   &F_{(\zeta-m,m,d)} = \frac{(-1)^{-m} 2^{m-\frac{1}{2}} \sin \left(\frac{\pi  d}{2}\right) \Gamma (-d-1) \Gamma (\zeta ) \Gamma (m) \Gamma (d-m+3) \Gamma (-m+\zeta +1)}{\pi ^{3/2} \Gamma (-d+2 m-3)}.
\end{aligned}
\end{equation}
By further inspection of small-$n\equiv \zeta-m$ cases, we found the following recursion relation in $n$ for arbitrary $m, d$:
\begin{equation}
\begin{aligned}
f_{(0,m)}(w) &= 1, \quad f_{(1,m)}(w) = w + \alpha_{(0,m)}\\
\alpha_{(\zeta-m,m)} &=2 \zeta ^2-4 m+6, \quad \beta_{(\zeta-m,m)} = \zeta  (\zeta -1) (\zeta -m) (\zeta -1+m)\\
f_{(n+1,m)}(w)&\equiv \underbrace{\left(\alpha _{(n,m)}+w\right)}_{=2\zeta^2+(d+2)(d+3-2m)} f_{(n,m)}(w)  -\beta _{(n,m)} f_{(n-1,m)}(w). \label{eq:recursion}
\end{aligned}
\end{equation}
This recurrence relation looks like the formula for the determinant of a tridiagonal matrix\footnote{For a tridiagonal matrix $M_n$ with diagonals $(a_i,D_i,c_i)$, performing a cofactor expansion on the last row gives $f_n \equiv \det(M_n) = D_n f_{n-1} - a_{n-1} c_{n-1} f_{n-2}$. }. %
For integer $n$, we can write the general solution of this integral as
\begin{equation}\label{eq:fsolution}
f_{(n,m)}(w) = \det \begin{pmatrix} w+\alpha_{(0,m)} & \beta_{(1,m)} \\ 1 & w+\alpha_{(1,m)} & \beta_{(2,m)} \\ & \ddots & \ddots & \ddots \\ && 1 & w+\alpha_{(n-2,m)} & \beta_{(n-1,m)}\\ &&& 1 & w+\alpha_{(n-1,m)}\end{pmatrix}. 
\end{equation}
This also gives a solution for a new family of integrals of Gegenbauers, which to the authors' knowledge is not related to any of the known identities \cite{Laursen:1980gm,dongOverlapIntegralThree2002,alisauskasIntegralsInvolvingTriplets2005,szmytkowskiIntegralsSeriesInvolving2011}.

Let us now prove that our $f$ \eqref{eq:fdef} indeed satisfies the recursion relation. 
Substituting $f$ into \eqref{eq:recursion} for $n=\zeta-m$, we see that the recursion relation holds if the following integral vanishes for $\lambda = m-d-\tfrac{5}{2}$.
\begin{equation}
\begin{aligned}
    0 &\overset{?}{=} I_r \equiv \int_{-1}^1\odif{c}\, (1-c^2)^{\frac{m}{2}-\frac{3}{4}} P_{-\half}^\lambda(-c) g(c),\\
    g(c) &\equiv C_{-1+\zeta}^{(1)}(c) \left((m-1) m-2 \zeta^2+\tfrac{1}{4}-\lambda ^2\right) C_{-m+\zeta}^{(m)}(c)\\
    & +\zeta (m+\zeta-1) C_{-2+\zeta}^{(1)}(c) C_{-1-m+\zeta}^{(m)}(c)+\zeta (-m+\zeta+1) C_\zeta^{(1)}(c) C_{1-m+\zeta}^{(m)}(c).
\end{aligned}
\end{equation}
The integrand itself clearly does not vanish, so one immediate conclusion might be that it is a total derivative.
Noticing the distinct presence of $\lambda^2$, and factors reminiscent of the rule for derivatives of Gegenbauers, we recall the equation satisfied by the Legendre function $h=P_{-1/2}^{\lambda}(-c)$:
\begin{equation}
\mathcal{L}[h] \equiv \odv{}{c} \left((1-c^2) \odv{h}{c} \right)-\left(\frac{1}{4}+\frac{\lambda^2}{1-c^2}\right)h = 0.
\end{equation}

Thus, we look for a function $G(c)$ which gives $(1-c^2)^{\tfrac{m}{2}-\tfrac{3}{4}}  g(c)$ on application of $\mathcal{L}$, and indeed we find that for integer $m$ and $\zeta$,
\begin{equation}
    (1-c^2)^{(\tfrac{m}{2}-\tfrac{3}{4})}  g(c) = \mathcal{L}\left[G(c)\right], \quad G(c) \equiv (1-c^2)^{\frac{m}{2}+\frac{1}{4}} C_{\zeta-1}^{(1)}(c) C_{\zeta-m}^{(m)}(c).
\end{equation}
This identity follows from eliminating any explicit factors of $c$ using the Gegenbauer recursion relations and derivative rules.
Hence, integrating by parts, and using $\mathcal{L}[P_{-\half}^\lambda(-c)] = 0$, we find
\begin{equation}
    I_r \equiv \int_{-1}^1 \odif{c}\, \mathcal{L}\left[G(c)\right] P_{-\half}^\lambda(-c) = \left[(1-c^2)\big(P_{-\half}^\lambda(-c) G'(c) -G(c) P_{-\half}^{\lambda\,\,\prime}(-c)\big)\right]_{c=-1}^{c=1}=0,
\end{equation}
which vanishes assuming that $n=\zeta-m$ is an integer, which is always true given that we take $\zeta$ integer. 
Thus, the recursion relation \cref{eq:recursion} is proven.

\subsubsection{An attempt at summing the results}

Dividing out by the overall factor, we find that we obtain \eqref{eq:desiredIresult} if
\begin{equation}\label{eq:mustBeTrue}
\Gamma (\zeta +1) \frac{\left(\frac{d}{2}+1\right)_{\zeta }}{\left(-1-\frac{d}{2}\right)_{\zeta+1}} \overset{?}{=} %
 \sum_{m=1}^\zeta \frac{(-1)^m \Pochhammer{-d-1}{2(m-1)} f_{(\zeta-m,m)}(d(d+5-2m))}{\left(m-1-\dotwo \right) \Gamma (m) \Gamma (\zeta -m+1)}
\end{equation}
is true; this equation is easily checked numerically for various $\zeta$s using our explicit formula \eqref{eq:fsolution} for $f$.
Ideally, we would now like a proof.
Unfortunately, we are only able to offer the following observation.

The summands on the right-hand side have: numerators that are manifestly polynomials in $d$; and linear denominators $(m-1-\dotwo)$. 
Hence, the right-hand side has precisely the pole structure in $d$ that we expect.
This inspires the following: let us isolate only the $d=2(m-1)$ pole in each term, using
\begin{equation}
    \lim_{\dotwo \to m-1} \frac{\Pochhammer{-d-1}{2(m-1)} f_{(\zeta-m,m)}(d(d+5-2m))}{m-1-\dotwo } =\frac{\Gamma(2m) f_{(\zeta-m,m)}(6(m-1))}{m-1-\dotwo  }.\label{eq:polecontributions}
\end{equation}
The recurrence relation for $f_{(\zeta-m,m)}(6(m-1))$ is solvable: after substituting $w=6(m-1)$, we see that $\alpha_{(n,m)} + w = 2((m+n)^2 +m)$.
Then \eqref{eq:recursion} is trivially solved by $f_{(n,m)}(6(m-1))= n! \Pochhammer{2m}{n} \binom{n+m}{m}$.
Adding together the pole contributions \eqref{eq:polecontributions}, we find a sum which can be performed:
\begin{equation}
 \sum_{m=1}^\zeta \frac{(-1)^m \Gamma(2m) f_{(\zeta-m,m)}(6(m-1))}{\left(m-1-\dotwo \right) \Gamma (m) \Gamma (\zeta -m+1)} = \Gamma (\zeta +1) \frac{\left(\frac{d}{2}+1\right)_{\zeta }}{\left(-1-\frac{d}{2}\right)_{\zeta+1}},
\end{equation}
which is exactly what we wanted.
So it remains only to prove that the part that is polynomial in $d$ vanishes, which we leave to more talented wranglers of polynomials.

\section{Agreement with Juhl's formulae for the GJMS operators}\label{sec:Juhl}

\textit{We take $\zeta$ integer in this section, and therefore ignore the nonlocal primary solutions}.

In this section, we compare our flat-space $T^{\mu\nu}$ with that which would arise from the known construction of Weyl-covariant differential operators $\WeylOp_{\zeta}$.
We find them to be identical.
Indeed, the GJMS formulae could have been used to separately derive our $T^{\mu\nu}$ (for integer $\zeta$); however, doing it that way would have involved some tricky sums.

We choose to work with $\WeylOp_{\zeta} \equiv (-1)^{\zeta} P_{2\zeta}$, making $\WeylOp_{\zeta}$ a positive-definite operator.
Ideally, we would like an explicit formula for $\WeylOp_{\zeta}$.
Unfortunately, one is not known in general \cite{caseFactorizationGJMSOperators2023}.
On Einstein manifolds, where $\cR_{\mu\nu} \equiv (d-1) \lambda g_{\mu\nu}$, we know the exact form:
\begin{equation}
    \begin{aligned}\label{eq:EinsteinGJMS}
\WeylOp_{\zeta} \rvert_{g=\text{Einstein}} &= \prod_{j=1}^{\zeta} (-\Box_g + (\dotwo +j-1)(\dotwo-j) \lambda)\\
&= (-\lambda )^{\zeta } \left(\thalf-\sqrt{\left(\tfrac{d-1}{2}\right)^2+\tfrac{-\Box_g}{\lambda }}\right)_{\zeta } \left(\thalf + \sqrt{\left(\tfrac{d-1}{2}\right)^2+\tfrac{-\Box_g}{\lambda }}\right)_{\zeta }.
    \end{aligned}
\end{equation}
This second definition clearly extends to noninteger values of $\zeta$ (where it should be interpreted as usual in terms of eigenvalues of $\Box_g$).
In flat space $\Box_g = \partial^2$, and it manifestly becomes the $(-\partial^2)^\zeta$ used above.
This does not suffice to determine $T^{\mu\nu}$, as to find it we need to perform an arbitrary variation of $g_{\mu\nu}(x)$.
For arbitrary $g_{\mu\nu}$, only a recursive formula is available for the GJMS operators with arbitrary integer $\zeta$, due to Juhl \cite{Juhl:2011ua,feffermanJuhlsFormulaeGJMS2012}.
We shall review this formula, and then use it to confirm that it gives the same flat-space EM tensor that we constructed.

\subsection{A review of Juhl's formula}

Let us now review Juhl's construction of the GJMS operators. 
Following Fefferman and Graham,
\cite[(1.6)]{feffermanJuhlsFormulaeGJMS2012} gives the recursive definition of $\WeylOp_{\zeta}$ in terms of (1) a sum over \href{https://en.wikipedia.org/wiki/Composition_(combinatorics)}{integer compositions} of $\zeta$, where each term is made of multiple lower-order $\WeylOp_{\zeta}$s and (2) a particular second-order differential operator $\cM_{2\zeta}$:
\begin{equation}\label{eq:JuhlRecursion}
    \WeylOp_{\zeta} = - \sum_{\substack{\abs{I}=\zeta \\ {I \neq (\zeta)}}} m_I \WeylOp_I + (-1)^\zeta\cM_{2\zeta}.
\end{equation}
Here $I=(I_1,\ldots,I_\ell)$ is an \textit{ordered} list of positive integers with sum $\abs{I} \equiv \sum_{i=1}^\ell I_i$, which has an associated differential operator $\WeylOp_I$ and a number $m_I$:
\begin{align}
    \WeylOp_I &= \WeylOp_{I_1}\circ \WeylOp_{I_2} \circ \cdots \circ  \WeylOp_{I_\ell},\\
    m_I &\equiv (-1)^{\ell+1} \frac{\abs{I}! (\abs{I}-1)!}{\prod_{j=1}^\ell I_j! (I_j-1)!} \prod_{j=1}^{\ell-1} \frac{1}{I_j + I_{j+1}}.
\end{align}
For example, $\WeylOp_{(1,2,3)} \phi = \WeylOp_{1} \circ \WeylOp_{2} \circ \WeylOp_{3} \phi$.
This is not the same as $\WeylOp_{(2,1,3)} \phi$, because $\WeylOp_\zeta$s do not commute.

The second-order differential operators $\cM_{2\zeta}$ are not straightforward to obtain, and will require some technology (also reviewed in physicists' language in \cite[\S B]{Diaz:2008hy}).
Take $g$ to be our $d$-dimensional metric on the manifold $M$. 
Then a \enquote{Poincaré metric in normal form relative to $g$} is a metric $g_+$ on the $(d+1)$-dimensional space $M\times (0,\epsilon)$ of the form 
\begin{equation}
    g_+ = r^{-2} (\odif{r}^2 + g_r), \quad g_r|_{r=0} = g,
\end{equation}
with $g_r$ a smooth family of metrics on $M$ parametrised by $r$, specified by the demand that the metric be asymptotically Euclidean AdS in the sense that $\mathrm{Ric}[g_+] + d\, g_+$ is asymptotically zero at $r=0$.

Then we can find the $\cM_{2\zeta}$s by expanding the following $\cM(r)$ with respect to $r$.
$\cM(r)$ is a continuous family of second-order operators on the original manifold $M$ parametrised by $r$,
\begin{equation}\label{eq:Mexpansion}
    \cM (r) = \delta(g_r^{-1} \odif{\,}) - U(r) = \sum_{k \geq 1} \cM_{2k} \frac{1}{(k-1)!^2}\left(\frac{r^2}{4}\right)^{k-1},
\end{equation}
where (i) $\delta$ is the divergence of a $d$-dimensional vector field with respect to the original $g_0 \equiv g$, i.e. $\delta v = \nabla_\mu v^\mu$; (ii)
 $g_r^{-1}\mathrm{d}  f$ is simply $g_r^{\mu\nu}\nabla_\nu f$, so $\delta(g^{-1}_r \odif{\,}) f = \nabla_\mu(g_r^{\mu\nu}\nabla_\nu f)$;
 and (iii) $U(r)$ is the function
\begin{equation}\label{eq:UWdef}
U(r) \equiv \frac{\left[\partial^2_r - \tfrac{(d-1)}{r}\partial_r + \delta(g^{-1}_r \odif{\,})\right]W(r)}{W(r)}, \quad W(r) \equiv \left(\frac{\det g_r}{\det g_0}\right)^{1/4}.
\end{equation}

Juhl's formula yields, for the first case $\zeta=1$, (minus) the Yamabe operator,
\begin{align}\label{eq:minusYamabe}
    \WeylOp_1 &\equiv -\cM_2 = -\nabla^2 + \frac{d-2}{4(d-1)} \cR,
\end{align}
which indeed yields \eqref{eq:TForzetaOne} on variation.
For $\zeta=2$ we find the Paneitz operator, and so on.

\subsection{Using our \TtextOrPDF in Juhl's formula}

One interesting question is: could our $T^{\mu\nu}$s have been derived in a completely different way, using the known recursive implementation of the GJMS operators?
The answer is yes. 
The required sums are not trivial to perform, and it is easiest to work backwards from our known answer.
Essentially:
\begin{itemize}
    \item In this subsection, we compute the linearization of $\cM(r)\rvert_{g=\gflat+h}$ that is found using our formulae for $T^{\mu\nu}$; 
    \item In subsection \ref{sec:findingMdirectly}, we confirm that this $\cM(r)$ is the same as the one obtained from Juhl's construction.
\end{itemize}
Together, these two facts show that our $T^{\mu\nu}$ can also be obtained by varying the GJMS operators with respect to the metric.

Let us now consider $\delta_g \WeylOp_{\zeta}$. Plugging in Juhl's formula \eqref{eq:JuhlRecursion}, we obtain
\begin{equation}\label{eq:varWeylOp}
    \delta_g \WeylOp_{\zeta} = -\sum_{\substack{\abs{I}= \zeta\\I\neq(\zeta)}} m_I \sum_{k=1}^\ell \WeylOp_{I_1}\circ \cdots\circ \WeylOp_{I_{k-1}} \circ \delta_g\WeylOp_{I_k} \circ \WeylOp_{I_{k+1}}\circ\cdots \circ \WeylOp_{I_{\ell}} + (-1)^\zeta \delta_g \cM_{2\zeta}.
\end{equation}
Restricting to flat space and Fourier transforming, $(\delta_g \WeylOp_{\zeta})f(x)$ should become a function $E_\zeta^{\mu\nu}(q,r)$ of the momenta of $f$ and $h_{\mu\nu}$, i.e.
\begin{equation}
    (\delta_g \WeylOp_{\zeta}(x)) f(x) |_{g=\gflat}= \int_{q,r} e^{i x\cdot(q+r)} E_\zeta^{\mu\nu}(q,r) f(q) h_{\mu\nu}(r)
\end{equation}
(recall that we distinguish position-space and momentum-space fields by their argument only).
Substituting \eqref{eq:varWeylOp} into $\int_x \phi (\delta_g \WeylOp_{\zeta}(x)) \phi |_{g=\gflat}$, we see
\begin{equation}
\begin{aligned}
    &\int_{xpqr} \phi(p) e^{i x\cdot(p+q+r)} E_\zeta^{\mu\nu}(q,r) \phi(q) h_{\mu\nu}(r) \\
    = &\int_{p,q,r}  \delta(p+q+r) \phi(p) F^{\mu\nu}_\zeta(q,r)\phi(q) h_{\mu\nu} (r) + \int_x \phi ((-1)^\zeta \delta_g \cM_{2\zeta}) \phi|_{g=\gflat},
\end{aligned}
\end{equation}
where
\begin{align}
    F^{\mu\nu}_\zeta(q,r) = & -\sum_{\substack{\abs{I}= \zeta\\I\neq(\zeta)}} m_I \sum_{k=1}^{\ell} (q+r)^{2\sum_{i=1}^{k-1} I_i} E^{\mu\nu}_{I_k}(q,r)\, q^{2\sum_{i=k+1}^{\ell} I_i},
\end{align}
whose form is clear when we see that the operators $\WeylOp_{I_{k+1,\cdots, \ell}}$ do not differentiate $h$, so just become $q^{2I_i}$; by contrast, the operators $\WeylOp_{I_{1, \cdots, k-1}}$ do differentiate $h$, and so become $(q+r)^{2I_i}$.
Note that unlike in the rest of the paper, we cannot symmetrise with respect to $p$ and $q$, because we now have three momenta in the game.

\subsubsection{Finding \texorpdfstring{$E$}{E}}

Let us now compute $E_\zeta - F_\zeta$, which \eqref{eq:varWeylOp} tells us should yield a term $\sim\delta_g\cM_{2\zeta}$.
To do so, we first need to find $\delta_g \WeylOp_{I_k}\rvert_{g=\gflat}$.
This is not quite the same thing as $S^{\mu\nu}(p,q)$ from \eqref{eq:fullMomSpace}, as the operator between the $\phi$s in the action is $\sqrt{g} \WeylOp_{\zeta}$.
Consider $S_W \equiv \half \int \odif[d]{x}\, \sqrt{g} \phi \WeylOp_{\zeta} \phi$, where crucially $\WeylOp_\zeta$ is just a differential operator, so the left-hand $\phi$ is not differentiated.
Performing the variation and then setting $g=\gflat$ after the variation, we can do so as follows.
\begin{equation}
    \int_x\, T^{\mu\nu} h_{\mu\nu} = -2\, \delta_g S_W |_{g=\gflat}= - \int_x h_{\mu\nu} \frac{\gflat^{\mu\nu}}{2} \phi(-\partial^2)^\zeta \phi -\int_x \phi (\delta_g \WeylOp_{\zeta}) \phi |_{g=\gflat}
\end{equation}
Substituting \eqref{eq:fullMomSpace}, we find
\begin{align}
\int_{p,q,r}  \delta(p+q+r) \phi(p) \left[S^{\mu\nu}(p,q) + \frac{\gflat^{\mu\nu}}{2} q^{2\zeta} + E_\zeta^{\mu\nu}(q,r)\right] \phi(q) h_{\mu\nu}(r) =  0.
\end{align}
We need to be careful to map $h_{\mu\nu} \phi (-\partial^2)^\zeta \phi \,\mapsto\, q^{2\zeta}$ here, as the derivative acts only on the right-hand $\phi$ by assumption.
This gives us an explicit formula for $E^{\mu\nu}(q,r)$: it is just $-S^{\mu\nu}(p,q)$, but replacing $p=-q-r$ and removing the $q^{2\zeta}$ from $\half \gflat^{\mu\nu} \phi (-\partial^2)^\zeta \phi$ (which itself came from the variation of $\sqrt{g}$):
\begin{equation}
    E^{\mu\nu}_\zeta(q,r) = -S^{\mu\nu}(-q-r,q) -\frac{\gflat^{\mu\nu}}{2} q^{2\zeta}.
\end{equation}
This is unambiguous: the left-hand $\phi$ is not differentiated, so $E_\zeta$ must be independent of $p$ -- this will prove important later when identifying $\delta_g \cM$.

We now want to compute $E-F$. 
It should be second-order in $q$, because $\cM_{2\zeta}$ is a second-order operator.
For compactness, we set $p=-q-r$ in all equations below.

\subsubsection{Writing \texorpdfstring{$F$}{F} in a more convenient form} 

First, we find a better form for $F$: we single out all terms in $F$ containing $E_k$ for some $k$, and work out the polynomial in $p^2$ and $q^2$ that sits in front of it.
\begin{subequations}
\begin{equation}
    F^{\mu\nu}_\zeta(q,r) = \sum_{k=1}^{\zeta-1} f_{\zeta,k} E^{\mu\nu}_k(q,r), \quad f_{\zeta,k} =\sum_{j=0}^{\zeta-k} a_{\zeta,k}^{(j)} (p^2)^{j} (q^2)^{\zeta-k-j};
\end{equation}
$a_{\zeta,k}^{(j)}$ is clearly $-\sum_{I\in S_{k}^{(j)}} m_I$, where $S_{k}^{(j)}$ is the set of all $\ell$-tuples with $I_m =k$, $\sum_{i=1}^{m-1} I_i = j$, and $\sum_{i=m+1}^{\ell} I_i = \zeta-k-j$ for any $\ell$ and $m$.
For $I=(A,k,B)$, this is equivalent to: a sum over $\abs{A}=j$ (for any length of $A$); followed by a sum over $\abs{B} = \zeta-k-j$.
We see that $m_{(A,k,B)}$ then factorises, 
\begin{align}\label{eq:mtildeDef}
    m_{(A,k,B)} =\frac{\zeta! (\zeta-1)!}{k!(k-1)!} \tilde{m}_{A,k}  \tilde{m}_{B,k}, \quad \tilde{m}_{A,k} = \frac{1}{\prod_{a \in A}a!(a-1)!}\frac{(-1)^{\mathrm{len(A)}}}{A_{\mathrm{len}(A)}+k} \prod_{i=1}^{\mathrm{len}(A)-1} \frac{1}{A_i + A_{i+1}} 
\end{align}
where $\tilde{m}_{(),k}=1$ for the empty tuple $A=()$.
Defining $\cC(j)$ as the set of \href{https://en.wikipedia.org/wiki/Composition_(combinatorics)}{integer compositions} of $j$ (i.e. ordered integer partitions of $j$ of any length, $\cC(j) = \{ A \in \cup_{\ell=0}^j \mathbb{N}_+^\ell \, \rvert \, \sum_{i=1}^\ell A_i = j\}$), we see that the sum in $a_{\zeta,k}^{(j)}$ factorises into a pair of tractable sums, which we can just do:
\begin{equation}\begin{aligned}\label{eq:aValue}
    a_{\zeta,k}^{(j)} &=-\sum_{I\in S_{k}^{(j)}} m_I = -\frac{\zeta! (\zeta-1)!}{k!(k-1)!} \sum_{A \in \cC(j)} \tilde{m}_{A,k} \sum_{B \in \cC(\zeta-k-j)} \tilde{m}_{B,k}\\
    &= -(-1)^{\zeta-k} \frac{k}{\zeta} \binom{\zeta}{k+j}\binom{\zeta}{j}, 
\end{aligned}\end{equation}\end{subequations}
which satisfies $a_{\zeta,k}^{(j)}= a_{\zeta,k}^{(\zeta-k-j)}$.
The sum used in \eqref{eq:aValue} is
\begin{align}\label{eq:mtildesum}
    \sum_{A \in \cC(j)} \tilde{m}_{A,k} = (-1)^j \frac{k!}{j!(k+j)!},
\end{align}
valid for $j\ge 0$,
which is easily proven inductively (see \cref{app:mtildeProof}).

\subsubsection{Finding the linearized \texorpdfstring{$\cM(r)$}{M(r)} from our \TtextOrPDF} %
We are computing
\begin{equation}\label{eq:EminusF}
    \int_x \phi ((-1)^\zeta \delta_g \cM_{2\zeta}) \phi|_{g=\gflat} = \int_{p,q,r} e^{i x\cdot(p+q+r)}\phi(p)\left(E_\zeta^{\mu\nu}(q,r) - F_\zeta^{\mu\nu}(q,r)\right) \phi(q) h_{\mu\nu}(r).
\end{equation}
Plugging in our expressions from above, $E$ and $F$ cancel for $\zeta \geq 3$ up to a simple remnant:
\begin{equation} \label{eq:EminusFvalue}
\begin{aligned}
    & \quad E^{\mu\nu}_\zeta(q,r)-F_\zeta^{\mu\nu}(q,r) \\
    =&\sum_{m=2}^\infty \frac{m-1}{m-1-\dotwo} (r^2)^{m-3} (-1)^{\zeta -m} \binom{\zeta-1}{m-1} (r^2)^{\zeta-m} (\cD_{\mathrm{TT}}^{\mu\nu\rho\sigma}|_{s=-r}) \frac{(-2q-r)_\rho (-2q-r)_\sigma}{4}\\
    =&- \frac{(\zeta-1)!}{\prod_{l=1}^{\zeta-1} (\dotwo -l)} (r^2)^{\zeta -3} (\cD_{\mathrm{TT}}^{\mu\nu\rho\sigma}|_{s=-r}) q_\rho q_\sigma,
\end{aligned}
\end{equation}
where we dropped the terms $\propto r_{\rho},r_{\sigma}$ because $\cD_{\mathrm{TT}}^{\mu\nu\rho\sigma}$ gives zero when contracted with $s$ on any index.
This simple form is derived by noting that the first two of the three structures in $E^{\mu\nu}$ perfectly cancel in $E-F$, except for some low-$\zeta$ cases, because
\begin{align}\label{eq:perfectCancellation}
   e_\zeta(q,r) =\sum_{k=1}^{\zeta-1} f_{\zeta,k} e_k(q,r) \quad\text{for}\quad e_\zeta(q,r) \in \left\{ \zeta \frac{p^{2\zeta} +  q^{2\zeta}}{2}, \quad p^{2\zeta} - q^{2\zeta}\right\};
\end{align}
where in \cref{app:EmFidentities} we show the first to hold for $\zeta\geq 3$, and the second for $\zeta\geq 2$.
The remaining structure is $M^{\zeta-m} C^{(m)}_{\zeta-m}$, which gives for all $\zeta\geq 1$ the remnant
\begin{align}\label{eq:remnant}
    \sum_{k=1}^{\zeta-1} f_{\zeta,k} \, \mg^{k-m} C_{k-m}^{(m)}(c) = \mg^{\zeta-m} C_{\zeta-m}^{(m)}(c) - (-1)^{\zeta-m} \binom{\zeta-1}{m-1} (s^2)^{\zeta-m},
\end{align}
which is proven in \cref{app:remnantCproof}.

Using this to compute \eqref{eq:EminusF}, we determine that the action of $\delta_g \cM_{2\zeta}$, $\zeta \geq 3$, on a function $f(x)$ is\footnote{
    We note that this appears to be related to the linearized extended obstruction tensors $\Omega^{(k-1)}$ \cite[(1.8)]{acheObstructionflatAsymptoticallyLocally2011}.
}
\begin{align}\label{eq:deltagM}
    ((-1)^\zeta \delta_g \cM_{2\zeta}) f = \frac{(\zeta-1)!}{\prod_{l=1}^{\zeta-1} (\dotwo -l)} ((-\partial^2)^{\zeta -3}\cD_{\mathrm{TT}}^{\mu\nu\rho\sigma}  h_{\mu\nu})\, \rptl_\rho \rptl_\sigma f + O(h^2), %
\end{align}
where $(-\partial^2)^{\zeta-3}\cD_{\mathrm{TT}}$ acts only on $h_{\mu\nu}$ and $\rptl$ acts on $f(x)$.
Note that we have not integrated by parts, since the first $\phi$ should not be differentiated, and we have indeed found a second-order differential operator (i.e. a quadratic in $q$).
The remaining cases are
\begin{itemize}
    \item $\zeta=1$: we find for $E_1 - F_1 = E_1$ that
\begin{equation}\label{eq:M2}
    E_1^{\mu\nu}= -q^\mu q^\nu + \frac{1}{2} \left(\gflat^{\mu  \nu } q\cdot r-q^{\nu } r^{\mu}-q^{\mu} r^{\nu }\right)+\frac{d-2}{4(d-1)}\left(r^2 \gflat^{\mu \nu}-r^{\mu} r^{\nu}\right).
\end{equation}
Since $\cM_2 = P_2 = -\WeylOp_1 = \nabla^2 - \frac{d-2}{4(d-1)} \cR$, this matches
    \begin{equation}
        \cM_2 f
= \partial^2 f - h^{\mu\nu}\partial_\mu\partial_\nu f
+\Big(\thalf\partial^\nu h - \partial_\mu h^{\mu\nu}\Big)\partial_\nu f
+\frac{d-2}{4(d-1)}\Big(\partial_\mu\partial_\nu h^{\mu\nu}-\partial^2 h\Big) f
+O(h^2),
    \end{equation}
so we can confirm \eqref{eq:EminusF} for this case.
\item $\zeta=2$: we find
\begin{equation}\label{eq:M4}
    E_2^{\mu\nu} -F_2^{\mu\nu} = -2 \cD^{\mu\nu\rho\sigma}\rvert_{s=-r} \frac{(-2q-r)_\rho (-2q-r)_\sigma}{4}.
\end{equation}
For compactness, we have here used $\cD_{\mathrm{TT}}^{\mu\nu\rho\sigma}/(1-\dotwo) = -2(s^2 \cD^{\mu\nu\rho\sigma}-\cP^{\mu\nu} s^\rho s^\sigma)$ for $\cD$ defined in \cite[(18)]{Stergiou:2022qqj}.
\end{itemize}
We now have an explicit form for the linearization of each term in \eqref{eq:Mexpansion}.

\subsection{Finding \texorpdfstring{$\cM(r)$}{M(r)} using Juhl's definition} \label{sec:findingMdirectly}

Let us now find $\cM(r)$ for $g_0=\gflat+h$.
Moving to $\rho = r^2$, and denoting $\rho$ derivatives by primes, we indicate the deviation of $g_r$ from the flat metric by $\gamma$. 
Thus, we write 
\begin{equation}\label{eq:perturbationForH}
\mathbf{g}_{\mu\nu}\equiv [g_r]_{\mu\nu} = \gflat_{\mu\nu} + \gamma_{\mu\nu}(\rho,x), \qquad \gamma_{\mu\nu}(\rho=0,x)=h_{\mu\nu}(x),
\end{equation}
with Greek indices running only from $1$ to $d$. 
Expanding \eqref{eq:UWdef}, we have
\begin{equation}
    W(r) = 1 + \frac{\gamma-h}{4} + O(h^2), \quad
    U(r) \equiv \rho \gamma'' - \tfrac{d-2}{2} \gamma' + \partial^2 \left(\tfrac{\gamma-h}{4}\right) + O(h^2),
\end{equation}
where $\gamma = \gflat^{\mu\nu} \gamma\indices{_{\mu\nu}}$ and $h = \gflat^{\mu\nu} h\indices{_{\mu\nu}}$ at this order.
Then \eqref{eq:Mexpansion} becomes
\begin{align}
    \cM(r)f %
    =& \partial^2 f +(\thalf \partial_\nu h-\partial^\mu \gamma_{\mu\nu}) \partial^\nu f -\gamma_{\mu\nu} \partial^\mu \partial^\nu f - \left[\rho \gamma'' - \tfrac{d-2}{2} \gamma' + \partial^2 \left(\tfrac{\gamma-h}{4}\right)\right]f +O(h^2). \label{eq:expandM}
\end{align}
where we have used $\nabla_\mu V^\mu = \partial_\mu V^\mu + \Gamma^\mu_{\mu \nu} V^\nu$ with $\Gamma^\mu_{\mu \nu} = \half \partial_\nu h+O(h^2)$. 
Our goal is to solve for $\gamma_{\mu\nu}$ to all orders in $r$ and to leading order in $h_{\mu\nu}$.

\subsubsection{Solving for the asymptotically AdS metric} 

The $(d+1)$-dimensional Einstein equations for the asymptotically AdS metric $g_+ = r^{-2} (\odif{r}^2 + \mathbf{g})$ give (from \cite[(2.5)]{deHaro:2000vlm}, but using the opposite sign convention for Riemann) %
\begin{equation}
\begin{aligned}
    &\rho [2 \mathbf{g}'' - 2\mathbf{g}' \mathbf{g}^{-1} \mathbf{g}' + \Tr(\mathbf{g}^{-1} \mathbf{g}') \mathbf{g}'] - \text{Ric}[\mathbf{g}] -(d-2) \mathbf{g}' -\Tr(\mathbf{g}^{-1} \mathbf{g}') \mathbf{g} =0,\\
 &\nabla_\mu \Tr(\mathbf{g}^{-1} \mathbf{g}') - \nabla^\nu \mathbf{g}_{\mu\nu}'=0,\\
 &\Tr(\mathbf{g}^{-1} \mathbf{g}'') - \half \Tr (\mathbf{g}^{-1} \mathbf{g}' \mathbf{g}^{-1} \mathbf{g}') =0,
\end{aligned}
\end{equation}
which linearize under \eqref{eq:perturbationForH} to
\begin{align}\label{eq:LinearizedEinstein}
    &2\rho \gamma_{\mu\nu}''  - \cR_{\mu\nu}^\text{lin}[\delta+\gamma] -(d-2) \gamma_{\mu\nu}' -\gamma' \gflat_{\mu\nu} =0,\quad \partial_\mu \gamma' - \partial^\nu \gamma_{\mu\nu}'=0, \quad \gamma'' =0.
\end{align}
Then we have 
\begin{align}
    \cR_{\mu\nu}^\text{lin}[\delta+\gamma] &= \partial_\alpha \Gamma^\alpha_{\mu\nu} - \partial_\mu \Gamma^\alpha_{\alpha\nu}=  \half(-\partial^2 \gamma_{\mu\nu} + \partial_\mu \partial_\alpha \gamma\indices{_\nu^\alpha}+ \partial_\nu \partial_\alpha\gamma\indices{_\mu^\alpha} - \partial_\mu \partial_\nu \gamma),\\
    \cR^\text{lin}[\delta+\gamma] &= \partial_\mu \partial_\alpha \gamma^{\mu\alpha}-\partial^2 \gamma.
\end{align}%

It is straightforward to compute the trace and the divergence of $\gamma_{\mu\nu}$ by plugging the second and third equations into the first, since both are only linear in $\rho$.
$\gamma''=0$ tells us that we need only a linear term in $\rho$. Taking the trace of \eqref{eq:LinearizedEinstein} therefore gives
\begin{equation}
    \gamma'=\frac{-1}{2(d-1)}\cR^\text{lin}[\delta+\gamma] \quad \implies \quad \gamma -h = \frac{-\rho}{2(d-1)}\left(\partial_\mu \partial_\alpha h^{\mu\alpha}-\partial^2 h\right). \label{eq:traceOfGam}
\end{equation}
We can use this with $\partial^\mu \gamma_{\mu\nu}' = \partial_\nu \gamma'$ to find the (exact) divergence
\begin{align}
   & \partial^\mu (\gamma_{\mu\nu}-h_{\mu\nu}) =  \frac{-\rho \partial_\nu}{2(d-1)}\left(\partial_\mu \partial_\alpha h^{\mu\alpha}-\partial^2 h\right) \label{eq:divergenceOfGam}. 
\end{align}

Having found the trace and transverse parts, which are at most linear in $\rho$, we now construct the remaining (transverse-traceless) part. Since it should be TT and linear in $h$, we can presume that it should be $\propto \cD_{\mathrm{TT}}^{\mu\nu\rho\sigma} h_{\rho\sigma}$.
We can now compute the first derivative at $\rho=0$ using \eqref{eq:LinearizedEinstein}, as $\gamma_{\mu\nu}''$ drops out
\begin{align}
    &\gamma_{\mu\nu}'|_{\rho=0}  =\frac{-1}{d-2} \left(\cR_{\mu\nu}^\text{lin}[\delta+h] + \frac{\gflat_{\mu\nu}}{2(d-1)}\cR^\text{lin} [\delta+h]\right).
\end{align}
Then, we guess a solution of the form 
\begin{equation}
    \gamma^{\mu\nu} =  h^{\mu\nu} + \rho (\gamma^{\mu\nu})'|_{\rho=0} + f(\rho (-\partial^2)) \frac{\cD_{\mathrm{TT}}^{\mu\nu\rho\sigma}}{(-\partial^2)^2} h_{\rho\sigma} + O(h^2),
\end{equation}
which isolates only the transverse-traceless part as the unknown variable, and automatically solves \eqref{eq:traceOfGam} and \eqref{eq:divergenceOfGam}. Taking $x=\rho (-\partial^2)$, this turns \eqref{eq:LinearizedEinstein} into the equation
\begin{equation}
    2 (d-2) f'-4 x f''+f+1=0
\end{equation}
for $f(x)$, which has Bessel or hypergeometric solutions
\begin{align}
    f(x)+1 &= c_1 I_{-\frac{d}{2}}\left(\sqrt{x}\right)+c_2 x^{d/4} I_{\frac{d}{2}}\left(\sqrt{x}\right)\\
    &= \tilde{c}_1 \, {}_0F_1\left(;1-\tfrac{d}{2};\tfrac{x}{4}\right) + \tilde{c}_2 \, x^{d/2} \, {}_0F_1\left(;\tfrac{d}{2}+1;\tfrac{x}{4}\right).
\end{align}
Demanding regularity at $\rho=0$, we set $\tilde{c}_2=c_2=0$; demanding that $\gamma_{\mu\nu} \rvert_{\rho=0}=h_{\mu\nu}$ means that $\tilde{c}_1=1$, and so:
\begin{align}
    \gamma^{\mu\nu} &= h^{\mu\nu} +\frac{\rho}{2(d-1)} \frac{\partial^\mu \partial^\nu}{(-\partial^2)} \left(\partial_\alpha \partial_\beta h^{\alpha\beta}-\partial^2 h\right) +\left[{}_0F_1\left(;1-\tfrac{d}{2};\tfrac{\rho  (-\partial^2)}{4}\right) -1\right]\frac{\cD_\mathrm{TT}^{\mu\nu\rho\sigma}}{(-\partial^2)^2} h_{\rho\sigma} + O(h^2).
\end{align}

 \subsubsection{Assembling \texorpdfstring{$\cM(r)$}{M(r)} and confirming that it is the same}

 Putting together the pieces, we find for $f(x) = {}_0F_1\left(;1-\frac{d}{2};\frac{x}{4}\right)-1$ that \eqref{eq:expandM} becomes
 \begin{equation}
 \begin{aligned} \label{eq:Msolution}
    \cM(r)f %
    &= \rptl^2 f- \left[\frac{f(r^2 (-\partial^2)) \cD_{\mathrm{TT}}^{\mu\nu\rho\sigma}}{(-\partial^2)^2} h_{\rho\sigma} \right]\rptl_\mu \rptl_\nu f -h^{\mu\nu}(\lptl_\mu+\rptl_\mu)\rptl_\nu f\\
    &\quad + \cP^{\mu\nu}h_{\mu\nu} \left(\frac{2-d}{4}-\frac{r^2\left((-\lptl^2)-2 \lptl \cdot \rptl \right)^2}{8 (-\lptl^2)}\right)f + O(h^2).
\end{aligned}
\end{equation}
Recalling that $\cM(r) = \cM_2 + \frac{\cM_4 r^2}{4} + O(r^4)$, we find that the first two terms match \eqref{eq:M2} and \eqref{eq:M4} perfectly.
Using
\begin{equation}
    f(x) = \sum_{k=2}^\infty 
        \frac{\Gamma \left(1-\frac{d}{2}\right)}{\Gamma \left(k-\frac{d}{2}\right)} \frac{1}{(k-1)!} \left(\frac{x}{4}\right)^{k-1}
\end{equation}
we find that the order $r^{\geq 4}$ terms in \eqref{eq:Msolution} are
\begin{equation}
    \cM(r) \supset \left[\sum_{k=2}^\infty  \frac{1}{(k-1)!^2} \left(\frac{r^2 }{4}\right)^{k-1}
    \frac{(k-1)!\Gamma \left(1-\frac{d}{2}\right)}{\Gamma \left(k-\frac{d}{2}\right)} (-\partial^2)^{k-3}\cD_{\mathrm{TT}}^{\mu\nu\rho\sigma} h_{\rho\sigma} \right]\rptl_\mu \rptl_\nu f,
\end{equation}
so we can match all the higher terms to \eqref{eq:deltagM} after using $\prod_{\ell=1}^{k-1}(\dotwo - \ell) =(-1)^{k-1} \tfrac{\Gamma(k-\dotwo)}{\Gamma(1-\dotwo)}$.
Hence, our EM tensors satisfy the first variation of Juhl's formula, because \eqref{eq:varWeylOp} yields the correct $\cM(r)$.
As expected, then, for integer $\zeta$ we find that our EM tensors are identical to the ones obtained from varying Juhl's GJMS operators.

\appendix
\section{Conventions}\label{app:conventions}

We work in $d$-dimensional Euclidean space, and write
\begin{equation}
    \int_x \equiv \int \odif[d]{x} \sqrt{g}
\end{equation}
though we specialise to flat space $g_{\mu\nu} = \gflat_{\mu\nu}$ except in \cref{sec:WeylConf,sec:Juhl}.
However, our constructions in this paper apply identically to Lorentzian signature (as in \cite{Stergiou:2022qqj}), as the four conditions \ref{prop1}--\ref{prop4} are identical.
The Euclidean conformal symmetry group is $\SO(d+1,1)$, which, as discussed in \cref{sec:WeylConf}, can be embedded into $\text{Diff}\times\text{Weyl}$\footnote{We note that the group of Weyl transformations in $d=2$ is larger than the (infinite-dimensional) Virasoro group, as Virasoro is the harmonic subset of Weyl transformations (with a compensating diffeomorphism)
\begin{equation}
\mathrm{Vir} \subset \mathrm{Diff} \times \{ \Omega \subset \mathrm{Weyl}\, \rvert\,   \Box \Omega=0\}.
\end{equation}
$\Omega$ must be harmonic because the Weyl factor in $d=2$ is holomorphic \cite[footnote 3]{Farnsworth:2017tbz}.}.

When working in flat space, we define the momentum-space integral
\begin{equation}
\int_p \equiv \int \frac{\odif[d]{p}}{(2\pi)^d}.
\end{equation}
Similarly, we write our $d$-dimensional delta functions as $\delta(p+q) \equiv (2\pi)^d \delta^{(d)}(p+q)$, and $\delta(x-y) = \delta^{(d)}(x^\mu-y^\mu)$.
We distinguish fields and their Fourier transforms by argument alone, i.e.
\begin{equation}
\phi(p) \equiv \int_x e^{ip \cdot x} \phi(x), \quad \phi(x) \equiv \int_p e^{ip \cdot x} \phi(p);
\end{equation}
and any field without argument is in position space at the point $x$.

As discussed in \cref{sec:momSpace}, we freely move between bilinears in position space and functions of two momenta in momentum space.
This means that differential operators $\cD_O(-i\partial)$ that act on an entire bilinear just become functions of $s=p+q$, i.e.
\begin{equation}\begin{aligned}
    \cD_O(-i\partial) &\,\left[ \phi(x) \left(f(-i(\lptl+\rptl), -i(\lptl-\rptl))\right) \phi(x)\right]\\
    =& \int_{p,q} \cD_O(-i\partial) \phi(p) e^{i p\cdot x} f(-i(\lptl+\rptl), -i(\lptl-\rptl)) e^{iq \cdot x} \phi(q) \\
    =& \int_{p,q} e^{i s\cdot x} \cD_O(s) \phi(p) f(s, \di) \phi(q).
    \end{aligned} \end{equation}
We occasionally use a simple function arising from the general Fourier transform of a traceless monomial in momentum or position space \cite{Karateev:2018oml},
\begin{equation}\begin{aligned}\label{eq:FlamsDef}
&\cF_{\lambda, s} \equiv i^{-s}2^{d-\lambda}\pi^{d/2} \frac{\Gamma(\frac{d+s-\lambda}{2})}{\Gamma(\frac{\lambda+s}{2})},\\ %
\text{where } & \cF_{\lambda,s} (p^{\mu_1} \cdots p^{\mu_s} -\text{traces})\, p^{-(d-\lambda)-s}=\int_x e^{-i p\cdot x} (x^{\mu_1}\cdots x^{\mu_s}-\text{traces})\, x^{-\lambda-s}.
\end{aligned}\end{equation}
Finally: all powers of vectors are to be interpreted in the obvious way, i.e. $p^{2\zeta} = (p^2)^\zeta$; and whenever symmetrising or antisymmetrising indices we do so with weight one, i.e. $M^{(\alpha\beta)} =\half(M^{\alpha\beta}+ M^{\beta\alpha})$ and $M^{[\alpha\beta]} =\half(M^{\alpha\beta}- M^{\beta\alpha})$.

\subsection{EM tensor sign}

Our stress tensor definition is as given in \eqref{eq:TdefDeriv}.
In Lorentzian signature, this is typically taken so that $T_{00}$ is positive (and so also $T^{00}$). %
Operationally, the Wick rotation involves the identification $\tau_E = i t_M$ such that
\begin{equation}
    i S_M = - S_E,
\end{equation}
and therefore if \cite[(5)]{Stergiou:2022qqj} $S_M = - \half \int \odif{t_M} \odif[d-1]{x}  \,\sqrt{\abs{g}} \phi \WeylOp_{\zeta} \phi$ (where we have used $\sqrt{-g} = \sqrt{\abs{g}}$), then 
\begin{equation}
    S_E = -i S_M = \half \int \odif{ i t_M} \odif[d-1]{x} \,\sqrt{g} \phi \WeylOp_{\zeta} \phi = \half \int \odif[d]{x} \sqrt{g} \phi \WeylOp_{\zeta} \phi.
\end{equation}
In Lorentzian signature we define \cite[(11)]{Stergiou:2022qqj}
\begin{equation}
    T_M^{\mu\nu} \equiv \frac{2}{\sqrt{-g}} \fdv{S_M}{g_{\mu\nu}}, \quad T_{M,\mu\nu} =  - \frac{2}{\sqrt{-g}} \fdv{S_M}{g_{\mu\nu}};
\end{equation}
then, for the EM tensors to look the same, in Euclidean space this must differ by a sign:
\begin{equation}
    T^{\mu\nu} \equiv -\frac{2}{\sqrt{\abs{g}}}\fdv{S_E}{g_{\mu\nu}}, \quad T_{\mu\nu} = +\frac{2}{\sqrt{\abs{g}}}\fdv{S_E}{g^{\mu\nu}}.
\end{equation}
Note that when we vary with respect to the inverse metric, the sign changes.
The functional derivative with respect to a symmetric tensor is defined \href{https://physics.stackexchange.com/questions/405495/matrix-derivative-of-a-matrix-with-constraints}{as usual} (see \href{https://physics.stackexchange.com/questions/149066/variation-of-the-metric-with-respect-to-the-metric}{also}), which ensures that the metrical EM tensor is symmetric by definition.

\section{Correlation functions at separated points}
\label{sec:separatedPoints}

The reader might recall that we said the equation of motion applies only inside correlation functions at separated points. 
This is indeed true -- but how then did we apply it to the equations 
\begin{equation}
\partial_\mu T^{\mu\nu} = -(\partial^\nu \phi) (-\partial^2)^\zeta \phi \underset{?}{\overset{\text{eom}}{=}} 0, \quad \gflat_{\mu\nu} T^{\mu\nu} = -(\dotwo -\zeta) \phi (-\partial^2)^\zeta \phi \underset{?}{\overset{\text{eom}}{=}} 0,
\end{equation}
where the left-hand $\phi$ field is at the same point as the right-hand field?

The short answer is \enquote{normal ordering}; the long answer is that, as ever, our QFT is secretly regularized, and there are terms which we have neglected. 
Being somewhat orthogonal to the main argument, we do not discuss this in the bulk of the text.
Everything we have written above is implicitly normal ordered: \enquote{$\phi \WeylOp_\zeta \phi$} is not equal to the product of $\phi$ and $\WeylOp_\zeta \phi$. 
This is a standard convention, but in this section we make it explicit.

We note that secretly, we require (local) curvature counterterms, to ensure the finitude of the correlator.
\begin{equation}\label{eq:SgWithC}
S^g[\phi] = \half \int_x \phi \WeylOp_\zeta \phi + \int_x c[g].
\end{equation}
From the perspective of the correlators, these yield contact terms, in the sense that they provide corrections to correlators containing metrics at coincident points.
Recall that the path integral is invariant under a change of integration variables $\phi\mapsto \phi+\delta\phi$ (there is no Jacobian):
\begin{align}
    \int \cD{\phi}\,e^{-S^g[\phi]} \cO_1 \cO_2 \cdots =\int \cD{\phi}\,e^{-S^g[\phi]}\left(1+\int_x\delta \phi\left[-\fdv{S^g}{\phi}\cO_1 \cO_2 + \fdv{\cO_1}{\phi}\cO_2 \cdots + \cO_1\fdv{\cO_2}{\phi} \cdots\right]\right).
\end{align}
Hence, the equation of motion holds up to terms of the form $\fdv{\cO_n(x_n)}{\phi(x)} \propto \delta(x-x_n)$.
Writing $\delta(0)$ to indicate the divergence arising from coincident points (somewhat formally -- we make this more rigorous by introducing a regulator in the next section)
\begin{equation}
    \expval{\phi(x) \WeylOp_{\zeta} \phi(x) \cO_1(x_1) \cdots} = \frac{1}{\sqrt{g(x)}}\expval{\delta(x-x) \cO_1(x_1) \cdots}=\expval{\frac{\delta(0)}{\sqrt{g}} \cO_1(x_1) \cdots},
\end{equation}
where we take $x_i \neq x$.
Normal ordering amounts to 
\begin{align}
    :\phi(x) \WeylOp_{\zeta} \phi(x): &\equiv \phi(x) \WeylOp_{\zeta} \phi(x)-\expval{\phi(x) \WeylOp_{\zeta} \phi(x)}  =\phi(x) \WeylOp_{\zeta} \phi(x)- \frac{\delta(0)}{\sqrt{g}}.
\end{align}
This applies even to the nonlocal cases; hence \eqref{eq:hypersingular} should also be normal-ordered in this way.
Restricting to flat space to avoid any Weyl anomalies, we find that
\begin{equation}
    \expval{:\phi(x) \WeylOp_{\zeta} \phi(x): \cO_1(x_1) \cO_2(x_2)\cdots} = 0;
\end{equation}
Hence, the correct version of our demand \ref{prop3} is
\begin{equation}
    T\indices{^\mu_\mu} \equiv -(\dotwo -\zeta) :\phi(x) \WeylOp_{\zeta} \phi(x):
\end{equation}
the trace of the stress tensor with this particular value of the cosmological constant $\propto \delta(0)$ vanishes as an operator in flat space away from coincident points ($x\neq x_i$).
Since in flat space the only available symmetric tensor that is independent of $\phi$ is $g_{\mu\nu}$, we conclude that (again formally)
\begin{equation}
    T^{\mu\nu} = T^{\mu\nu}_\text{naive} + \frac{\dotwo -\zeta}{d} \gflat^{\mu\nu} \delta(0), \quad \expval{T\indices{^\mu_\mu}(x)}\rvert_{g=\delta} = -(\dotwo -\zeta) [\expval{\phi(x) \WeylOp_{\zeta}\phi(x)}  - \delta(0)]=0,
\end{equation}
where $ T^{\mu\nu}_\text{naive}$ is as defined in \eqref{eq:fullT}.
\ref{prop1} is of course unchanged; \ref{prop2} is as well, because the derivative of the metric is zero; and so is \ref{prop4}, because the identity operator vanishes under the action of $\hat{K}_b$.
However, we still want this stress tensor to be the one inserted when we vary with respect to the metric. 
Performing the variation with respect to $g_{\mu\nu}$, we find that 
\begin{align}
    T^{\mu\nu}_W &= -\frac{2}{\sqrt{g}}\fdv{S^g}{g_{\mu\nu}} =  T^{\mu\nu}_\text{naive}  -\frac{2}{\sqrt{g}}\fdv{\int \odif[d]{x}\sqrt{g} c[g]}{g_{\mu\nu}}\\
    &=  T^{\mu\nu}_{\text{naive}} - \left(g^{\mu\nu}c[g] + 2 \underbrace{\fdv{\int \odif[d]{x} \, c[g]}{g_{\mu\nu}}}_{c'[g]^{\mu\nu}} \right)
\end{align}
In flat space, $c[g=\gflat] = \Lambda$ only, so $c'[g=\gflat]^{\mu\nu} = 0$.
We define our preferred value of $c[g]$ to be the one that makes $\expval{T\indices{^\mu_\mu}}_{g=\gflat} =\expval{T\indices{_W^\mu_\mu}}_{g=\gflat} = 0$.
Thus, we need to define the divergent
\begin{equation}
    c[g] \equiv -\frac{\dotwo -\zeta}{d} \delta(0).
\end{equation}
If we do so, then when varying \eqref{eq:SgWithC} we find that indeed our correct $T^{\mu\nu}$ is the response of the action to variation, i.e.
\begin{equation}
    T^{\mu\nu} = -\frac{2}{\sqrt{g}}\fdv{S^g}{g_{\mu\nu}} \rvert_{g=\gflat} = T^{\mu\nu}_\text{naive} + \frac{\dotwo -\zeta}{d} \gflat^{\mu\nu} \delta(0), 
\end{equation}
as desired, which satisfies all of our conditions:
\begin{equation}
\begin{aligned}
    T^{[\mu\nu]}&=0; \quad \partial_\mu T^{\mu\nu} = -:(\partial^\nu \phi) (-\partial^2)^\zeta \phi: \overset{\text{eom}}{=} 0;\\
    \gflat_{\mu\nu} T^{\mu\nu} &=-(\dotwo -\zeta) :\phi (-\partial^2)^\zeta \phi: \overset{\text{eom}}{=} 0;\quad \hat{K}_\rho T^{\mu\nu}(0) =0.
\end{aligned}
\end{equation}
We end this section by noting that higher-order curvature terms would generically be required in the action to act as contact terms cancelling UV divergences in higher-point correlation functions of EM tensors.

\subsection{Point-splitting regularization}

To do this more properly, we must introduce a regulator.
One such is point-splitting.
\begin{equation}
    S_\epsilon = \half \int_x \phi(x) \WeylOp_{\zeta,\epsilon} \phi(x) + \int_x c_\epsilon[g],
\end{equation}
where $c_\epsilon[g]$ is a collection of curvature counterterms that depend explicitly on the metric and the regulator. 
We shall assume for the moment that we have $c_\epsilon[g] = \Lambda + O((\partial g)^2) \neq 0$.
For example, we could take $\phi(x) \WeylOp_{\zeta,\epsilon} \phi(x) =\phi(x+\epsilon/2) \WeylOp_{\zeta,\epsilon} \phi(x-\epsilon/2)$, (usually symmetrising over $\epsilon$ directions).
We then normal order by subtracting the divergent value of $\expval{\phi(x+\epsilon/2) \WeylOp_{\zeta,\epsilon} \phi(x-\epsilon/2)}$ from the operator $\phi(x+\epsilon/2) \WeylOp_{\zeta,\epsilon} \phi(x-\epsilon/2)$.
One potential complication here is the conservation condition,
\begin{equation}
    \expval{\nabla_\mu [T_{\mathrm{naive}, \epsilon}]\indices{^{\mu\nu}}}_{g=\gflat} = \expval{(-\nabla^\nu \phi)\WeylOp_{\zeta,\epsilon} \phi} = -\partial_y^\nu \delta_\epsilon(y-x)\rvert_{y=x}
\end{equation}
Assuming a symmetric point-split regularization of the delta function $\delta_\epsilon$, we use the fact that $\delta_\epsilon'$ is odd, so $\delta_\epsilon'(0)=0$, and this vanishes. 
This is good, because otherwise there would exist no flat-space term to cancel this divergence, because $\partial_\mu( \gflat^{\mu\nu}c[\gflat]) = 0$.

\section{Gegenbauer polynomials}\label{app:Gegenbauer}

The Gegenbauer polynomials, also called ultraspherical polynomials, are conventionally defined in terms of their generating function
\begin{equation}\label{eq:GegenbauerGenerator}
    {\frac {1}{(1-2xt+t^{2})^{\alpha }}}=\sum _{n=0}^{\infty }C_{n}^{(\alpha )}(x)t^{n},\qquad |x|<1,\; |t|\leq 1,\;\alpha >0,
\end{equation}
which manifestly also defines them for noninteger $\alpha$.
As discussed in the main text, for integer $n$ they are polynomials of degree $n$, which are even for $n$ even and odd for $n$ odd.
However, the extension to noninteger $n$ is not as obvious.

Additionally, they are particular solutions of the Gegenbauer differential equation
\begin{equation}
   (1-x^{2})y''-(2\alpha +1)xy'+n(n+2\alpha )y=0.
\end{equation}
We can write the solutions to this as (terminating) hypergeometric functions,
\begin{equation}
    C_{n}^{(\alpha )}(x) \equiv {\frac {(2\alpha )_{n}}{n!}}\,_{2}F_{1}\left(-n,2\alpha +n;\alpha +{\frac {1}{2}};{\frac {1-x}{2}}\right). \label{eq:GChypergeometric}
\end{equation} %
These are of course only usually defined for integer $n$, but by analytic continuation, \eqref{eq:GChypergeometric} serves to define the \href{https://dlmf.nist.gov/15.9#iii}{\textbf{Gegenbauer functions}} for arbitrary $n$; and they still solve the Gegenbauer DE.
Observe also that the Gegenbauer function vanishes identically for negative integer $n$.
Differentiating either of \eqref{eq:GegenbauerGenerator} or \eqref{eq:GChypergeometric} with respect to $x$ gives
\begin{equation}\label{eq:GegenbauerDiff}
    \pdv[m]{}{x} C^{(\alpha)}_n(x) = 2^m \Pochhammer{\alpha}{m} C^{(\alpha+m)}_{n-m}(x).
\end{equation}
Plugging this into the recursion relation, we find
\begin{equation}
    4k(k+1)(1-x^{2})C^{(k+2)}_{\zeta-k-2} -2k(2k+1)x C^{(k+1)}_{\zeta-k-1} +(\zeta-k)(\zeta+k) C_{\zeta-k}^{(k)}=0. \label{eq:recFormOfGDE}
\end{equation}
We can also use it to expand
\begin{equation}\begin{aligned}\label{eq:GegenbauerExpandedInApp}
C_{n}^{(\alpha)}(\cos\theta + y) &= \sum_{k=0}^\infty \frac{y^k}{k!} \odv[k]{}{w} C^{(\alpha)}_{n}(w) \rvert_{w=\cos\theta} = \sum_{k=0}^\infty \frac{\Pochhammer{\alpha}{k}}{k!} (2y)^k C^{(\alpha+k)}_{n-k}(\cos\theta),
\end{aligned}\end{equation}
which obviously converges absolutely for integer $n$ because the sum is finite. 
For noninteger $n$, the radius of convergence is the distance from $\cos\theta$ to the nearest branch cut.
We know that $C^{(\alpha)}_{n}(x)$ is analytic in the $x$-plane cut along $(-\infty,-1]$ \cite[below (1.4)]{durandExpansionFormulasAddition1976}.
This agrees with the branch cut in \eqref{eq:GChypergeometric} \cite[(15.2.1)]{NIST:DLMF} along $\frac{1-x}{2} \in (1,\infty)$. 
Thus, the radius of convergence is $\abs{y}< 1+\cos\theta$.

It is worth noting here that Gegenbauer polynomials $C^{(\tfrac{d}{2}-1)}_J$ are often seen in the context of rank-$J$ symmetric traceless tensors $n_{\mu_1}\cdots n_{\mu_J} - \text{traces}$, e.g. \cite{Kotikov:1995cw} or \cite[(2.12)]{Hogervorst:2013sma}. 
However, in our case the parameter of the polynomial is an integer $k$, rather than being related to the spacetime. %

\section{\texorpdfstring{$\hat{K}_b$}{\^K\_b} as a derivative operator} \label{app:Kbhat}

The full operators that implement $\hat{K}_\rho$ acting on bilinears are collected here, as their precise form is slightly adjacent to the main discussion. 
As noted in \cref{footnote:Knorm}, the normalization of $\hat{K}_b$ that we use is unconventional, being adapted for momentum space.
In the following, for readability, we separate the action of $\hat{K}$ on each of the different tensor structures.
Hence,
\begin{equation}
\hat{K}_b A_{\mu\nu} f(P,Q,R) = k_{\mu\nu}(p,q) \equiv [\hat{K}_b^A f(P,Q,R)]_{\mu\nu},
\end{equation}
for an arbitrary spin-$2$ structure $A_{\mu\nu} \in \{\gflat_{\mu\nu},\, p_\mu p_\nu, \, q_\mu q_\nu, \, p_\mu q_\nu,\, q_\mu p_\nu\}$,
is taken to be the compact notation for
\begin{equation}
    b^\rho \hat{K}_\rho \int_{p,q} e^{i (p+q)\cdot x} \phi(p) A_{\mu\nu} f(P,Q,R) \phi(q) \Big\rvert_{x=0} = \int_{p,q} e^{i (p+q)\cdot x} \phi(p)k_{\mu\nu}(p,q) \phi(q) \Big\rvert_{x=0},
\end{equation}
where $\phi$ has scaling dimension $\Delta \equiv \dotwo -\zeta$, and so $k_{\mu\nu}$ will depend explicitly on $\zeta$. 
We find:
\begin{equation}
\begin{aligned}\label{eq:KbonMetric}
    &\hat{K}_b \,\gflat_{\mu\nu} f(P,Q,R)\\
    &= \gflat_{\mu  \nu } \left[b \cdot  q \left(\left(\dotwo -\zeta \right) f^{\{0,0,1\}}+\left(R-\tfrac{P}{2}\right) f^{\{0,0,2\}} +2 Q f^{\{0,2,0\}}+2 P f^{\{1,0,1\}} -2 (\zeta -1) f^{\{0,1,0\}}\right)\right.\\
    &\left.+b \cdot  p \left(\left(\dotwo-\zeta \right) f^{\{0,0,1\}}+\left(R-\tfrac{Q}{2}\right) f^{\{0,0,2\}} + 2 P f^{\{2,0,0\}} +2 Q f^{\{0,1,1\}}-2 (\zeta -1) f^{\{1,0,0\}}\right)\right],
    \end{aligned}
\end{equation}
\begin{equation}
\begin{aligned}
    &\hat{K}_b \,p_{\mu}p_{\nu} f(P,Q,R)\\
&=(b^{\mu } p^{\nu }+b^{\nu } p^{\mu }) \left[R f^{\{0,0,1\}}+2 P f^{\{1,0,0\}} +\left(\dotwo-\zeta +1\right) f\right]\\
&-(p^{\mu } q^{\nu }+p^{\nu } q^{\mu }) (b \cdot  p) f^{\{0,0,1\}} - \gflat^{\mu  \nu }  (b \cdot  p)f\\
&+ p^{\mu } p^{\nu } \left[b \cdot  q \left(\left(\dotwo -\zeta +2\right) f^{\{0,0,1\}}+\left(R-\tfrac{P}{2}\right) f^{\{0,0,2\}}+2 Q f^{\{0,2,0\}} +2 P f^{\{1,0,1\}} -2 (\zeta -1) f^{\{0,1,0\}}\right)\right.\\
&\left.+b \cdot  p \left(\left(\dotwo -\zeta \right) f^{\{0,0,1\}}+\left(R-\tfrac{Q}{2}\right) f^{\{0,0,2\}}+2 P f^{\{2,0,0\}} +2 Q f^{\{0,1,1\}} -2 (\zeta -1) f^{\{1,0,0\}}\right)\right],
\end{aligned}
\end{equation}
\begin{equation}
\begin{aligned}
    &\hat{K}_b \,p_{\mu}q_{\nu} f(P,Q,R)\\
    &=b^{\nu } p^{\mu } \left(R f^{\{0,0,1\}}+2Q f^{\{0,1,0\}} +\left(\dotwo-\zeta \right) f\right)+b^{\mu } q^{\nu } \left(R f^{\{0,0,1\}}+2P f^{\{1,0,0\}} +\left(\dotwo-\zeta \right) f\right)\\
    &- \left(p^{\mu } p^{\nu } b \cdot q +q^{\mu } q^{\nu } b \cdot p \right) f^{\{0,0,1\}}\\
    & + p^{\mu } q^{\nu } \left[(b \cdot  q) \left(\left(\dotwo-\zeta +1\right) f^{\{0,0,1\}}+\left(R-\tfrac{P}{2}\right) f^{\{0,0,2\}}+2Q f^{\{0,2,0\}} +2P f^{\{1,0,1\}} -2 (\zeta -1) f^{\{0,1,0\}}\right)\right.\\
    &\left.+(b \cdot  p) \left(\left(\dotwo-\zeta +1\right) f^{\{0,0,1\}}+\left(R-\tfrac{Q}{2}\right) f^{\{0,0,2\}}+2P f^{\{2,0,0\}} +2Q f^{\{0,1,1\}} -2 (\zeta -1) f^{\{1,0,0\}}\right)\right],
\end{aligned}
\end{equation}
using the shorthand $f^{\{a,b,c\}} = \partial^a_P \partial^b_Q \partial^c_R f(P,Q,R)$.
The analogous results for $q_\mu q_\nu$ and $p_\nu q_\mu$ are obtained by switching $p$ and $q$ in the expressions above (including in the derivatives).

\section{Putting \TtextOrPDF{} in Osborn-Vacca-Zanusso form}\label{app:OSform}

The Gegenbauer expansions \eqref{eq:diffPowsAsGegenbauers} and \eqref{eq:sumPowsAsGegenbauers} let us write \eqref{eq:fullMomSpace} as an infinite sum,
\begin{equation}
\begin{aligned}
    4S^{\mu\nu}(p,q)&= -2 \gflat^{\mu\nu}  \frac{p^{2\zeta}+q^{2\zeta}}{2} \\
    &-\sum_{k=2}^\infty (s^2)^{k-3} \cD_{\mathrm{TT}}^{\mu\nu\rho\sigma}  f_k \di_\rho \di_\sigma \mg^{\zeta -k} C_{\zeta-k}^{(k)}(c)\\
    &+\sum_{k=1}^\infty (s^2)^{k-2} (s^2 \gflat^{\mu(\rho} \gflat^{\sigma)\nu}- \cP^{\mu\nu} \gflat^{\rho\sigma})\di_\rho \di_\sigma \mg^{\zeta -k} C_{\zeta-k}^{(k)}(c)\\
    &+ \sum_{k=0}^\infty(s^2)^{k-1} \frac{2\zeta^2 \cP^{\mu\nu}}{k} \mg^{\zeta -k} C_{\zeta-k}^{(k)}(c),
    \end{aligned}
\end{equation}
where $f_k\equiv \frac{k-1}{k-1-\dotwo}$.
Separating out the first terms of each of these, we find
\begin{equation}
\begin{aligned}
    4S^{\mu\nu}(p,q)&= -2 \gflat^{\mu\nu}  \frac{p^{2\zeta}+q^{2\zeta}}{2} \\
    &-\sum_{k=3}^\infty (s^2)^{k-1} \frac{\cD_{\mathrm{TT}}^{\mu\nu\rho\sigma}}{s^4}  f_k \di_\rho \di_\sigma \mg^{\zeta -k} C_{\zeta-k}^{(k)}(c)\\
    &+\sum_{k=2}^\infty (s^2)^{k-1} (\gflat^{\mu(\rho} \gflat^{\sigma)\nu}- \frac{\cP^{\mu\nu}}{s^2} \gflat^{\rho\sigma})\di_\rho \di_\sigma \mg^{\zeta -k} C_{\zeta-k}^{(k)}(c)\\
    &+ \sum_{k=1}^\infty(s^2)^{k-1} \frac{2\zeta^2 \cP^{\mu\nu}}{k} \mg^{\zeta -k} C_{\zeta-k}^{(k)}(c)\\
        &-\frac{\cD_{\mathrm{TT}}^{\mu\nu\rho\sigma}}{s^2}  f_2 \di_\rho \di_\sigma \mg^{\zeta -2} C_{\zeta-2}^{(2)}(c)\\
    &+ (\gflat^{\mu(\rho} \gflat^{\sigma)\nu}- \frac{\cP^{\mu\nu}}{s^2} \gflat^{\rho\sigma})\di_\rho \di_\sigma \mg^{\zeta -1} C_{\zeta-1}^{(1)}(c)\\
    &+ (s^2)^{-1} 2\zeta^2 \cP^{\mu\nu} \mg^{\zeta} \tilde{C}_{\zeta}^{(0)}(c)
    \end{aligned}
\end{equation}
We can rewrite just these last three terms by using the operator $\cD$ mentioned above, defined in \cite[(18)]{Stergiou:2022qqj}: $\cD_{\mathrm{TT}}^{\mu\nu\rho\sigma} f_2 = -2(s^2 \cD^{\mu\nu\rho\sigma}-\cP^{\mu\nu} s^\rho s^\sigma)$, so
\begin{equation}
\begin{aligned}
        &2(\cD^{\mu\nu\rho \sigma} - \frac{\cP^{\mu\nu}s^\rho s^\sigma}{s^2})\di_\rho \di_\sigma \mg^{\zeta -2} C_{\zeta-2}^{(2)}(c)\\
    &+ (\gflat^{\mu(\rho} \gflat^{\sigma)\nu}- \frac{\cP^{\mu\nu}}{s^2} \gflat^{\rho\sigma})\di_\rho \di_\sigma \mg^{\zeta -1} C_{\zeta-1}^{(1)}(c)\\
    &+ (s^2)^{-1} 2\zeta^2 \cP^{\mu\nu} \mg^{\zeta} \tilde{C}_{\zeta}^{(0)}(c)\\
    =&2\cD^{\mu\nu\rho \sigma} \di_\rho \di_\sigma \mg^{\zeta -2} C_{\zeta-2}^{(2)}(c) + \di^\mu \di^\nu \mg^{\zeta -1} C_{\zeta-1}^{(1)}(c)\\
    &+\frac{\cP^{\mu\nu}}{s^2} \left[ -2 (s\cdot \di)^2 \mg^{\zeta -2} C_{\zeta-2}^{(2)}(c) - \di^2 \mg^{\zeta -1} C_{\zeta-1}^{(1)}(c) +  2\zeta^2 \mg^{\zeta} \tilde{C}_{\zeta}^{(0)}(c)\right].
    \end{aligned}
\end{equation}
Now, $(s\cdot \di)^2 =s^2 \di^2 -4 \mg^2(1-\cos^2 \theta)$, and $\di^2 = s^2 + 4 \mg \cos\theta$, so this further simplifies to 
\begin{equation}
\begin{aligned}
    =&2\cD^{\mu\nu\rho \sigma} \di_\rho \di_\sigma \mg^{\zeta -2} C_{\zeta-2}^{(2)}(c) + \di^\mu \di^\nu \mg^{\zeta -1} C_{\zeta-1}^{(1)}(c)\\
    &-\cP^{\mu\nu} \left[2  \di^2 \mg^{\zeta -2} C_{\zeta-2}^{(2)}(c) + \mg^{\zeta -1} C_{\zeta-1}^{(1)}(c)\right]\\
     &+\frac{2\cP^{\mu\nu}}{s^2} \mg^{\zeta} \left[4 (1-c^2) C_{\zeta-2}^{(2)}(c) - 2 c C_{\zeta-1}^{(1)}(c) +  \zeta^2  \tilde{C}_{\zeta}^{(0)}(c)\right].
    \end{aligned}
\end{equation}
This last bracket is exactly the $k=0$ limit of the recursion form of the Gegenbauer differential equation \eqref{eq:recFormOfGDE}, and so it vanishes.

We conclude that the overall OSV-type form is 
\begin{equation}
    \begin{aligned}
         4S^{\mu\nu}(p,q)&= -2\gflat^{\mu\nu}  \frac{p^{2\zeta}+q^{2\zeta}}{2}-\cD_{\mathrm{TT}}^{\mu\nu\rho\sigma} \sum_{k=3}^\infty (s^2)^{k-3} f_k \di_\rho \di_\sigma \mg^{\zeta -k} C_{\zeta-k}^{(k)}(c)\\
    &+(s^2 \gflat^{\mu(\rho} \gflat^{\sigma)\nu}- \cP^{\mu\nu} \gflat^{\rho\sigma}) \sum_{k=2}^\infty (s^2)^{k-2} \di_\rho \di_\sigma \mg^{\zeta -k} C_{\zeta-k}^{(k)}(c)\\
    &+ 2\zeta^2 \cP^{\mu\nu} \sum_{k=1}^\infty(s^2)^{k-1} \frac{1}{k} \mg^{\zeta -k} C_{\zeta-k}^{(k)}(c)\\
    &+2\cD^{\mu\nu\rho \sigma} \di_\rho \di_\sigma \mg^{\zeta -2} C_{\zeta-2}^{(2)}(c) + \di^\mu \di^\nu \mg^{\zeta -1} C_{\zeta-1}^{(1)}(c)\\
    &-\cP^{\mu\nu} \left[2  \di^2 \mg^{\zeta -2} C_{\zeta-2}^{(2)}(c) + \mg^{\zeta -1} C_{\zeta-1}^{(1)}(c)\right],
    \end{aligned}
\end{equation}
which is manifestly local for integer $\zeta$, confirming again the results of \cref{sec:locality}.

\section{Identities for the GJMS matching}\label{app:GJMSidentities}
In this appendix we prove certain identities required to match our EM tensors to those arising from the GJMS operators. 
$\zeta$ is taken to be an integer throughout, as in \cref{sec:Juhl}.

\subsection{Inductive proof of the \texorpdfstring{$\tilde{m}$}{m\~} identity} \label{app:mtildeProof}

  We want to prove \eqref{eq:mtildesum}: that for $j \ge 0$
    \begin{align}\label{eq:MtildewantToInduct}
        \sum_{A \in \cC(j)} \tilde{m}_{A,k} = (-1)^j \frac{k!}{j!(k+j)!},
    \end{align}
    which is evidently true for $j=0,1$ (using $\cC(0)=\{()\}$ and $\tilde{m}_{(),k}=1$).
    We note that the right-hand side admits a partial fraction decomposition in $k$, in the form $\sum_{i} \frac{1}{k+i} f(i)$. 
    This motivates an inductive proof. 
    First, we observe that if we extract the last element $m$ of the ordered list $A = (B,m)$ in \eqref{eq:mtildeDef}, then
\begin{align}
   \tilde{m}_{A,k} =\tilde{m}_{(B,m),k} &= \frac{-1}{m+k} \frac{1}{m!(m-1)!} \tilde{m}_{B,m}.
\end{align}
    Now, assume that \eqref{eq:MtildewantToInduct} is true for $j=0,1,\cdots,n-1$.
    We want to prove it for $j=n$.
    Recalling that $m$ can take any value from $1$ to $|A|=n$,
    \begin{align}
       \sum_{A \in \cC(n)} \tilde{m}_{A,k} &= \sum_{m=1}^{n} \sum_{B \in \cC(n-m)} \tilde{m}_{(B,m),k} = \sum_{m=1}^{n} \frac{-1}{m+k} \frac{1}{m!(m-1)!} \sum_{B \in \cC(n-m)} \tilde{m}_{B,m}\\
       &=\frac{(-1)^n}{n!}\sum_{m=1}^{n} \frac{(-1)^{1-m}}{(m-1)! (m+k) (n-m)!} = (-1)^n \frac{k!}{n!(k+n)!},
    \end{align}
    which completes the proof.

  Incidentally, we note that there is a slightly more natural form for $a_{\zeta,k}^{(j)}$, which is just multinomial coefficients and beta functions.
  By comparison with the equation for $m_I$, we find that the $\tilde{m}_{A,k}$ sum becomes almost exactly an Euler beta function:
  \begin{equation}
     \sum_{A \in \cC(j)}  \abs{A}!(\abs{A}-1)!\, \tilde{m}_{A,k} = (-1)^j B(j,k+1).
  \end{equation}
Hence, we find
\begin{equation}
  \begin{aligned}
      a_{\zeta,k}^{(j)} &= -\frac{\zeta! (\zeta-1)!}{k!(k-1)!j(j-1)!(\zeta-k-j)!(\zeta-k-j-1)!} \\
      & \times \sum_{A \in \cC(j)} j! (j-1)! \tilde{m}_{A,k} \sum_{B \in \cC(\zeta-k-j)} (\zeta-k-j)!(\zeta-k-j-1)!\tilde{m}_{B,k}\\
      &= -(\zeta-2)(\zeta-1)(-1)^{\zeta-k} \binom{\zeta}{k,j,\zeta-k-j} \binom{\zeta-3}{k-1,j-1,\zeta-k-j-1}\\
      &\quad \times B(j,k+1)B(\zeta-j-k,k+1).
  \end{aligned}
\end{equation}

\subsection{Proof of cancellations in \texorpdfstring{$E-F$}{E-F}}\label{app:EmFidentities}

Use again $P=p^2$ and $Q=q^2$. 
We treat the two identities in \eqref{eq:perfectCancellation} together, defining $\sigma\in\{+1,-1\}$ and $\alpha_k^{(-1)}=1,\quad \alpha_k^{(+1)}=\frac{k}{2}$.
Then, using $f_{\zeta,\zeta}=-1$, we need to prove that
\begin{equation}
e_k^{(\sigma)}(P,Q):=\alpha_k^{(\sigma)}\bigl(P^k+\sigma Q^k\bigr), \quad S_\sigma \equiv \sum_{k=1}^{\zeta} f_{\zeta,k} e_k^{(\sigma)}=0,
\end{equation}
for $\zeta\ge 2$ when $\sigma=-1$, and for $\zeta\ge 3$ when $\sigma=+1$. 
Since
\begin{equation}
   S_\sigma= %
 \sum_{k=1}^{\zeta} e_k^{(\sigma)}\sum_{j=0}^{\zeta-k} a_{\zeta,k}^{(j)} P^{j} Q^{\zeta-k-j} = \sum_{k=1}^{\zeta} e_k^{(\sigma)}\sum_{j=0}^{\zeta-k} (-1)^{\zeta-k-1} \frac{k}{\zeta} \binom{\zeta}{k+j}\binom{\zeta}{j}P^{j} Q^{\zeta-k-j}
\end{equation}
is a homogeneous polynomial of degree $\zeta$, we can freely set $Q=1$ WLOG, so
\begin{equation}
     = \frac{(-1)^{\zeta-1}}{\zeta}\sum_{k=1}^{\zeta} \sum_{j=0}^{\zeta-k} \alpha_k^{(\sigma)} (-1)^{k} k \binom{\zeta}{k+j}\binom{\zeta}{j}P^{j}\bigl(P^k+\sigma\bigr).
\end{equation}
Consider the coefficient of $P^m$ in this sum,
\begin{align}
[P^m]S_\sigma =\frac{(-1)^{\zeta-1}}{\zeta}\binom{\zeta}{m} \left[\sum_{k=1}^{m} \alpha_k^{(\sigma)} (-1)^{k} k \binom{\zeta}{m-k}  +\sigma \sum_{k=1}^{\zeta-m} \alpha_k^{(\sigma)} (-1)^{k} k\binom{\zeta}{k+m}\right].
\end{align}
We can show this to vanish by rewriting the sums to use $p=m-k$ in the first and $p=m+k$ in the second.
The two sums in square brackets become
\begin{equation}
   \sum_{p=0}^{m} \alpha_{m-p}^{(\sigma)} (-1)^{m-p} (m-p) \binom{\zeta}{p}  +\sigma \sum_{p=m+1}^{\zeta} \alpha_{p-m}^{(\sigma)} (-1)^{p-m} (p-m)\binom{\zeta}{p},
\end{equation}
where we extended the first sum to $p=m$ (i.e. $k=0$) because the summand vanishes there.
Because $\alpha_{m-p}^{(\sigma)} (m-p) = \sigma \alpha_{p-m}^{(\sigma)} (p-m)$ for both values of $\sigma$, this is the single sum
\begin{equation}
    \sum_{p=0}^{\zeta} \alpha_{m-p}^{(\sigma)} \binom{\zeta}{p} (-1)^{m-p} (m-p)
\end{equation}
which vanishes for $\sigma=-1,+1$ if each of 
\begin{equation}
\sum_{p=0}^{\zeta} (-1)^{p} \binom{\zeta}{p}  (m-p), \quad \sum_{p=0}^{\zeta} (-1)^{p} \binom{\zeta}{p}  (m-p)^2
\end{equation} 
 vanish. 
As required, they vanish for $\zeta\ge 2$ and $\zeta \ge 3$ respectively, using the standard fact\footnote{The LHS is the $\zeta$-th finite difference of $\pi$ which decreases the order of a polynomial by $\zeta$ -- hence it must vanish for $\deg \pi < \zeta$. 
 This also follows from $(x\odv{}{x})^k (1-x)^\zeta = \sum_{p=0}^\zeta (-1)^p  \binom{\zeta}{p} p^k x^p$ and setting $x=1$.} that for any polynomial $\pi$ in $p$ with $\deg \pi<\zeta$,
\begin{equation}\label{eq:finDif}
\sum_{p=0}^{\zeta}(-1)^p\binom{\zeta}{p}\pi(p)=0.
\end{equation}

\subsubsection{The Gegenbauer identity}\label{app:remnantCproof}

It remains to prove \eqref{eq:remnant}, which we rewrite as
\begin{equation}\label{eq:wantToShowGegenbauerIdentity}
    S_C \equiv \sum_{k=1}^{\zeta} f_{\zeta,k} \, \mg^{k-m} C_{k-m}^{(m)}(c) \overset{?}{=} (-1)^{\zeta-m-1} \binom{\zeta-1}{m-1} (s^2)^{\zeta-m}.
\end{equation}
Defining $t^2 = P/Q$, we can use $f_{\zeta,0}=0$ to extend the sum to include $k=0$,
\begin{align*}
    S_C=Q^{\zeta-m} t^{-m}\sum_{k=0}^{\zeta} C_{k-m}^{(m)}(c) \sum_{j=0}^{\zeta-k} (-1)^{\zeta-k-1} \frac{k}{\zeta} \binom{\zeta}{k+j}\binom{\zeta}{j} t^{2j+k}
\end{align*}
Since $C_{k-m}^{(m)} =0$ for $k < m \le \zeta$, we change to $k=m+\ell$, giving
\begin{align*}
    S_C=Q^{\zeta-m}\sum_{\ell=0}^{\zeta-m} C_{\ell}^{(m)}(c)  (-1)^{\zeta-m-\ell-1} \frac{m+\ell}{\zeta} \sum_{j=0}^{\zeta-m-\ell}\binom{\zeta}{m+\ell+j}\binom{\zeta}{j} t^{2j+\ell}.
\end{align*}

Recall $s^2 = P+Q-2Mc$, so $(s^2)^{\zeta-m} = Q^{\zeta-m}(1- 2 t c +t^2)^{\zeta-m}$. 
Then we can consider the $t^n$ coefficient of each side. First,
\begin{align}\label{eq:LHSinapp}
    \frac{[t^n] S_C}{(-1)^{\zeta-m-1}Q^{\zeta-m}} &=\frac{(-1)^n}{2\zeta}\sum_{k=0}^{\floor{n/2}} \bigl(m+n-2k\bigr) C_{n-2k}^{(m)}(c) \binom{\zeta}{k}\binom{\zeta}{m+n-k}.
\end{align}
Then in the RHS of \eqref{eq:wantToShowGegenbauerIdentity}, using the Gegenbauer generating function \eqref{eq:GegenbauerGenerator}, we find
\begin{equation}
 [t^n]\binom{\zeta-1}{m-1} (1-2tc +t^2)^{\zeta-m} =\binom{\zeta-1}{m-1} C^{(m-\zeta)}_n(c),
\end{equation}
but this is identical to \eqref{eq:LHSinapp} by the standard Gegenbauer change-of-parameter formula \cite[(18.18.16)]{NIST:DLMF}, which proves \eqref{eq:wantToShowGegenbauerIdentity}.

\section{Generating code}\label{app:code}

We can write the generator of $S^{\mu\nu}(p,q)$ in dimension \lstinline|D| as follows, using \texttt{FeynCalc} \cite{Shtabovenko:2020gxv,Shtabovenko:2016sxi,Mertig:1990an}:
\begin{lstlisting}[language=Mathematica,numbers=none]
<<FeynCalc`;
Pmn=-1/(D-1) (FVD[s,m]FVD[s,n]-MTD[m,n]s2);
Dmnuv=-1/(D-2) (MTD[m,u]FVD[s,v]FVD[s,n]+MTD[n,u]FVD[s,v]FVD[s,m]-MTD[m,u]MTD[v,n]s2-MTD[m,n]FVD[s,u]FVD[s,v])-MTD[u,v]/(D-2) Pmn//FCSymmetrize[#,{u,v}]&;
DTTmnuv=(D-2)(Dmnuv s2-Pmn FVD[s,u]FVD[s,v]);
generateT[z_Integer]:=Module[{},
    {-(1/2 MTD[m,n]-z Pmn/s2)((p2^z+q2^z)/2),
        Sum[(MTD[m,u]MTD[v,n]s2^2-Pmn MTD[u,v] s2-DTTmnuv (k-1)/((k-1)-D/2)) s2^(k-3) 
        (FVD[t,u]FVD[t,v]/4) M^(z-k) GegenbauerC[z-k,k,c],{k,1,z}]}];
subCandM={Rule[c,-pq/Sqrt[p2 q2]], Rule[M,Sqrt[p2 q2]]};
genTfull[z_Integer]:=Contract[Total[generateT[z]]]/.subCandM/.Rule[s2,SPD[s]]/.{Rule[t,p-q], Rule[s,p+q]}/.{Rule[pq,SPD[p,q]], Rule[p2,SPD[p,p]], Rule[q2,SPD[q,q]]}//ExpandScalarProduct//FCE;
  \end{lstlisting}
This is the code for \cref{eq:PmomentumSpace,eq:DQmomentumSpace,eq:fullMomSpace}.
We can then programmatically check: the $\cD_{\mathrm{TT}}$ conditions \eqref{eq:DQconditions}; and the conditions \eqref{eq:Strace} and \eqref{eq:Sderiv} on $S^{\mu\nu}(p,q)$ for $\zeta = 7$. 
  \begin{lstlisting}[language=Mathematica,numbers=none]
Contract[{MTD[m,n],MTD[u,v],FVD[s,m],FVD[s,u]} DTTmnuv/.Rule[s2, SPD[s]]]//Simplify
Contract[{MTD[m,n],FVD[p+q,m],FVD[p+q,n]} genTfull[7]]//ExpandScalarProduct//Simplify
  \end{lstlisting}
By replacing $p \mapsto -i \lptl$ and $q\mapsto -i \rptl$ (and $s\mapsto$ an external $-i\partial$), this code also serves to symbolically generate a position-space $T^{\mu\nu}$ for arbitrary $\zeta$.
To add the extension \eqref{eq:RdefHom}, one simply adds into the list in \lstinline|generateT[]|:
\begin{lstlisting}[language=Mathematica,numbers=none]
    DTTmnuv/s2^2  (-FVD[t, u] FVD[t,v])/4 s2^(D/2) M^(z - D/2 - 1) (1 - c^2)^(-(D+1)/4)
        (C1 LegendreP[z-1/2, (D+1)/2, c] + C2 LegendreQ[z-1/2, (D+1)/2, c])
\end{lstlisting}

\bibliographystyle{JHEP}
{\raggedright  %
\bibliography{references-zot}    %
}              %
\end{document}